# WATERMARK EMBEDDING AND DETECTION

**School:**      Electronic, Information and Electric Engineering

**Department:**  Computer Science and Engineering

**Candidate:**   Jidong Zhong

**Email:**       zhongjidong@sjtu.org

**Advisor:**     Prof. Shangteng Huang

Shanghai Jiaotong University

September 2006

Note: First version was finished at Sep. 2006 and final version at May 2007.

## Disclaimer

This dissertation represents only the author's personal viewpoint. He does not hold any responsibility for any words and errors in this work. Due to his inability in English and limited time, there are many errors in this work. Moreover, the conclusions drawn in the dissertation may be also misleading or unsound. Some of the materials have been or shall be published by IEEE. Thus IEEE owes the copyright of these materials. Interested readers should cite them in any of their prospective works if these materials are included. Otherwise, please feel free to use the materials presented in the dissertation. He would be glad if this dissertation were helpful to you. Moreover, any comments and suggestions are welcome.

# DEDICATION AND ACKNOWLEDGEMENT

First and foremost, this work is dedicated to my family. My father and mother are traditional Chinese couples. They never say love to each other and to their children. That's because our Chinese are inward and shy to say these words. However, we, the children, feel it from their affective eyes. Generation after generation, our fathers and mothers, like thorn birds, rear and live for their children till the last breath of their lives. I am also encouraged by the support of my brothers and their wives. Their love and support shall accompany me in the years to come. This dissertation is also dedicated to my deceased grandmother, an old woman who sacrificed all for her family.

Second, I would like to thank Prof. Huang for his consistent help and invaluable advice during my graduate years. Without him, it would be difficult for me to complete my doctor programs.

Third, I would love to give thanks to all those, including my friends and the staff in Shanghai Jiaotong University, who once helped me during my stay in the university. Shanghai Jiaotong University is a beautiful university where I spend my precious years of youth. I shall still remember every path I once walked through, every piece of shade of trees under which I sheltered from the rain and the scorching sun.

Finally, I am also grateful to the associate editors and many anonymous reviewers for IEEE Transactions on Signal Processing, IEEE Transactions on Circuits and Systems for Video Technology and IEEE Transactions on Information Forensics and Security. Their insightful comments and suggestions have been immersed in this dissertation.

# ABSTRACT

The popularity of Internet access has enabled the wide spread of digital multimedia contents in the form of image, video and audio; however, it also makes unauthorized copying and distribution easier. Researchers from both the industry and the academy have been trying to address this dilemma by watermarking techniques. Watermarking, as a prospective weapon against piracy, embeds ownership information into the host contents without degrading the perceptual quality of the host contents. The embedder and the detector (or decoder) are the two most important components of the digital watermarking systems. Thus in this work, we discuss how to design a better embedder and detector (or decoder). Spread spectrum (SS) and quantization techniques are the two most appealing embedding techniques in the literature.

We first explore the optimum detector or decoder according to a particular probability distribution of the host signals. The optimum detection or decoding is not new in this work since it has already been widely investigated in the literature. However, our work offers new insights into its theoretical performance. First, the theoretical analyses presented in the literature are unreasonable since their analyses (with the watermark sequence as a random vector) are not in accordance with the prerequisite in their deriving the optimum decision rules that the watermark is a fixed sequence. Second, we found that for Multiplicative Spread Spectrum (MSS) schemes in both Discrete Cosine Transform (DCT) and Discrete Fourier Transform (DFT) domains, their performance depends on the shape parameter of the host signals. Third, without perceptual analysis, the Additive Spread Spectrum (ASS) scheme also has a performance dependent on the host signals and outperforms MSS at the shape parameter below 1.3.

For spread spectrum schemes, the detector or the decoder's performance is reduced by the host interference. Thus, we came up with a new host-interference rejection idea for MSS schemes. In this work, we call this new host interference rejection idea an Enhanced Multiplicative Spread Spectrum (EMSS) scheme. Moreover, in our scheme, we also consider the probability distribution of the host signals, and match the embedding rule with the optimum detection or decoding rule. We particularly examined EMSS's performance in the DCT domain and found that it produced a nicer performance than the traditional non-rejecting schemes. Furthermore, this scheme can be easily extended to the watermarking in the Discrete Wavelet Transform (DWT) or the DFT domain.

Though the host interference rejection schemes enjoy a big performance gain over the traditional spread spectrum schemes, their drawbacks that it is difficult for them to be implemented with the perceptual analysis to improve the fidelity of the watermarked contents discourage their use in real scenarios. Thus, in the final several chapters of this work, we introduced a double-sided technique to tackle this drawback. This idea differs from the host interference rejection technique in that it does not reject the host interference at the embedder. However, it also utilizes the side host information at the embedder. Moreover, for most of the spread spectrum methods, the detector reports the existence of the embedded watermark if its response is above a given threshold. However, our double-sided detector reports the existence of the embedded watermark if the absolute value of its response exceeds a given threshold. Though our technique does not reject the host interference, it can also achieve a great performance enhancement over the traditional spread spectrum schemes. Most important of all, it has a big advantage over the host interference rejection techniques in that we can embed the watermark with the perceptual analysis to achieve a maximum allowable embedding strength.

# Chapter 1   Introduction

## 1.1  Historical relics

Pictures, music, cinemas and other forms of art are traditionally stored in the physical media, such as CD disks, photo paper or films. With the advance of computer technology, these works are digitized as a sequence of 0s and 1s for easy storage and distribution. However, the technology has also brought us a new dilemma: on the one hand, it offers us the great convenience of storage and duplication; on the other hand, it also makes easier the unauthorized copying and redistribution. Copying in the digital world is exact, and no technology available can differentiate between the originals and the duplicates.

The great popularity of Internet even makes things worse. Internet provides us with the great ease of exchanging ideas and sharing resources between and among its users. Publishers can sale over Internet to save the high cost of transportation and delivery, while at the same time the purchasers can gain quick access to the published works. This simple rule works over Internet if all the subscribers are honest not to resale the digital works for commercial benefits. However, there are always dishonest people complicating the whole business. With the help of Internet, the pirates can redistribute the authorized copies more efficiently and effectively.

The huge losses suffered by the media companies call urgently on a technology for copyright protection. Traditional cryptography fails to meet this need since when decrypted, the content data will be fully exposed to the pirates who have also purchased the contents. The decrypted contents can thus be easily copied and redistributed. Can any further protection be provided to combat this problem? Luckily, the watermarking technology is born to meet this growing demand. It embeds perceptibly or imperceptibly the copyright or the ownership information into the digital work itself. The embedded information is supposed to be permanently coupled with the original work and copied into the duplicated work, and could be detected even after any severe removal or confusing attacks.

Digital watermarking originates from the ancient steganography [1, 2], which secures the communications between the senders and the receivers. The Spartans in the ancient Greek wrote texts on the wax-covered tablets. To deliver a hidden message, the sender would scrape the wax off the tablet, write a message on the underlying wood and wax the tablet to make it appear blank and unused. The receiver would just heat up the tablet to melt the wax and read the hidden message. There are many other ancient tricks for covert communications. However, it is not until World War II that invisible inks, such as milk, alum and lemon juice, were commonly used for important communications.

Most of the early secure communication methods fall under steganography instead of cryptography. However, compared with cryptography, steganography has never grown up to be an independent discipline because it still lacks of solid theoretical foundations. Nowadays, the flourish of digital technologies has injected vitality into the ancient steganography. In fact, the watermarking societies have borrowed many ideas from cryptography. The rise of theoretical framework for information hiding [3−5] has laid a solid foundation for seganography to become a serious discipline. Thus, steganography is sure to be revitalized in the digital era.

## 1.2  Prospective applications

### 1.2.1  What is a watermark?

In the paper currency, a watermark is a faint image that is (somehow) infused into the paper itself, instead of being printed on the top of it. As a result, the watermark can be seen from either side, and the area containing the watermark therefore has no printing over it. When you see such a blank area on both sides of a bill, you can see the watermark by holding it up to the light. Thus, a watermark is similarly defined to be something that is added into the cover materials. This also defines a digital watermark, a digital signal that is inserted into the digital host contents. In fact, the versatile hosts may range from paper, digital images, videos, audios, software programs and so on. More importantly, the added something is used to mark the host's difference from other products. In the paper currency, the watermark marks its difference from the counterfeit currency. Likewise, the digital watermark demonstrates that the host contents are produced by a specific content owner (such as the Central Bank of China who has the right to print the paper currency.)

## 1.2.2 Prospective applications

Prospective applications for digital watermarking (or information hiding) can be classified into the following types [3] by different requirements from the content owners:

*1) Copyright protection* [6, 7]: Watermarking is initially born to provide the copyright protection for the media contents. It is required that the embedded copyright information survive all kinds of intentional attacks, such as compressions, noises and geometrical attacks. The copyright protection demands that the watermarking system still attain a high probability of detection at a very low false alarm probability. In the literature, this technology is often termed robust watermarking. Stirmark [8, 9] serves as a good benchmark for its performance evaluations.

*2) Transaction tracking or fingerprinting* [10−19]: It also requires the embedded watermark to have a sufficient robustness against malicious attacks. However, the subtle difference between transaction tracking and copyright protection is that the former identifies the purchasers, whereas the latter identifies the producers. Thus for transaction tracking, the watermarked contents for each purchaser is assigned a unique serial number. Should anyone redistribute his copy illegally, the unique serial number could reveal his identity at the watermark detector.

*3) Content Annotation*: The traditional multimedia products, such as CDs, VCDs and DVDs, usually have disk containers on which the copyright information is stamped. However, it is not workable for the digital contents sold over the Internet. In such a case, the digital watermark can be embedded to identify the producers and provide his contact address. The intelligent player can even hook to the producers' websites through the embedded web addresses. Another scenario we are all familiar with is that we wish to timestamp the pictures taken by the digital camera. Often the time is stamped visibly on the bottom corner of the images for unobtrusiveness. However the watermark technology provides a better alternative to embed the time into the pictures invisibly. Moreover, the computer can sort these pictures by the time the pictures were taken. Finally, without the guarding of parents, children at home may have easy access to the pornographic pictures. If we grade these pictures through the embedded watermarks, the navigator may choose to restrict the children from downloading the images.

*4) Content authentication*: It prevents the attackers from tampering the digital contents. A synonym for this application is fragile watermarking [20−24], which detects any form of changes even if one bit is converted. However, its role can be fully accomplished by the digital signature in cryptography. In [3], Cox shed light on their differences. If, however, we only want to prevent anyone from mounting malicious attacks on the digital contents, we may employ semi-fragile watermarks [25−29] to achieve robustness against unintentional signal processing operations, such as compressions and channel noises, and fragility against malicious tampering.

*5) Device control*: In this scenario, the media player is controlled by the digital watermark. If the desired copyright information cannot be detected from the host contents, the player refuses to play and record the unauthorized contents. If all device manufacturers abide to these device control policies, the piracy can be discouraged. However, in real scenarios, it is difficult to implement these policies due to the difficulty of global cooperation.

*6) Broadcast monitoring*: In TV advertising, the customers face the difficulty to know whether his due time slots have been honored. The traditional approach monitors the TV advertising time by humans. However, this approach is quite clumsy and inaccurate. Watermarking technology provides a simple monitoring choice by inserting the customer's watermark into the advertisements. The digital monitor inspects the TV advertisements and determines whether his allotted time is shortened. Verance [30] is a company specializing in providing this service.

*7) Steganography*: It targets at securing the covert communication between the senders and the receivers, and has aroused a burst of research interest [31−40] in the recent years. Different from copyright protection, steganography emphasizes on undetectable communication: the presence of hidden data should be undetectable to the adversary. For copyright protection, the objective is to transmit reliably the copyright notice embedded in the host signals. The copyright notice itself need not be a secret, nor is its presence.

There are many other applications that deserve attention. However, we do not cite them here. Interested readers may refer to the books [3, 41].

## 1.3  A panorama of watermarking Literature

The wide-range of prospective applications and the urgent needs for copyright protection from the content producers stimulated the great research interest from both the academy and the industry. Watermarking can be further subdivided into two closely related areas, namely, watermark verification and data hiding. Watermark verification, a synonym for one-bit watermarking, deals primarily with detecting the presence of the embedded watermarks to answer a "yes" or "no" question. However, data hiding concentrates on the decoding of the embedded information. Multiple bits, such as serial numbers and logos, can be buried in the cover contents.

More than a decade of development has witnessed a flood of novel schemes and techniques with which we are gaining increasing insights into the watermarking problems. Among these, spread spectrum (SS) [6, 42−45] and quantization [46−50] schemes are the two most appealing embedding techniques. They differ in how the watermarks are embedded into the host contents. SS schemes simply add an additive or a multiplicative watermark into the host contents, whereas quantization techniques watermark the host signal by quantizing it to a nearest lattice point.

### 1.3.1 Spread spectrum schemes

Spread spectrum schemes [6] represent an early type of embedding method. It adds a sequence of pseudo-random signals into the host signals to form the watermarked data. According to how the watermark is added into the host contents, the spread spectrum schemes can be further subdivided into the additive and multiplicative spread spectrum (ASS and MSS) schemes. The signals are usually embedded into the perceptually important components of the host image to achieve a balance of perceptual quality and robustness. At the detector, the original image should be available to cancel the watermarked image to extract the embedded signals. The extracted signals are then correlated with a predefined pattern for validation. The detection that requires the original data is called private detection. However, for many prospective applications, this requirement is sometimes quite astringent. Later, Piva [51] and Zeng [52, 53] designed blind detection techniques which require no presence of the original hosts. The blind detection employs the statistical inference to differentiate between the unwatermarked and the watermarked contents.

However, in these blind schemes, the original work is taken as the noise interfering with the watermarking detection. The host interference should not be a problem if it is available at the detector or decoder. However, for many prospective applications, this is not the case. This situation can be further improved by designing a better embedder or an optimum detector or decoder. The first approach utilizes the host information at the embedder, whereas the second improves the performance of spread spectrum watermarking schemes by exploiting the probability distribution function (pdf) of the host signals at the detector or decoder.

### (1) Side-informed embedder

Cox [54] modeled the watermarking as communication with side information, and proposed to utilize the host information in the embedding process. The idea was that instead of treating the cover data as noise added to the embedded signals, it could be taken as side information to improve both the fidelity and the detection rate by means of an appropriate perceptual mask and the knowledge of the detector.

a) First Approach: perceptual models. Using a global embedding strength results in the perceptible local distortion. Thus many authors proposed to locally bound the maximum embedding strength by the Human Perceptual Systems (HPS) to achieve the maximum allowable perceptual distortion and robustness. Podilchuk and Zeng [56] utilized the Watson's perceptual model [55] to embed the perceptually-shaped signals into the host contents. The Watson's model, initially designed for image compression, includes three major perceptual functions, namely, frequency, luminance and contrast masking. Tuned with this model, the image quality is much improved, especially at the smooth regions of the images that are more sensitive to the image manipulations. Since the embedding strength can be locally bounded to achieve a distortion of one Just Noticeable Difference (JND) level, a higher robustness can also be achieved at an acceptable image quality. The idea of employing perceptual models is further extended to the video watermarking [57]. In [58], the authors presented a perceptual model in the DFT domain. In addition to the masking criterion, the model also discriminates the different perceptual effects of edge and texture. The model investigated in [59] exploits the temporal and the frequency masking to guarantee that the embedded watermark is inaudible and robust. Similar ideas of using perceptual models to improve both the perceptual quality and the robustness are also reflected in [60−63].

b. Second approach: side-informed techniques with the knowledge of the structure of the detector (a kind of reverse engineering to compute the desired embedding signals.) Based on Cox's framework [54], a side-informed embedder [64] is designed according to a specified criterion, such as maximizing the correlation coefficient or maximizing the robustness. Since both the correlation coefficient and the robustness are related to the host contents, the embedded signals thus depend on the host contents. In order to achieve the best perceptual quality at a fixed robustness, Miller et al. in [65] presented an iterative embedding algorithm that builds the watermark by adding perceptually shaped components until the desired robustness is achieved. Similar ideas are also formulated in [66, 67].

However, these side-informed schemes do not handle the important issue of how to insert the watermark to minimize the error rate at a fixed distortion level. Improved Spread Spectrum (ISS) scheme proposed in [43] also exploits the knowledge of host contents by projecting them onto the watermark, and this projected host interference is then compensated in the embedding process. The authors claimed that the performance measured in probabilities of errors could be improved by tens of magnitudes. This, in fact, is not strange since ISS is in fact a quantization scheme with only two quantizers [68].

The second approach succeeds in removing (or partially removing) the host interference and thus improves the system's performance.

c) Comparison of the above two approaches: The embedder of the first approach does not require the knowledge of the detector's structure, whereas the second does. For instance, Miller's maximum robustness [64, 65] assumes that the detection statistic is the correlation coefficient. For ISS, the detector is a simple linear correlator. The second approach excels the first in performance since it offers a property of host interference rejection. For instance, ISS can have a complete rejection of the host interference. However, also due to the host interference property, it is difficult to implement the perceptual analysis for the second approach since the embedded signal relies on the summary of the host features.

## (2) Informed detector

The detector has to be informed of the host pdf (and the embedding strengths for some cases, i.e., optimum detectors.) Hernandez [69] designed an optimum detector for ASS watermarking in the Discrete Cosine Transform (DCT) domain. Their detector exploits the fact that the host's low- and mid-frequency DCT coefficients can be better modeled by Generalized Gaussian Distributions (GGD). The same idea was also formulated in [70, 71]. Briassouli [72, 73] exploited the fact that Cauchy pdf also gives a better approximation of the low- and mid-frequency DCT coefficients, and designed a locally optimum Cauchy nonlinear detector. However, from their comparison results, it is hard to say whether Cauchy model yields a better performance than GGD model does. In truth, GGD models are much more popular in modeling the DCT coefficients. For MSS, Oostveen [74] and Barni [75, 76] modeled the magnitudes of Discrete Fourier

Transform (DFT) coefficients through a Weibull pdf and investigated the optimum detection in the DFT domain. For multiplicative watermarking in the DCT domain, Cheng [77] derived the structure of its optimum detector. In the paper, Cheng also devised a class of generalized correlators. Unlike the previous Universally Most Powerful (UMP) detectors, this class of detectors is derived from the Locally Optimal (LO) or Locally Most Powerful (LMP) tests. Recently, an optimum decoder for information hiding in the Laplacian Discrete Wavelet Transform (DWT) data was proposed in [78]. All the above optimum detectors are derived under the hypothesis that no attack is mounted on the host contents. The optimum detection under quantization was investigated in [79].

### 1.3.2 Quantization schemes

Quantization schemes embed the watermarks into the host contents by quantizing the host signals or its host features to the fixed lattice points. The early works [80−82] replace the least significant bits of the host features (such as pixel values) with the embedded information bits. However, these schemes cannot achieve a large payload of information and are also vulnerable to attacks. Chen [46, 47] proposed a pioneering dither-based quantization technique that opened a new era of active research on quantization schemes. Their Quantization Index Modulation (QIM) schemes are now still under wide investigation [83−88]. The QIM technique uses a set of quantizers modulated by the embedded message. The quantizers divide the signal space into sets of reconstruction points and the host signal is quantized to the nearest reconstruction point in the corresponding set. Since SS and QIM schemes have their own advantages and disadvantages, Chen also proposed a spreading technique [46, 47] (also termed spread transform) to bridge SS and QIM methods. The proposed Spread Transform Dither Modulation (STDM) incorporating the spreading technique has been shown to achieve great signal-to-noise (SNR) advantage over additive SS methods. The spread transform technique is further investigated in [4, 49, 89, 90]. For instance, borrowing the idea of STDM, Wong proposed SWE and MWE schemes [89] to embed one or multiple watermark sequences into the host image.

Chen's work also inspired Eggers to rediscover the work [91] and propose a suboptimal Scalar Costa Scheme [48]. In their work, the random codebook in [91] is approximated by a lattice-structured codebook for the ease of implementation. As a matter of fact, SCS is in essence a Distortion Compensated QIM (DC-QIM) scheme [46] and there are other possible implementations of Costa's results [92, 93, 65]. In the literature, the performance of quantization schemes is evaluated in terms of both the information capacity [94, 95] and the probability of errors [49, 50, 96].

The above quantization schemes use the uniform quantizers to embed the information bit since it is simple for both the embedder and the detector to encode and decode the embedded information. There are also schemes [97−99] employing non-uniform quantizers to improve the perceptual quality of the watermarked image or the robustness of the scheme.

The quantization schemes are inspected principally under data hiding scenarios. Eggers and Girod [100] contributed an early effort to introduce their SCS scheme into the watermark verification problems. Recently, Liu and Moulin [101] and Pérez-Freire *et al.* [102] also applied QIM schemes to watermark verification scenarios. These works can achieve a great performance gain over the additive spread spectrum (ASS) methods.

## 1.4  The task of this work

As stated in the above section, for spread spectrum schemes, the host interference hampers the detection of the embedded watermark, and there are basically two approaches to combating this problem.

The first approach utilizes the probability distributions of the host signals to design optimum detectors or decoders. However, up until now, no mathematically rigorous performance comparisons have been made to validate to what extent these detectors or decoders can improve the traditional correlation detectors. Moreover, the performance comparisons between these detectors and decoders also lead to another interesting question that which method, ASS or MSS, works better in the real scenarios? Therefore, we will answer these questions in Chapter 3 of this work.

The second approach improves the performances of the watermarking systems by canceling the host interference at the embedder. Improved Spread Spectrum (ISS) scheme proposed in [43] exploits the knowledge of the host content by projecting the host signal onto the watermark, and this projected host interference is then compensated in the embedding process. A problem with ISS is that it does not take the probability density function (pdf) of the host signals into account. In fact, the implicit assumption for ISS is that the host signals are normally distributed, and thus the optimum detection statistic is the linear correlation.

The above two approaches improve the performance of SS schemes in the embedding and detection process, respectively. Can we incorporate both ideas into the same watermarking scheme? We will answer this question in Chapter 4 and Chapter 5 where a host interference technique, similar to ISS, marring the optimum detection or decoding rules is presented.

Though the host interference rejection schemes (also including the quantization schemes) can achieve a large performance improvement over the traditional spread spectrum schemes, they also have a great drawback that they have not taken the perceptual quality into account. Their performance advantage, in truth, is largely due to the employed MSE metric. In contrast to the SS schemes, it is difficult to implement the perceptual models in the host interference rejection schemes. This difficulty thus discourages the use of host interference rejection schemes in real scenarios. Can we instead utilize the host information without it being cancelled at the embedder? Of course, a possible solution is to use the perceptual models for spread spectrum schemes. However, this solution still does not produce a large enough performance improvement. In this work, we instead answer this question by presenting a new informed embedder coupled with double-sided detection to achieve a large performance improvement over the traditional spread spectrum schemes. More importantly, the perceptual analysis can be easily implemented in the new embedder.

The rest of this work is organized as follows. In Chapter 2, we define the symbols used throughout this work and present the fundamentals for future discussions. Chapter 3 investigates and compares the performance of optimum detectors and decoders for both MSS and ASS. In Chapter 4 and 5, we marry the host interference technique with the optimum detection or decoding rules to improve the performance of traditional spread spectrum schemes. In Chapter 6 and 7, we present a new model of watermark detection − double-sided detection and validate its performance advantage through both theoretical analyses and extensive experiments. Its performance advantage when the perceptual analysis is implemented is demonstrated in Chapter 8. Finally, we summarize the work and present future research directions in the last chapter.

# Chapter 2   Problem Formulation

## 2.1   Notation

In this work, we denote random variables by italic capital letters and their realizations by italic small letters, such as $X$ and $x$. Random vectors and their realizations are denoted by boldface capital and small letters, respectively, such as $\mathbf{X}$ and $\mathbf{x}$. Other variables are written as italic letters, and vectors as small boldface letters.

Let $\mathbf{X} = \{X_1, X_2, \ldots, X_N\}$ be a collection of $N$ host data, where $X_1, X_2, \ldots, X_N$ are i.i.d. random variables with standard deviation $\sigma_X$. In this work, $N$ is supposed to be an even integer. Let $\mathbf{x} = \{x_1, x_2, \ldots, x_N\}$ be a particular realization of the host data. Similarly, the watermarked data and its particular realization are given by $\mathbf{S} = \{S_1, S_2, \ldots, S_N\}$ and $\mathbf{s} = \{s_1, s_2, \ldots, s_N\}$ respectively. The watermarked data may suffer from intentional or unintentional attacks. The characteristics of these attacks vary dramatically. For instance, additive noise attacks may be independent of the watermarked data, whereas valumetric scaling attacks depend on the watermarked data. Thus there is no simple mathematical framework into which all attacks can be successfully formulated. In this work, we denote the attack noise by $\mathbf{V} = \{V_1, V_2, \ldots V_N\}$, where all components $V_i$s are i.i.d. random variables with standard deviation $\sigma_V$. In real scenarios, such an assumption may fail to characterize the real attacks. However, it can simplify the performance comparisons. Under attacks, the attacked data $\mathbf{Y}$ is

$$\mathbf{Y} = \mathbf{S} + \mathbf{V}. \tag{2.1}$$

In the future discussions, we also drop the index to the vector elements when no specific element is concerned. For instance, $X$ may refer to any element $X_i$ in $\mathbf{X}$.

In this work, we denote the mean and the variance of a random variable by $E(\cdot)$ and $Var(\cdot)$. For instance, $E(X)$ and $Var(X)$ refer to the mean and the variance of $X$, respectively. Moreover, the probability of an event $A$ is written as $P(A)$.

Let $\mathbf{w} = \{w_1, w_2, \ldots, w_N\}$ be a bipolar watermark sequence (with $w_i = +1$ or $-1$) to be embedded in the host data. For information hiding problems, **w** is modulated by the message bit $b$ (+1 or −1). Of course, $w_i$ is not necessary to be +1 or −1. Instead, it can be any real number. For instance, many early works consider **w** to be a sequence of normal random variables [6, 54]. For the convenience of performance comparisons, we also assume throughout the thesis that

$$\sum_{i=1}^{N} w_i = 0 \, , \tag{2.2}$$

that is, the numbers of +1s and −1s are equal. This zero-mean sequence ensures that watermark embedding does not result in the change of statistical mean of the watermarked data. Moreover, as will also be observed in the following chapters, this assumption renders easier the performance analyses.

## 2.2  Distortion Measure

Watermark embedding incurs distortion on the host contents. Most watermarking applications require that the distortion be small enough to be imperceptible for humans. However, there is, up till now, no mathematically tractable metric that can fully evaluate the perceptual quality of the watermarked contents. Of course, the watermark embedder can interact with the viewer to select an appropriate parameter that achieves the maximum allowable perceptual distortion. Nevertheless, this approach is impracticable since it does not work for a large database of digital contents. This thesis will measure the distortion by the Mean Squared Errors (MSE). Most researchers have adopted this measure since it is both simple and mathematically tractable. Furthermore, it does characterize the quality of watermarked image for some cases. But it is still important to note that it fails to assess the perceptual quality of the attacked image when, for instance, the watermarked data suffer from geometrical attacks.

In this thesis, the embedding distortion is defined as

$$DT(\mathbf{s}, \mathbf{x}) = \frac{1}{N} \sum_{i=1}^{N} |s_i - x_i|^2 \, . \tag{2.3}$$

The expected embedding distortion $D_w$ is given by

$$D_w = E[DT(\mathbf{S}, \mathbf{X})] \tag{2.4}$$

The expected attacking distortion $D_a$ incurred by attacks can be similarly defined as

$$D_a = E[DT(\mathbf{Y}, \mathbf{S})] = \frac{1}{N} E\left[ \sum_{i=1}^{N} \left| Y_i - S_i \right|^2 \right]. \tag{2.5}$$

In the following discussions, if not explicitly stated otherwise, the distortion refers to the expected distortion $D_w$. For the convenience of performance comparisons, we also define Document to Watermark Ratio (DWR) and Watermark to Noise Ratio (WNR) as

$$\text{DWR} = 10\log_{10}[E(X^2)/D_w] \text{ and WNR} = 10\log_{10}(D_w/D_a). \tag{2.6}$$

## 2.3 Modeling of host signals

The host signal $\mathbf{X}$ can be pixel values, transform coefficients, projected data, or other features of the host data. Transform coefficients, such as DCT coefficients and DWT magnitudes, are widely used in the literature to carry the embedded information since the embedded watermarks in these data are proven to be more robust against many signal processing attacks. In this work, we are also primary concerned with the transform coefficients.

The transform coefficients can be assumed to follow some particular probability distribution function (pdf). In the early works, the host data are usually presumed to be normally distributed, which is not valid for natural images. For instance, DCT coefficients follow the Generalized Gaussian Distribution (GGD) [69−71, 103, 104] defined by

$$f_X(x) = Ae^{-|\beta x|^c}, \tag{2.7}$$

$$\text{where } \beta = \frac{1}{\sigma_X}\left( \frac{\Gamma(3/c)}{\Gamma(1/c)} \right)^{1/2}, \quad A = \frac{\beta c}{2\Gamma(1/c)}, \tag{2.8}$$

$\sigma_X$ is the standard deviation of the host signal, and $\Gamma(\cdot)$ is the Gamma function defined as

$$\Gamma(x) = \int_0^{\infty} t^{x-1} e^{-t} dt. \tag{2.9}$$

In the above formula, $\beta$ is the scale parameter and $c$ the shape parameter. Gaussian and Laplacian distributions are just special cases of (2.7) with $c = 2.0$ and $c = 1.0$, respectively.

In [74−76], the DFT magnitudes are successfully modeled by a Weibull distribution

$$f_X(x) = \frac{\delta}{\theta}\left(\frac{x}{\theta}\right)^{\delta-1}\exp\left[-\left(\frac{x}{\theta}\right)^{\delta}\right], \tag{2.10}$$

where $\theta$, the scale parameter and $\delta$, the shape parameter, are positive constants controlling the mean, variance and shape of the distribution.

For natural images, the parameters $c$, $\sigma_X$, $\theta$ and $\delta$ can be estimated from the images. In [69, 103], the authors estimated $c$ and $\sigma_X$ by Moment or Maximum Likelihood Estimation (MLE) methods. The $\theta$ and $\delta$ can also be estimated by MLE methods. The likelihood function $L(\theta, \delta)$ for such a case is

$$L(\theta,\delta) = \ln f_X(x_1, x_2, ..., x_N) = N\ln\delta - N\delta\ln\theta + (\delta-1)\sum_{i=1}^{N}\ln x_i - \frac{1}{\theta^\delta}\sum_{i=1}^{N}x_i^\delta . \tag{2.11}$$

Thus, to make the partial derivative $\partial L/\partial\theta = 0$ and $\partial L/\partial\delta = 0$, we must have

$$\frac{\partial L}{\partial\theta} = -\frac{N\delta}{\theta} + \frac{\delta}{\theta^{\delta+1}}\sum_{i=1}^{N}x_i^\delta = 0 , \tag{2.12}$$

$$\frac{\partial L}{\partial\theta} = \frac{N}{\delta} - N\ln\theta + \sum_{i=1}^{N}\ln x_i + \frac{\ln\theta}{\theta^\delta}\sum_{i=1}^{N}x_i^\delta - \frac{1}{\theta^\delta}\sum_{i=1}^{N}x_i^\delta\ln x_i = 0 . \tag{2.13}$$

Thus, (2.12) leads to

$$\theta^\delta = \frac{1}{N}\sum_{i=1}^{N}x_i^\delta . \tag{2.14}$$

Substituting (2.14) into (2.13), we obtain

$$\frac{N}{\delta} + \sum_{i=1}^{N}\ln x_i - \frac{N\sum_{i=1}^{N}x_i^\delta\ln x_i}{\sum_{i=1}^{N}x_i^\delta} = 0 . \tag{2.15}$$

Solving the above equation numerically gives the estimated $\delta$. The estimated $\theta$ can be computed from the estimated $\delta$ by (2.14).

## 2.4  Embedding rules for spread spectrum schemes

The embedding rules for watermark verification and data hiding problems are a bit different. Thus we first review the embedding rules for data hiding problems and then extend them to the watermark verification scenarios.

### 2.4.1  Data hiding

In this thesis, we focus our attention mainly on SS and quantization schemes. Spread Spectrum (SS) [6, 42, 69−79] watermarking schemes add an additive or a multiplicative watermark into the host signals to form the watermarked data. Thus, the embedding rule for the spread spectrum schemes can be given by

$$s_i = x_i + b \cdot a_i \cdot g(x_i) \cdot w_i .$$  (2.16)

where $i = 1, 2, \ldots, N$ and $g(.)$ is a function of $x$. Perceptual analysis may help determine the embedding strength $a_i$ adaptively for each host feature $x_i$. However, if the perceptual analysis is not implemented, we may set $a_i$ to be the same value $a$ (with $a > 0$ for simplicity). If $g(x_i) = 1$, we have the rule for the additive spread spectrum (ASS) schemes

$$s_i = x_i + baw_i .$$  (2.17)

However, if we assume that $g(x_i) = x_i$, the multiplicative spread spectrum (MSS) schemes can be formulated as

$$s_i = x_i + bax_iw_i .$$  (2.18)

In (2.18), we notice that the second term scales proportionally to the value of the host signal. As a matter of fact, the multiplicative embedding automatically implements a simple contrast masking of Watson's perceptual model [55]. Thus, multiplicative embedding rules achieve a better perceptual quality. Barni's multiplicative embedding rule [42] can also be derived by letting $g(x_i) = |x_i|$, that is,

$$s_i = x_i + ba|x_i|w_i .$$  (2.19)

Other embedding functions can also be derived from (2.16). However, in this thesis, we concentrate primarily on the additive and multiplicative embedding rules.

### 2.4.2  Watermark verification

In the above subsection, we reviewed the embedding rules for the data hiding problems. In fact, if we fix $b$ to 1, these rules can be automatically adapted to the watermark detection problems. Thus we similarly have

$$s_i = x_i + aw_i \,, \tag{2.20}$$

$$s_i = x_i + ax_i w_i \,, \tag{2.21}$$

$$s_i = x_i + a|x_i|w_i \,. \tag{2.22}$$

for ASS, MSS and Barni's schemes, respectively.

## 2.5  Embedding rules for quantization schemes

### 2.5.1  Data hiding

In the spread spectrum schemes (2.16), we observe that the watermarked signal comprises of two parts, the host signal and the embedded signal. Usually, in order to keep the embedding distortion imperceptible, the embedded signal should be weak enough. The problem of watermark detection thus investigates how to detect the hidden weak signals in the strong host signals. However, in the spread spectrum schemes, the host signals interfere with the successful detection of the embedded watermark. Hence Chen [46, 47] proposed a pioneering class of QIM techniques that can have a complete rejection of the host interference. Its embedding rule is given by

$$s = x + aw \,, \text{ where } a = q_b(x) - x \,. \tag{2.23}$$

In the above equation,

$$q_b(x) = q_\Delta(x - d[b]) + d[b] \text{ and } q_\Delta(x) = \Delta \cdot \lfloor x/\Delta + 0.5 \rfloor, \tag{2.24}$$

where $\lfloor \cdot \rfloor$ is the floor function, $d[b]$s are the dithers and $\Delta$ is the quantization step size. For symmetric dithers, $d[1] = -d[0] = \Delta/4$. Usually in the literature, (random and key-dependent) non-symmetric dithers are employed to prevent the attackers from decoding the embedded data. However, non-symmetric dithers incur a larger distortion than do symmetric dithers. In [46], Chen also proved that Distortion-Compensated Quantization Index Schemes (DC-QIM) is optimal in the sense of capacity-achieving. Chen's work also inspired Eggers to rediscover the work [91] and propose a suboptimal Scalar Costa Scheme (SCS) [48]. As a matter of fact, SCS is indeed the same as DC-QIM. Its embedding rule can be expressed by scaling the

quantization noise as

$$s = x + \lambda a w, \text{ where } a = q_b(x) - x.$$  (2.25)

where $\lambda$ is called the rejecting strength since it controls the strength of the host interference being rejected.

SS and quantization schemes have their own advantages and disadvantages. Chen [46, 47] bridged these two schemes by Spread Transform (ST) techniques. With the idea of spread transform, the spread spectrum and quantization methods can be formulated into the same framework. This new technique is also termed Quantization Projection (QP) in [49, 50]. Since quantization schemes (with $w = 1$) can be expressed as

$$s = x + q_b(x) - x = x + [q_b(xw) - xw]w,$$  (2.26)

we see that they are just special cases of spread spectrum schemes. Compared with quantization schemes, spread spectrum schemes instead use $w$ to secure the embedded data $b$.

For $N > 1$, we define the projected $\mathbf{x}$ on $\mathbf{w}$ as

$$\overline{x} = \sum_{i=1}^{N} x_i w_i \Big/ N.$$  (2.27)

$\overline{X}, s, \overline{s}$ are similarly defined. Equation (2.27) can also be interpreted as projecting the host signals on a selected pseudo-random sequence, which is the key idea of spread transform techniques. This also gives the reason why spread transform dither modulation (STDM) [46] is called quantization projection (QP) [49]. Projecting both sides of (2.20) on $\mathbf{w}$, we obtain

$$\overline{s} = \overline{x} + a,$$  (2.28)

If we let

$$\overline{s} = q_b(\overline{x}),$$  (2.29)

then we have

$$a = q_b(\overline{x}) - \overline{x}.$$  (2.30)

Substituting (2.30) into (2.20), we see that the quantization scheme is a special case of the spread spectrum schemes.

### 2.5.2 Embedding rules for watermark detection

It is also easy to obtain the embedding rules for the watermark verification problems. Pérez-Freire *et al.* [102] applied STDM scheme to the watermark verification scenarios. Its embedding rule is obtained by substituting $a = q_\Lambda(\overline{x}) - \overline{x}$ into the additive rule (2.20). In the above rule, $q_\Lambda(\cdot)$ is a Euclidean scalar quantizer of step size $\Delta$ whose centroids are defined by the points in the shifted lattice $\Lambda \triangleq \Delta\mathbb{Z} + \Delta/2$ (the $\Delta/2$ is chosen by symmetry reasons). Thus, the embedding rule is

$$s_i = x_i + [q_\Lambda(\overline{x}) - \overline{x}]w_i, \quad \text{where} \quad q_\Lambda(x) = q_\Delta(x - d) + d \;, \tag{2.31}$$

where $i = 1, 2, \ldots, N$, $q_\Delta(\cdot)$ is as defined in (2.24) and $d$ is a dither. For symmetrical dithers, $d = \Delta/2$. The best performance is achieved at the symmetrical dithers.

## 2.6 Random Number Generation

In this thesis, we validate the theoretical results by both Monte-Carlo simulations and experiments on real images. In order to conduct simulations, we have to generate a sequence of pseudo-random host data that follow a specific probability distribution. Since the Gamma distribution is often referred to in this work, we also formulate its pdf as

$$f_X(x) = \frac{x^{\alpha-1}\exp(-x/\beta)}{\Gamma(\alpha)\beta^\alpha} \tag{2.32}$$

where $\alpha$ and $\beta$ are shape and scale parameters, respectively. Usually we denote a Gamma random variable by Gamma$(\alpha, \beta)$.

### 2.6.1 Generation of GGD data

Lemma 2.1: Let $W$ be a random variable with $P(W = 1) = P(W = 0) = 0.5$ and $E$ a Gamma random variable with shape parameter $1/c$ and scale parameter $1/\beta^c$. Moreover, $W$ is independent of $E$. Thus the random variable $X$ defined by

$$X = \begin{cases} E^{1/c}, & \text{if } W = 1; \\ -E^{1/c} & \text{if } W = 0. \end{cases} \qquad (2.33)$$

has a GGD pdf given by (2.7).

Proof: It is easy to see that

$$F_X(x) = P(X \le x) = P(E^{1/c} \le x \mid W = 1)P(W = 1) + P(-E^{1/c} \le x \mid W = 0)P(W = 0)$$

$$= 0.5 \cdot P(E^{1/c} \le x) + 0.5 \cdot P(E^{1/c} \ge -x)$$

$$= \begin{cases} 0.5 \cdot P(E \le x^c) + 0.5, & \text{if } x \ge 0; \\ 0.5 \cdot P[E \ge (-x)^c], & \text{if } x < 0, \end{cases}$$

where the second equality follows from conditioning on $W$, the third from (2.33) and the fourth from the fact

that $E$ is nonnegative. Thus, we obtain

$$f_X(x) = \frac{d}{dx} F_X(x) = \begin{cases} 0.5cx^{c-1} f_E(x^c) & \text{if } x \ge 0 \\ 0.5c(-x)^{c-1} f_E[(-x)^c], & \text{if } x < 0 \end{cases}$$

$$= \begin{cases} 0.5cx^{c-1} \dfrac{(x^c)^{1/c-1} \exp(-x^c \beta^c)(\beta^c)^{1/c}}{\Gamma(1/c)}, & \text{if } x \ge 0 \\ 0.5c(-x)^{c-1} \dfrac{[(-x)^c]^{1/c-1} \exp[-(-x)^c \beta^c](\beta^c)^{1/c}}{\Gamma(1/c)}, & \text{if } x < 0 \end{cases}$$

$$= \begin{cases} \dfrac{\beta c \exp(-x^c \beta^c)}{2\Gamma(1/c)}, & \text{if } x \ge 0; \\ \dfrac{\beta c \exp[-(-x)^c \beta^c]}{2\Gamma(1/c)}, & \text{if } x < 0, \end{cases}$$

which is the pdf given by (2.7).

 

With this lemma, a Gamma random number generator can generate the GGD data. It is possible, though a bit hard, to generate a Gamma random number with any given scale and shape parameter. However, in this thesis, we focus on three typical cases, i.e., $c = 2.0$, $1.0$ and $0.5$ since these typical parameters are sufficient for validating the theoretical results obtained in the thesis.

At $c = 2.0$, the GGD reduces to a Gaussian distribution. There are many ways, for instance, Box-Muller

method (See page 261 in Chapter 5 of [114]), to generate Gaussian random numbers. At $c = 1.0$, $E$ has a

Gamma distribution with shape parameter 1.0, which indeed is an exponential distribution. We may refer to the transformation method (See page 247 in Chapter 4 of [107]) to generate exponentially distributed random numbers. At $c = 0.5$, $E$ has a Gamma distribution with shape parameter 2.0 and scale parameter $1/\beta^{0.5}$. Thus, if two independent random variables $E_1$ and $E_2$ follow the same exponential distribution with the same scale parameter $1/\beta^{0.5}$, then $E_1 + E_2$ has a Gamma distribution with shape parameter 2 and scale parameter $1/\beta^{0.5}$. Thus, it is easy to generate $E$ at $c = 0.5$.

### 2.6.2  Generation of Weibull data

Lemma 2.2: Let $Z$ be an exponential random variable with scale parameter $\theta^{\delta}$. Then the random variable defined by $X = Z^{1/\delta}$ has a Weibull pdf with shape parameter $\delta$ and scale parameter $\theta$.

Proof: Since $P(X \leq x) = P(Z^{1/\delta} \leq x) = P(Z \leq x^{\delta})$, we obtain

$$f_X(x) = \frac{d}{dx} P(X \leq x) = \delta x^{\delta-1} f_Z(x^{\delta}) = \delta x^{\delta-1} \frac{\exp(-x^{\delta}/\theta^{\delta})}{\theta^{\delta}},$$

which is a Weibull pdf with shape parameter $\delta$ and scale parameter $\theta$ (See 2.10).

Thus with this lemma, we can also generate weibull-distributed data from exponential data.

# Chapter 3   Performance Comparisons between MSS and ASS

## 3.1  Introduction

For the early spread spectrum schemes, the host signal hinders the detection or decoding of the embedded watermark. This situation can be improved by designing a better embedder or an optimum detector or decoder. The first approach utilizes the host information at the embedder, whereas the second improves the performance of spread spectrum watermarking schemes by exploiting the pdf of the host signals at the detector or decoder.  In this chapter, we concentrate on the second approach, while delaying the first one to the next chapter.

Based on the fact that the GGD well models the low- and mid-frequency DCT coefficients, many optimum detectors and decoders [69−71, 77] have been proposed. However, in these works, no rigorous performance analysis has been given. Thus this chapter will investigate how the performances react to the different shape parameters under both no attack and attacks. This will further lead to our highlight on the performance comparisons between MSS and ASS since there is still no solid performance comparison between these two schemes. The work [108] compared them under the assumption that the host signals are normally distributed, which is not the case for natural images.

We find through comparisons an interesting result that on the contrary to MSS, ASS yields a better performance as the shape parameter of host DCT data decreases. Moreover, MSS outperforms ASS at the shape parameter above 1.3.

The rest of this work is organized as follows. Section 3.2 investigates the performance of optimum decoders for both ASS and MSS in the DCT domain. The performance of optimum detectors shall be examined in Section 3.3. Moreover, in Section 3.3, we also discuss a case where both $\mathbf{X}$ and $\mathbf{w}$ are assumed to be random variables. Finally, we conclude this chapter in Section 3.4.

## 3.2 Data Hiding Scenario

### 3.2.1 Optimum decoding statistic for ASS

The embedding rule for ASS is given by $s_i = x_i + b \cdot a \cdot w_i$, where $i = 1, 2, \ldots, N$. In order to decode the embedded information bit $b$, we formulate two alternative hypotheses as

$$H_0: S_i = X_i - a \cdot w_i \text{ and } H_1: S_i = X_i + a \cdot w_i. \tag{3.1}$$

Based on the classical decision theory, the Bayes test can be formulated as [69, 105],

$$l(\mathbf{s}) = \frac{f_\mathbf{s}(\mathbf{s} \mid b=1)}{f_\mathbf{s}(\mathbf{s} \mid b=-1)} = \frac{\prod\limits_{i=1}^{N} f_{X_i}(s_i - aw_i)}{\prod\limits_{i=1}^{N} f_{X_i}(s_i + aw_i)} = \frac{\exp(-\sum\limits_{i=1}^{N}|\beta(s_i - aw_i)|^c)}{\exp(-\sum\limits_{i=1}^{N}|\beta(s_i + aw_i)|^c)} \underset{b=-1}{\overset{b=1}{\gtrless}} \psi, \tag{3.2}$$

where the pdf of $X_i$ is given by (2.7) and $\psi$ is the decision threshold given by

$$\psi = \frac{p_0(c_{10} - c_{00})}{p_1(c_{01} - c_{11})}, \tag{3.3}$$

where $p_0 = p(b=-1)$ and $p_1 = p(b=1)$. Please see [105] for a clear definition of other parameters in (3.3). If we assume that $c_{00} = c_{11} = 0$ and $c_{01} = c_{10} = 1$, the Bayes test minimizes the total probability of error. If we further assume that $p_0 = p_1 = 0.5$, then $\psi = 1$. By taking logarithm on both sides of (3.2), we obtain an optimum decision statistic [69]

$$L(\mathbf{S}) = \frac{1}{N}\sum_{i=1}^{N}|S_i + aw_i|^c - |S_i - aw_i|^c \underset{b=-1}{\overset{b=1}{\gtrless}} 0. \tag{3.4}$$

In the above equation, we have divided both sides by a factor $N$ for the convenience of performance analyses. This would not make any difference on the decoder's performance. Thus, an equivalent decision statistic is given by

$$L(\mathbf{S}) = \frac{1}{N}\sum_{i=1}^{N}|S_i + aw_i|^\xi - |S_i - aw_i|^\xi. \tag{3.5}$$

In the above equation, we have replaced $c$ with $\xi$ for future discussion convenience. In this work, we call this form of decoder as $\xi$-order decoder and $\xi$ is the order parameter (OP). In the following discussions, we will show that its best performance is achieved at $\xi = c$. For $\xi = 2.0$, the above rule is reduced to the linear correlation, that is,

$$L(\mathbf{S}) = \frac{1}{N} \sum_{i=1}^{N} S_i w_i .$$  (3.6)

## 3.2.2 Optimum decoders for MSS

The embedding rule for MSS is $s_i = x_i + b \cdot a \cdot x_i \cdot w_i$, where $i = 1, 2, \ldots, N$. Similarly, we decides between a pair of hypotheses

$H_0$: $S_i = X_i - a \cdot X_i \cdot w_i$ and $H_1$: $S_i = X_i + a \cdot X_i \cdot w_i$.  (3.7)

Thus, the optimum test can be similarly formulated as

$$l(\mathbf{s}) = \frac{f_{\mathbf{S}|H_1}(\mathbf{s} \mid H_1)}{f_{\mathbf{S}|H_0}(\mathbf{s} \mid H_0)} = \frac{\prod_{i=1}^{N} \frac{1}{1 + aw_i} f_X[s_i /(1 + aw_i)]}{\prod_{i=1}^{N} \frac{1}{1 - aw_i} f_X[s_i /(1 - aw_i)]} = \frac{\left[ \prod_{i=1}^{N} \frac{1}{1 + aw_i} \right] \cdot \exp\left[ -\sum_{i=1}^{N} \frac{|\beta s_i|^c}{|1 + aw_i|^c} \right]}{\left[ \prod_{i=1}^{N} \frac{1}{1 - aw_i} \right] \cdot \exp\left[ -\sum_{i=1}^{N} \frac{|\beta s_i|^c}{|1 - aw_i|^c} \right]} .$$  (3.8)

By taking logarithm on both sides, we have

$$\ln(l(\mathbf{s})) = \sum_{i=1}^{N} \ln \frac{1 - aw_i}{1 + aw_i} + \sum_{i=1}^{N} \left[ \frac{|\beta S_i|^c}{(1 - aw_i)^c} - \frac{|\beta S_i|^c}{(1 + aw_i)^c} \right] .$$  (3.9)

In the above equation, we have used the fact that $| 1 + aw_i | = 1 + aw_i$ and $| 1 - aw_i | = 1 - aw_i$ since $a << 1$. Under the same assumption for ASS decoders, the decision threshold $\psi$, also given by (3.3), is set zero to minimize the probability of error. Moreover, since $\sum_{1 \le i \le N} w_i = 0$, we have

$$\sum_{i=1}^{N} \ln[(1 - aw_i)/(1 + aw_i)] = 0 .$$  (3.10)

Hence, an equivalent decision statistic for (3.9) is

$$L(\mathbf{S}) = \frac{1}{N} \sum_{i=1}^{N} |S_i|^c \left[ \frac{1}{(1 - aw_i)^c} - \frac{1}{(1 + aw_i)^c} \right] \underset{H_0}{\overset{H_1}{\gtrless}} 0 .$$  (3.11)

If $a$ is small, then we have the following approximations

$(1 + a)^x \approx 1 + ax$ and $1 - a^2 \approx 1$ ,  (3.12)

The first approximation in (3.12) can be easily derived from Taylor Series expansion. Then we can

approximate (3.11) by

$$L(\mathbf{S}) = \frac{1}{N} \sum_{i=1}^{N} |S_i|^c \left[ \frac{(1+aw_i)^c}{(1-a^2)^c} - \frac{(1-aw_i)^c}{(1-a^2)^c} \right] \approx \frac{1}{N} \sum_{i=1}^{N} |S_i|^c \, 2caw_i \; . \tag{3.13}$$

Therefore, an equivalent decision statistic is

$$L(\mathbf{S}) = \frac{1}{N} \sum_{i=1}^{N} |S_i|^{\xi} \, w_i \; . \tag{3.14}$$

In (3.14), we have replaced $c$ with $\xi$ for future discussion. The above decision statistic is called $\xi$-order decoder and $\xi$ is the order parameter (OP), which closely resembles the generalized correlator [77] derived by the locally optimal test. In this work, if the order parameter is not specified, we may also call a $\xi$-order decoder a generalized correlator.

### 3.2.3  Performance of ASS

For the ease of discussions, we denote $m_0, m_1, \sigma_0$ and $\sigma_1$ as the mean and standard deviation of the decision statistic under $H_0$ and $H_1$, i.e.,

$$m_0 = E[L(\mathbf{S} \,|\, H_0)], \, m_1 = E[L(\mathbf{S} \,|\, H_1)], \sigma_0^2 = Var[L(\mathbf{S} \,|\, H_0)] \text{ and } \sigma_1^2 = Var[L(\mathbf{S} \,|\, H_1)] \, . \tag{3.15}$$

These notations will used throughout this thesis and also apply to the scenarios where the attacks are involved, for which case $\mathbf{S}$ should be replaced by $\mathbf{Y}$. In this thesis, we characterize the performance of data hiding systems by the total probability of error ($p_e$). Since the decision statistics, such as (3.5), (3.6) and (3.14), are often expressed by a sum of i.i.d. random variables, they can thus be approximated by Gaussian random variables due to the central limit theorem. However, it is also important to note that the approximation may be not accurate if $N$ is not sufficiently large or the pdf of the constituent random variable is not smooth enough. The approximation error is bounded by Berry-Esseen Theorem. Therefore we have

$$p_e = p_0 p[L(\mathbf{S} \,|\, H_0) > 0] + p_1 p[L(\mathbf{S} \,|\, H_1) < 0] \approx 0.5 Q(-m_0 / \sigma_0) + 0.5 Q(m_1 / \sigma_1) \, , \tag{3.16}$$

where $Q(x) = \frac{1}{\sqrt{2\pi}} \int_x^{\infty} \exp(-t^2 / 2) dt$ . \hfill (3.17)

We first derive the performance of the correlator which is optimum for Gaussian host data. It is easy to see that $m_1 = -m_0 = a$ and $\sigma_0^2 = \sigma_1^2 = Var(X)/N = \sigma_X^2$. Thus the performance is given by

$$p_e = Q\left[\frac{a\sqrt{N}}{\sigma_X}\right].  \tag{3.18}$$

Under zero-mean noise attacks $\mathbf{V}$, the attacked signal $\mathbf{Y}$ is given by $\mathbf{Y} = \mathbf{S} + \mathbf{V}$, where $\mathbf{V}$ is independent of $\mathbf{S}$.

The decision statistic is then expressed as

$$L(\mathbf{Y}) = \frac{1}{N}\sum_{i=1}^{N} Y_i w_i = \frac{1}{N}\sum_{i=1}^{N} (X_i w_i + V_i w_i) + ba .  \tag{3.19}$$

Therefore, substituting $m_1 = -m_0 = a$ and $\sigma_0^2 = \sigma_1^2 = [Var(X) + Var(V)]/N = (\sigma_X^2 + \sigma_V^2)/N$ into (3.16), we obtain

$$p_e = Q\left[a\sqrt{N/(\sigma_X^2 + \sigma_V^2)}\right].  \tag{3.20}$$

The same result is also reported in Barni's work [108].

The performance of the optimal decoder (3.5) has to be derived numerically. It is easy to see that

$$L(\mathbf{S}\,|\,b=1) = \frac{1}{N}\sum_{i=1}^{N} |X_i + 2aw_i|^\xi - |X_i|^\xi \text{ and } L(\mathbf{S}\,|\,b=0) = \frac{1}{N}\sum_{i=1}^{N} |X_i|^\xi - |X_i - 2aw_i|^\xi .  \tag{3.21}$$

Since $X$ has a symmetric pdf, $E(|X + 2a|^\xi - |X|^\xi) = E(|X - 2a|^\xi - |X|^\xi)$ and $Var(|X + 2a|^\xi - |X|^\xi) = Var(|X - 2a|^\xi - |X|^\xi)$. Thus the optimum decoder's performance is determined by

$$m_1 = -m_0 = E\left(|X + 2a|^\xi - |X|^\xi\right) \text{ and } \sigma_1^2 = \sigma_0^2 = \frac{1}{N}Var\left(|X + 2a|^\xi - |X|^\xi\right).  \tag{3.22}$$

Moreover, since

$$m_1 = \sigma_X^\xi E\big[|X/\sigma_X + 2a/\sigma_X|^\xi - |X/\sigma_X|^\xi\big],  \tag{3.23}$$

$$\sigma_1^2 = \frac{E[(|X + 2a|^\xi - |X|^\xi)^2] - m_1^2}{N} = \frac{\sigma_X^{2\xi} E[(|X/\sigma_X + 2a/\sigma_X|^\xi - |X/\sigma_X|^\xi)^2] - m_1^2}{N},  \tag{3.24}$$

and the pdf of $X/\sigma_X$ does not depend on $\sigma_X$, the performance for ASS depends only on $a/\sigma_X$ or DWR. The performance of ASS is drawn in Fig. 3.1. In this figure, we see that the best performance is achieved at $\xi = c$. Moreover, all curves intersect at $\xi = 2.0$ since the optimum decoder (3.5) at $\xi = 2.0$ reduces to a correlator whose performance is invariant to the shape parameter $c$.

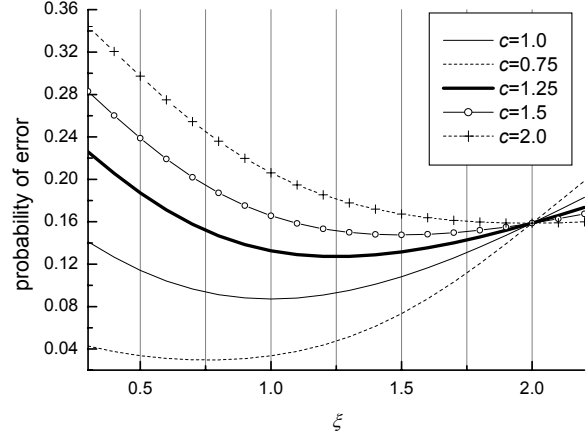

Fig. 3.1. Performance of ASS under no attack (with DWR = 20dB and $N = 100$).

Now we resort to numerical convolutions to obtain the decoder's performance under zero-mean noise attacks. Let $U = X + V$. Thus $f_U(u) = f_X(x)*f_I(v)$, where * represents convolution. Therefore,

$$m_1 = -m_0 = E(|U + 2a|^{\xi} - |U|^{\xi}),$$
(3.25)

$$\sigma_1^2 = \sigma_0^2 = E[(|U + 2a|^{\xi} - |U|^{\xi})^2] / N - m_1^2 / N.$$
(3.26)

With (3.25) and (3.26), the performance of ASS can be easily computed. Moreover, since

$$m_1 = \sigma_X^{\xi} E\left(\left|\frac{X}{\sigma_X} + \frac{V}{\sigma_X} + \frac{2a}{\sigma_X}\right|^{\xi} - \left|\frac{X}{\sigma_X} + \frac{V}{\sigma_X}\right|^{\xi}\right),$$
(3.27)

$$\sigma_1^2 = \frac{\sigma_X^{2\xi}}{N} E\left\{\left[\left|\frac{X}{\sigma_X} + \frac{V}{\sigma_X} + \frac{2a}{\sigma_X}\right|^{\xi} - \left|\frac{X}{\sigma_X} + \frac{V}{\sigma_X}\right|^{\xi}\right]^2\right\} - \frac{m_1^2}{N}$$
(3.28)

and the pdf of $V/\sigma_X$ depends only on $\sigma_X/\sigma_V$, the performance of ASS thus depends only on DWR and WNR.

### 3.2.4  Performance of MSS

Substituting $S_i = X_i(1 + baw_i)$ into (3.14) and via the approximation (3.12), we readily have

$$L(\mathbf{S}) \approx \frac{1}{N} \sum_{i=1}^{N} |X_i|^{\xi} (1 + \xi baw_i) w_i = \frac{1}{N} \sum_{i=1}^{N} |X_i|^{\xi} w_i + \frac{\xi ba}{N} \sum_{i=1}^{N} |X_i|^{\xi}.$$
(3.29)

It thus follows that

$$m_1 = -m_0 \approx \xi a E(|X|^{\xi}).$$
(3.30)

$$\sigma_1^2 = \sigma_0^2 = \frac{(1+a)^{2\xi} + (1-a)^{2\xi}}{2N} Var\left(|X|^{\xi}\right) \approx \frac{Var\left(|X|^{\xi}\right)}{N}.$$  (3.31)

Substituting (3.30) and (3.31) into (3.16), we see that

$$p_e = Q\left[a\sqrt{N}\xi E\left(|X|^{\xi}\right)\Big/\sqrt{Var\left(|X|^{\xi}\right)}\right].$$  (3.32)

We define the Mean-Variation Ratio (MVR) as

$$\text{MVR}(\xi) = \xi E\left(|X|^{\xi}\right)\Big/\sqrt{Var\left(|X|^{\xi}\right)}.$$  (3.33)

For GGD host signals with shape parameter $c$, we have

$$\begin{aligned}
E\left(|X|^{\xi}\right) &= \int_{-\infty}^{+\infty} |x|^{\xi} A e^{-|\beta x|^{c}} dx \\
&= 2A\int_{0}^{+\infty} x^{\xi} e^{-\beta^{c} x^{c}} dx \\
&= \frac{2A}{c} \int_{0}^{+\infty} y^{\frac{\xi+1}{c}-1} e^{-\beta^{c} y} dy \quad \left(\text{Let } x = y^{\frac{1}{c}}\right) \\
&= \frac{2A}{c\beta^{\xi+1}} \Gamma[(\xi+1)/c].
\end{aligned}$$  (3.34)

Thus the variance is given by

$$\begin{aligned}
\text{var}\left(|X|^{\xi}\right) &= E\left(|X|^{2\xi}\right) - \left[E\left(|X|^{\xi}\right)\right]^2 \\
&= \frac{2A\Gamma[(2\xi+1)/c]}{c\beta^{2\xi+1}} - (2A/c)^2 \frac{\{\Gamma[(\xi+1)/c]\}^2}{\beta^{2\xi+2}} \\
&\overset{(a)}{=} (2A/c)^2 \frac{\Gamma[(2\xi+1)/c]\Gamma(1/c)}{\beta^{2\xi+2}} - (2A/c)^2 \frac{\{\Gamma[(\xi+1)/c]\}^2}{\beta^{2\xi+2}} \\
&= (2A/c)^2 \frac{\Gamma[(2\xi+1)/c]\Gamma(1/c) - \{\Gamma[(\xi+1)/c]\}^2}{\beta^{2\xi+2}}
\end{aligned}$$  (3.35)

where ($a$) follows from (2.8). Consequently, we have

$$\text{MVR}(\xi) = \frac{\xi\Gamma[(\xi+1)/c]}{\sqrt{\Gamma[(2\xi+1)/c]\Gamma(1/c) - \{\Gamma[(\xi+1)/c]\}^2}}.$$  (3.36)

We see from (3.36) that MVR does not depend on $\sigma_X$, and thus the decoder's performance is completely determined by the shape parameter $c$. It is difficult to maximize the MVR analytically. However, we found through the numerical search that (3.36) is maximized at $\xi = c$ for all $c$ in [0.1, 2.5], proving that the optimum test does yield the best performance. The MVR shown in Fig. 3.2 clearly indicates that the maximum is achieved at $\xi = c$. What's more, for any given shape parameter, MVR($\xi$) increases as $\xi$ increases. Finally, plugging $\xi = c$ into (3.36), we obtain

$$
\begin{aligned}
[\mathrm{MVR}(c)]^2 &= \frac{c^2 \Gamma(1+1/c)\Gamma(1+1/c)}{\Gamma(2+1/c)\Gamma(1/c) - [\Gamma(1+1/c)]^2} \\
&= \frac{c^2 \Gamma(1+1/c)\Gamma(1+1/c)}{(1+1/c)\Gamma(1+1/c)\Gamma(1/c) - [\Gamma(1+1/c)]^2} \\
&= \frac{c^2 \Gamma(1+1/c)}{(1+1/c)\Gamma(1/c) - \Gamma(1+1/c)} \\
&= \frac{c\Gamma(1/c)}{(1+1/c)\Gamma(1/c) - (1/c)\Gamma(1/c)} \\
&= c.
\end{aligned}
\tag{3.37}
$$

It therefore follows that

$$
\mathrm{MVR}(c) = \sqrt{c} \, . \tag{3.38}
$$

In the above derivation, we have used the fact that

$$
\Gamma(x+1) = x\Gamma(x) \, . \tag{3.39}
$$

Therefore, the larger the $c$, the better the performance.

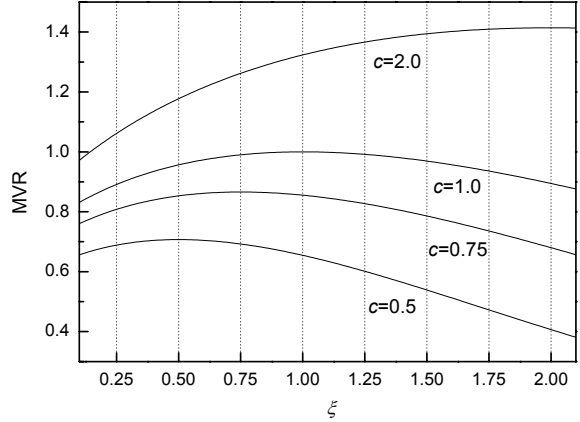

Fig. 3.2. MVR for different shape parameters.

Now we derive the performance of MSS under noise attacks. At $\xi = 2.0$, we have

$$
L(\mathbf{Y}) = \frac{1}{N}\sum_{i=1}^{N} Y_i^2 w_i = \frac{1}{N}\sum_{i=1}^{N}[X_i^2(1+baw_i)^2 + 2V_i X_i(1+baw_i) + V_i^2]w_i \, . \tag{3.40}
$$

Hence, with some simple algebraic manipulations, we obtain

$$m_1 = -m_0 = 2a\sigma_X^2 \,, \tag{3.41}$$

$$\sigma_0^2 = \sigma_1^2 = \frac{[(1+a)^4 + (1-a)^4] Var(X^2)}{2N} + \frac{2[(1+a)^2 + (1-a)^2] E(X^2) E(V^2) + Var(V^2)}{N} \,. \tag{3.42}$$

Since $a << 1$, we see that

$$\sigma_1^2 = \sigma_0^2 \approx \frac{Var(X^2) + 4E(X^2)E(V^2) + Var(V^2)}{N} \,, \tag{3.43}$$

which depends on the type of attacking noise. For Gaussian noise, we obtain

$$\sigma_1^2 = \sigma_0^2 \approx [2\sigma_X^4 + 4\sigma_X^2 \sigma_V^2 + 2\sigma_V^2] / N \,. \tag{3.44}$$

Therefore, the performance for MSS is given by

$$p_e = Q[a\sqrt{2N}\sigma_X^2 / (\sigma_X^2 + \sigma_V^2)] \,. \tag{3.45}$$

At $\xi \neq 2.0$, the decoder's performance has to be obtained numerically. Similarly, we have

$$m_1 = -m_0 = \{E(|X(1+a)+V|^\xi) - E(|X(1-a)+V|^\xi)\}/2 \,, \tag{3.46}$$

$$\sigma_1^2 = \sigma_0^2 = \{Var(|X(1+a)+V|^\xi) + Var(|X(1-a)+V|^\xi)\}/(2N) \,, \tag{3.47}$$

where the pdfs of $X(1+a)+V$, $X(1-a)+V$ and $X+V$ have to be obtained through numerical convolutions. As we did for ASS under attacks, we can prove similarly that the performance of MSS depends solely on DWR and WNR.

### 3.2.5  Performance comparisons under no attack

In this chapter, the performance comparisons are made at the same level of embedding distortion $D_w$. For ASS, it is trivial to see that

$$D_w = a^2 \,, \tag{3.48}$$

and for MSS,

$$D_w = a^2 \sigma_X^2 \,. \tag{3.49}$$

### (1)  Gaussian hosts

The correlation decoder is optimal for Gaussian hosts and yields a performance

$$p_e = Q[\sqrt{ND_w}/\sigma_X]. \tag{3.50}$$

For MSS, substituting (3.49) into (3.32), we obtain

$$p_e = Q[\sqrt{ND_w}\text{MVR}(\xi)/\sigma_X]. \tag{3.51}$$

It thus follows from (3.38) that the best performance for Gaussian host signals ($c = 2.0$) is

$$p_e = Q[\sqrt{2DN}/\sigma_X], \tag{3.52}$$

which outperforms its counterpart ASS (3.50).

### (2)  Non-Gaussian hosts

For $c = 1.0$, (3.51) produces the best performance

$$p_e = Q[\sqrt{DN}/\sigma_X] \tag{3.53}$$

since MVR(1.0) = 1.0. We find that (3.53) is identical to (3.50). However, at $c \neq 2.0$, the correlation decoder is not optimal and the optimum performance must be better than (3.50). Since MSS yields a better performance as the shape parameter $c$ increases, MSS's performance at $c < 1.0$ is worse than (3.53) or (3.50). Therefore, for $c \leq 1.0$, ASS surpasses MSS in performance. At which shape parameter do ASS and MSS achieve the same performance? Fig. 3.3 answers this question. Fig. 3.3(a) depicts the performance of MSS and ASS at the same DWR. An astounding result is that on the contrary to MSS, ASS yields a better performance as $c$ decreases (also see Fig. 3.1). Furthermore, MSS outperforms ASS at $c$ above 1.3. At a small $c$, for instance $c = 0.5$, ASS produces a far better performance. Fig. 3.3(b) displays that at $c = 1.3$, MSS behaves almost as well as ASS does. Please note that in Fig. 3.3(b), the embedding strength used for MSS is $a/\sigma_X$ (see (3.48) and (3.49)) to achieve the same distortion.

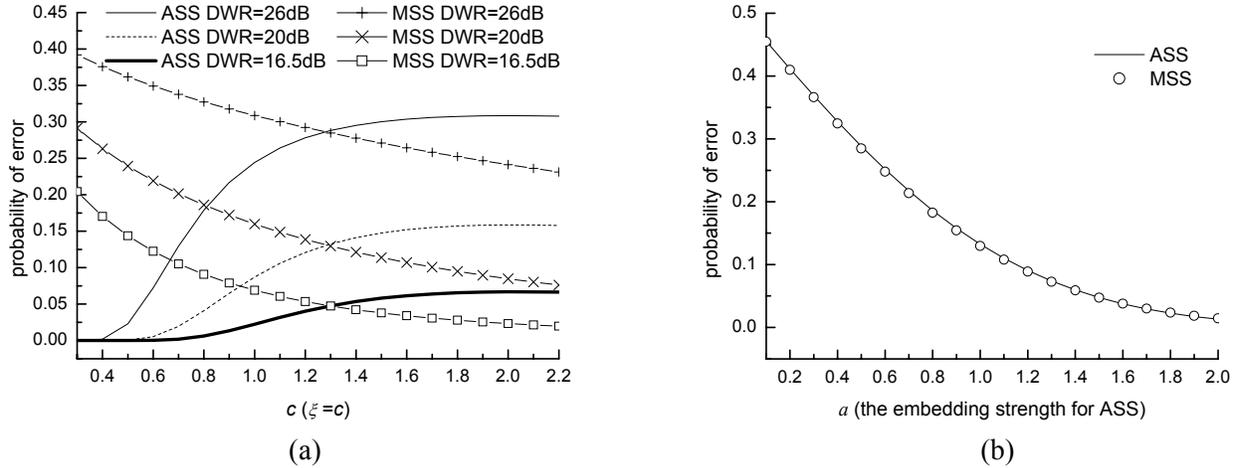

(a)                                                              (b)

Fig. 3.3. (a) MSS versus ASS under no attack (with $N = 100$). (b) MSS versus ASS at $c = 1.3$ (with $N = 100$, $\sigma_X = 10$, and the embedding strength for MSS is $a/\sigma_X$).

### 3.2.6  Performance comparisons under attacks

#### (1)  Gaussian hosts

Likewise, we first compare MSS with ASS for Gaussian hosts and attacks. Substituting (3.48) into (3.20) leads to

$$p_e = Q\left[ \sqrt{ND_w/(\sigma_X^2 + \sigma_V^2)} \right].\tag{3.54}$$

Similarly, the performance for MSS is evaluated by

$$p_e = Q\left[ \sqrt{2DN}\, \frac{\sigma_X}{\sigma_X^2 + \sigma_V^2} \right].\tag{3.55}$$

Comparing (3.54) with (3.55), we find that MSS outperforms ASS if $\sigma_V < \sigma_X$.

#### (2)  Non-Gaussian hosts

If the host signals are not Gaussian-distributed, we have to resort to the numerical approaches to compute their performance under noise attacks. In Fig. 3.4(a), we compared MSS with ASS under noise attacks. At $c = 1.3$, MSS outperforms ASS when the noise power is relatively small; ASS surpasses MSS in performance when the watermarked data suffer from strong attacks. For $c < 1.3$, for instance $c = 0.5$ and $c = 1.0$, a better performance can be obtained by ASS. In Fig. 3.4(b), we observe that for noise attacks at WNR= 0dB or WNR = −14dB, MSS exceeds ASS in performance at a large $c$ ($c > 1.3$). However, for very strong noise attacks, ASS can achieve a better performance.

Under noise attacks, the best $\xi$ may not be achieved at $c$. Thus, we search for the best $\xi$ under noise attacks and compare their best performances. The results are displayed in Fig. 3.5. Fig. 3.5(a) shows that for attacks at WNR = 0dB or WNR = −14dB, MSS outperforms ASS at $c$ above 1.3. Another interesting result is that even under noise attacks, the smaller $c$ implying the better performance still holds for ASS. Thus for ASS, we should select the DCT coefficients with a smaller $c$, whereas MSS should embed information in the DCT data with a larger $c$.

The above experiments make performance comparisons under Gaussian noise attacks. Fig. 3.5(b) displays the comparison results under GGD noise attacks, where the attack noise is assumed to follow a GGD distribution with a shape parameter $ac$. The comparison results show that as $ac$ increases, ASS yields a poorer performance and thus Gaussian noise attacks are not the strongest type of attacking noise. The underlying reason is that the attacked data (after noise attacks with a larger $ac$) have less heavy tails and thus can be approximated by a GGD with a larger shape parameter. Therefore, the best $\xi$ would be larger and hence the poorer performance since ASS yields a poorer performance at larger shape parameters. This is also clearly reflected in Fig. 3.9(a), where the best $\xi$ for $ac = 2.0$ is larger than that for $ac = 1.0$. However, for MSS, the best performance is almost invariant to the type of attacking noise. Moreover, the same conclusion that MSS outperforms ASS at $c$ above 1.3 can also be drawn from this figure.

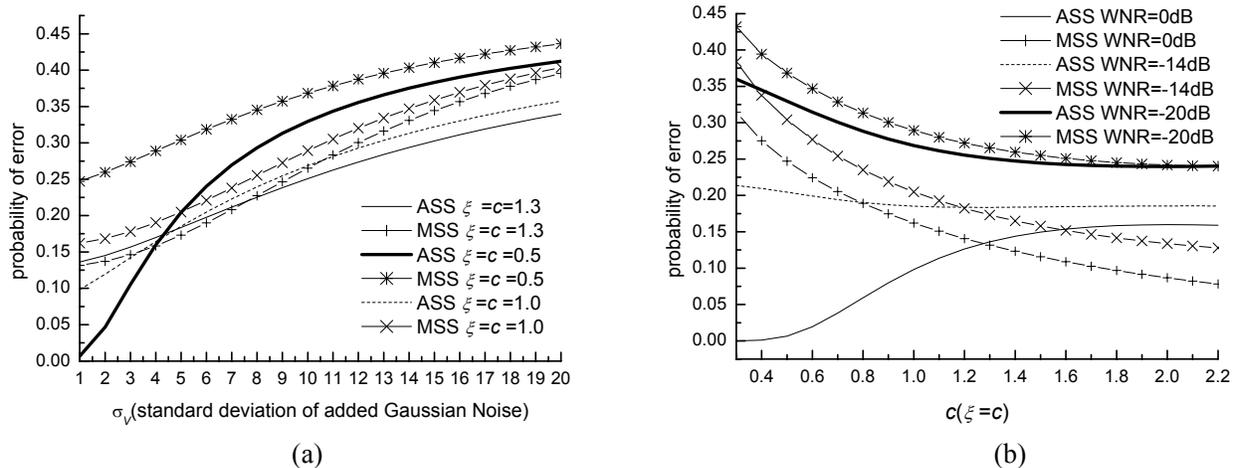

(a)                                                          (b)

Fig. 3.4. MSS versus ASS under Gaussian noise attacks with $N = 100$. (a) At different attack levels (with DWR = 20dB, $a = 1.0$ for ASS and $a = 0.1$ for MSS, $\sigma_X = 10$). (b) At different $c$ (with DWR = 20dB).

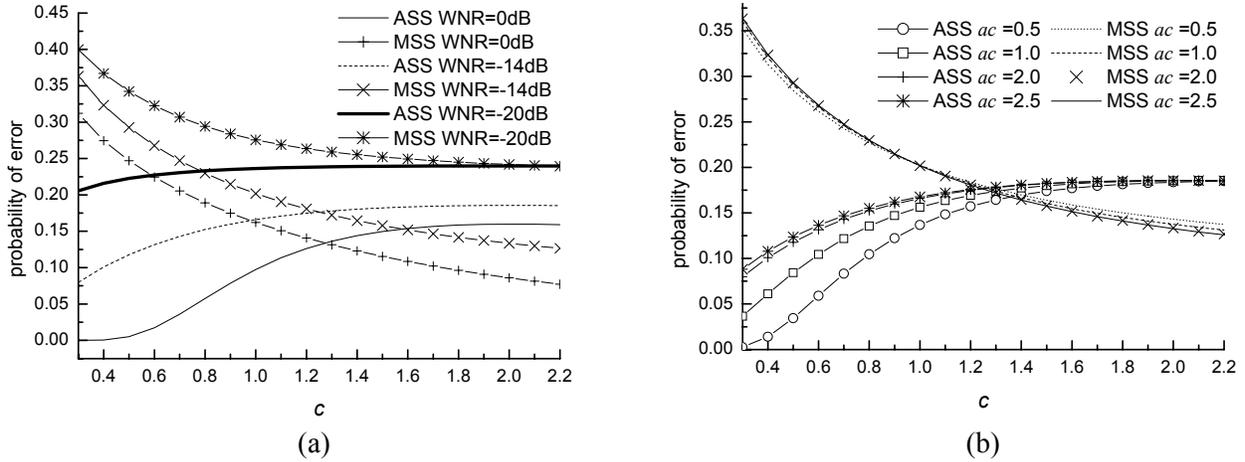

Fig. 3.5. Performance of the best $\xi$-order decoder (at DWR = 20dB and $N$ = 100). (a) Under zero-mean Gaussian noise attacks. (b) Under zero-mean GGD noise attacks (with the shape parameter $ac$ and WNR = −14dB).

### 3.2.7 Monte-Carlo Simulations

In this section, Monte-Carlo simulations are made to verify the theoretical results derived in the previous sections. The shape parameters taken in all experiments are 0.5 and 1.0, which are typical of many natural images. Please also note that in the legend of the figures, "E" stands for Empirical results obtained through Monte-Carlo simulations, "T" for Theoretical results, and "OP" for the optimum decoder (3.11).

In the first experiment, we validate the theoretical performances of ASS and MSS under no attack, and the experimental results are demonstrated in Fig. 3.6 and Fig. 3.7. Fig. 3.6(a) shows that the theoretical performance does match well with the simulated results. Moreover in Fig. 3.6(b), the best performance is achieved at $\xi = c$, though this is not clearly reflected for $c = 0.5$. In Fig. 3.7(a), we observe that though the generalized correlator is only suboptimal, it achieves almost the same performance with the optimum decoder. Fig. 3.7(b) demonstates that though the theoretical results can well predict the real performance of the optimum decoder at small decoding order parameters, they underestimate the real performance at larger $\xi$s. This performance underestimation is largely due to the inaccuracy of approximation by the central limit theorem since $|X|^{\xi}$ has a larger dispersion at a larger $\xi$ (especially for the host data with a smaller shape parameter.) This thus also explains why the theoretical performance for $c = 0.5$ displays a larger deviation from the experimental results (since the host data for $c = 0.5$ are more dispersed than that for $c = 1.0$.) The second experiment is done to verify the theoretical performance of both ASS and MSS under noise attacks.

The results in Fig. 3.8 demonstrate the nice agreement between the theoretical and the experimental performance.

In the previous section, we find the best $\xi$ through searching. However, in real scenarios, the parameters of the added noise and thus the best $\xi$ may not be known. Instead, the best $\xi$ can be roughly approximated by the shape parameter of the attacked data. In this experiment, we will verify this. Please note that the shape parameter is estimated by the moment method [69]. The comparison results between the theoretically best $\xi$ and the estimated shape parameters are shown in Fig. 3.9. In the figures, "T" stands for the theoretically best $\xi$, and "E" for estimated parameters. The estimated parameters in both figures can give a rough approximation of the best $\xi$.

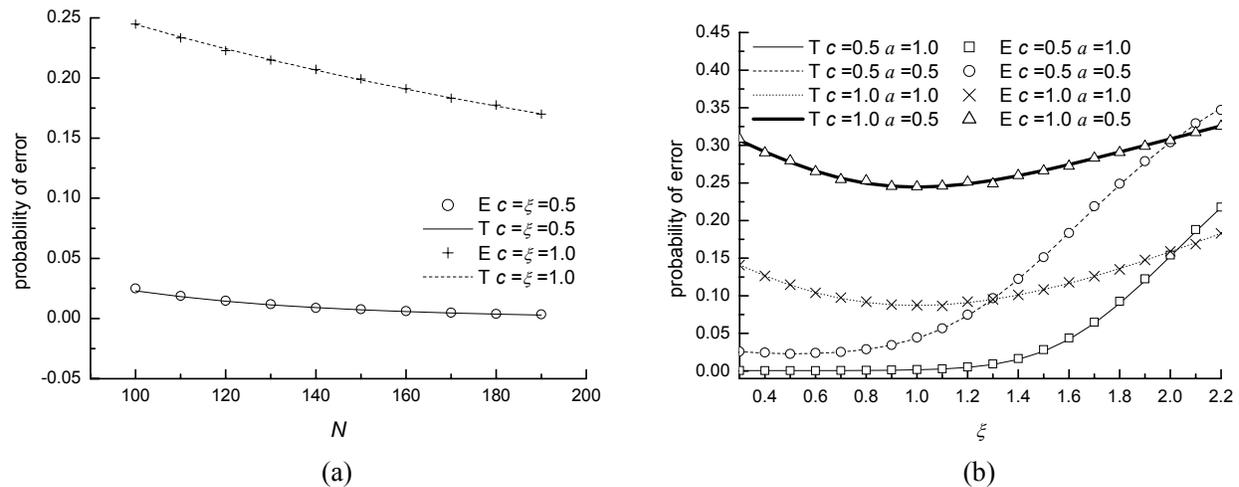

(a)                                              (b)

Fig. 3.6. Theoretical and simulated performance of ASS with $N = 100$, $\sigma_X = 10$. Empirical results are obtained on 100,000 groups of data. (a) Performance at different $N$s with $a = 0.5$. (b) Performance at different $\xi$s.

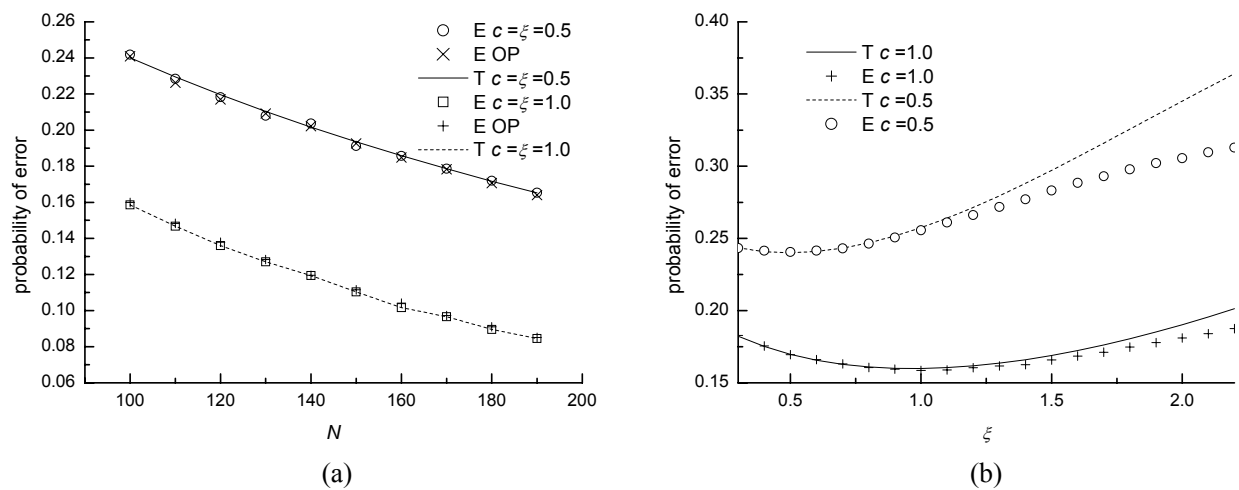

(a)                                              (b)

Fig. 3.7. Theoretical and simulated performance of MSS with $N = 100$, $\sigma_X = 10$ and $a = 0.1$. Empirical results are obtained on 100,000 groups of data. (a) Performance at different $N$s. (b) Performance at different $\xi$s.

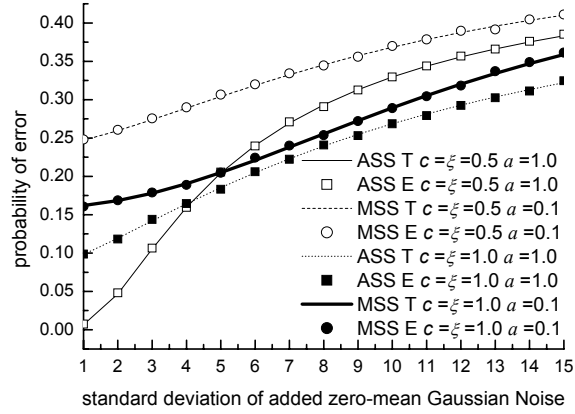

Fig. 3.8. Theoretical and simulated performance under zero-mean Gaussian noise attacks at $N = 100$, $\sigma_X = 10$. Empirical results are obtained on 100,000 groups of data.

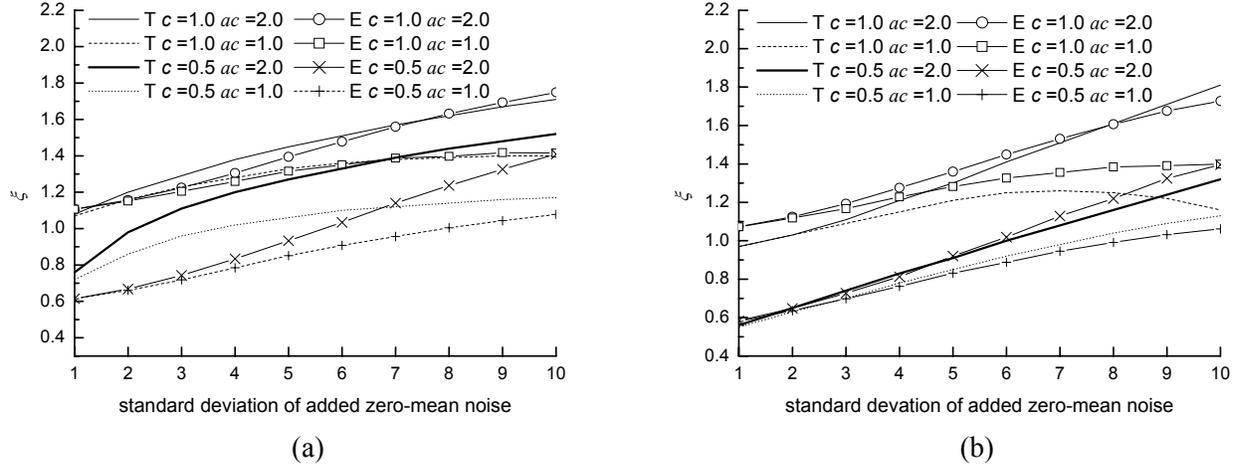

(a)                                                    (b)

Fig. 3.9. The best $\xi$ and the estimated shape parameter of the attacked data for ASS on 10,000 groups of data (with $\sigma_X = 10$, $N = 100$, the shape parameter $ac$ of the attacking noise). (a) For ASS with $a = 1.0$ (b) For MSS with $a = 0.1$.

## 3.3  Watermark verification scenarios

In the previous section, we investigate the performance of MSS and ASS under data hiding scenarios. In this section, we examine their performances for watermark verification problems.

### 3.3.1  Optimum detection rules for ASS

For ASS, the watermark detection can be formulated as a binary hypothesis test between $H_0$: $S_i = X_i$ and $H_1$: $S_i = X_i + a\, w_i$. Thus the likelihood ratio test (LRT) leads to

$$l(\mathbf{s}) = \frac{f_{\mathbf{S}|H_1}(\mathbf{s})}{f_{\mathbf{S}|H_0}(\mathbf{s})} = \frac{\prod_{i=1}^{N} f_{X_i}(s_i - a w_i)}{\prod_{i=1}^{N} f_{X_i}(s_i)} = \frac{\exp(-\sum_{i=1}^{N}|s_i - a w_i|^c)}{\exp(-\sum_{i=1}^{N}|s_i|^c)} \quad . \tag{3.56}$$

Taking logarithms on both sides, we immediately have an optimum decision statistic [69, 70]

$$L(\mathbf{S}) = \sum_{i=1}^{N}|S_i|^c - \sum_{i=1}^{N}|S_i - a w_i|^c \quad . \tag{3.57}$$

Thus an equivalent statistic is given by

$$L(\mathbf{S}) = \frac{1}{N}(\sum_{i=1}^{N}|S_i|^\xi - \sum_{i=1}^{N}|S_i - a w_i|^\xi) \quad . \tag{3.58}$$

In the above rule, we have also replaced $c$ with $\xi$ for discussion convenience. It can also be verified that the best performance is achieved at $\xi = c$. In this thesis, $\xi$ is called the order parameter (OP). At $\xi = 2.0$, (3.58) is reduced to a linear correlation

$$L(\mathbf{S}) = \frac{1}{N}\sum_{i=1}^{N} S_i w_i \quad . \tag{3.59}$$

The linear correlation, though optimal only for Gaussian host signals, is widely used in the spread spectrum schemes. It offers an advantage over (3.58) in that the detector requires no knowledge of the embedding strength $a$. For the above decision statistic, the decision rule is

$$L(\mathbf{S}) > \psi \Rightarrow H_1; \quad L(\mathbf{S}) < \psi \Rightarrow H_0, \tag{3.60}$$

where $\psi$ is the decision threshold.

### 3.3.2  Optimum detection rules for MSS

For MSS, the detector makes a choice between $H_0$: $S_i = X_i$ and $H_1$: $S_i = X_i (1 + a\, w_i)$. Therefore, the likelihood ratio test (LRT) is

$$l(\mathbf{s}) = \frac{f_{\mathbf{S}|H_1}(\mathbf{s})}{f_{\mathbf{S}|H_0}(\mathbf{s})} = \frac{\left[\prod_{i=1}^{N}\frac{1}{1+a w_i}\right] \cdot f_{X_i}\left[\frac{s_i}{1+a w_i}\right]}{\prod_{i=1}^{N} f_{X_i}(s_i)} = \frac{\exp\left[-\sum_{i=1}^{N}\frac{|\beta s_i|^c}{|1+a w_i|^c}\right]\prod_{i=1}^{N}\frac{1}{1+a w_i}}{\exp(-\sum_{i=1}^{N}|\beta s_i|^c)} \quad . \tag{3.62}$$

By taking logarithm on both sides of (3.62) and since $a \ll 1$, we obtain

$$\ln(l(\mathbf{s})) = \sum_{i=1}^{N} \ln \frac{1}{1+aw_i} + \sum_{i=1}^{N} |\beta S_i|^c \left[ 1 - \frac{1}{(1+aw_i)^c} \right]. \tag{3.63}$$

Thus, given any watermark sequence, the decision statistic can be written as

$$L(\mathbf{S}) = \frac{1}{N} \sum_{i=1}^{N} |S_i|^c \left[ 1 - 1/(1+aw_i)^c \right], \tag{3.64}$$

The decision rule (3.60) can be employed to decide between $H_0$ and $H_1$. If $a$ is small, the approximation technique (3.12) yields

$$L(\mathbf{S}) = \frac{1}{N} \sum_{i=1}^{N} |S_i|^c \left[ 1 - \frac{(1-aw_i)^c}{(1-a^2)^c} \right] \approx \frac{1}{N} \sum_{i=1}^{N} |S_i|^c \, caw_i \ . \tag{3.65}$$

Therefore, an equivalent one is

$$L(\mathbf{S}) = \frac{1}{N} \sum_{i=1}^{N} |S_i|^\xi \, w_i \ . \tag{3.66}$$

In (3.66), we have replaced $c$ in (3.65) with $\xi$ for the convenience of future discussion. The above decision statistic is called $\xi$-order detector in this work and $\xi$ is the order parameter. The same form of detector can also be found in [77] where it was derived from the LO test and called the generalized correlator. The $\xi$-order detector is suboptimal; however, it achieves almost the same performance as the optimal detector does since $a$ is rather small.

### 3.3.3 Performance of ASS

The performance of the watermark detection system is characterized by the receiver operating curves (ROC) plotting the probability of detection ($p_d$) or the probability of miss ($p_m$) at a specified false alarm probability ($p_{fa}$), where $p_d$, $p_m$ and $p_{fa}$ are defined as

$$p_{fa} = P(\text{say } H_1 \mid H_0 \text{ is true}), \ \ p_m = P(\text{say } H_0 \mid H_1 \text{ is true}) \text{ and } p_d = P(\text{say } H_1 \mid H_1 \text{ is true}). \tag{3.67}$$

It is important to note that $p_m = 1 - p_d$. Since the decision statistics are often expressed by a sum of a large number of independent random variables, they can be approximated by Gaussian random variables due to the central limit theorem. Thus we obtain

$$p_{fa} = Q[(\psi - m_0)/\sigma_0] \text{ and } p_m = 1 - Q[(\psi - m_1)/\sigma_1], \tag{3.68}$$

Substituting $p_{fa}$ into $p_m$, we have alternatively

$$p_m = 1 - Q\{[Q^{-1}(p_{fa})\sigma_0 + m_0 - m_1]/\sigma_1\}. \tag{3.69}$$

In this subsection, we first examine the performance of correlation detectors. Substituting $m_0 = 0$, $m_1 = a$ and $\sigma_0^2 = \sigma_1^2 = \sigma_X^2/N$ into (3.68), we find that the detector's performance under no attack is depicted by [71]

$$p_m = 1 - Q[Q^{-1}(p_{fa}) - a\sqrt{N}/\sigma_X]. \tag{3.70}$$

Under attacks, the detector decides between $H_0$: $Y_i = X_i + V_i$ and $H_1$: $Y_i = X_i + aw_i + V_i$. Its performance is evaluated by [101, 102]

$$p_m = 1 - Q\left[Q^{-1}(p_{fa}) - a\sqrt{N/(\sigma_X^2 + \sigma_V^2)}\right], \tag{3.71}$$

since $m_0 = 0$, $m_1 = a$ and $\sigma_0^2 = \sigma_1^2 = (\sigma_X^2 + \sigma_V^2)/N$.

We have to resort to numerical integrations to obtain the theoretical performance of (3.58) at $\xi \neq 2.0$. Under $H_0$ and $H_1$, we have $L(\mathbf{S}|H_0) = (\sum_{1 \leq i \leq N} |X_i|^\xi - |X_i - aw_i|^\xi)/N$ and $L(\mathbf{S}|H_1) = (\sum_{1 \leq i \leq N} |X_i + aw_i|^\xi - |X_i|^\xi)/N$, respectively. Since $X$ has a symmetric pdf, $E(|X+a|^\xi - |X|^\xi) = E(|X-a|^\xi - |X|^\xi)$ and $Var(|X+a|^\xi - |X|^\xi) = Var(|X-a|^\xi - |X|^\xi)$. Thus we obtain

$$m_1 = -m_0 = E(|X+a|^\xi - |X|^\xi), \tag{3.72}$$

$$\sigma_1^2 = \sigma_0^2 = Var(|X+a|^\xi - |X|^\xi)/N. \tag{3.73}$$

The above means and standard deviations have to be evaluated numerically. The performance for (3.58) can thus be evaluated by substituting (3.72) and (3.73) into (3.69). The detector's performance under attacks can be similarly characterized by

$$m_1 = -m_0 = E(|X+V+a|^\xi - |X+V|^\xi), \tag{3.74}$$

$$\sigma_1^2 = \sigma_0^2 = Var(|X+V+a|^\xi - |X+V|^\xi)/N, \tag{3.75}$$

where the pdf of $(X+V)$ is obtained through convolving the pdf of $X$ and the pdf of $V$. Similarly, we can prove that the performance of ASS under attacks is dependent on only DWR and WNR.

### 3.3.4  Performance of MSS

In this subsection, we inspect the performance of the optimal detector (3.64) and the generalized correlator (3.66). The decision has to be made between $H_0$: $S_i = X_i$ and $H_1$: $S_i = X_i$ $(1+aw_i)$. Thus, the performance of the optimum detector can be determined by

$$m_0 = \{[1-1/(1+a)^c] + [1-1/(1-a)^c]\}E(|X|^c)/2 ,  \tag{3.76}$$

$$\sigma_0^2 = \{[1-1/(1+a)^c]^2 + [1-1/(1-a)^c]^2\}Var(|X|^c)/(2N) ,  \tag{3.77}$$

$$m_1 = \{[(1+a)^c - 1] + [(1-a)^c - 1]\}E(|X|^c)/2 ,  \tag{3.78}$$

$$\sigma_1^2 = \{[(1+a)^c - 1]^2 + [(1-a)^c - 1]^2\}Var(|X|^c)/(2N) .  \tag{3.79}$$

For the generalized correlator (3.66), we similarly have

$$m_0 = 0 , \ \sigma_0^2 = Var(|X|^\xi)/N ,  \tag{3.80}$$

$$m_1 = \frac{E(|X|^\xi)[(1+a)^\xi - (1-a)^\xi]}{2} \approx \frac{E(|X|^\xi)[(1+\xi a)-(1-\xi a)]}{2} = \xi a E(|X|^\xi)  \tag{3.81}$$

$$\sigma_1^2 = [(1+a)^{2\xi} + (1-a)^{2\xi}]Var(|X|^\xi)/(2N) \approx Var(|X|^\xi)/N .  \tag{3.82}$$

Thus, substituting (3.80), (3.81) and (3.82) into (3.69), we obtain

$$p_m \approx 1 - Q\left[Q^{-1}(p_{fa}) - a\sqrt{N}\,\xi E(|X|^\xi)\Big/\sqrt{Var(|X|^\xi)}\right] = 1 - Q\left[Q^{-1}(p_{fa}) - a\sqrt{N}\mathrm{MVR}(\xi)\right] .  \tag{3.83}$$

In this thesis, we only examine the suboptimal detector's performance under noise attacks since it serves a benchmark for our informed schemes in our future discussions. The performance for the optimal detector can be similarly derived. Under attacks, the detector tests a pair of hypotheses $H_0$: $Y_i = X_i + V_i$ and $H_1$: $Y_i = X_i$ $(1+aw_i) + V_i$. At $\xi = 2.0$, the analytic performance is characterized by

$$m_0 = 0, \ \sigma_0^2 = [Var(X^2) + Var(V^2) + 4E(X^2)E(V^2)]/N, \ m_1 = 2a\sigma_X^2 \text{ and } \sigma_1^2 \approx \sigma_0^2 .  \tag{3.84}$$

Please see (3.40) to (3.43) for a better understanding of the above equation. For Gaussian hosts and zero-mean Gaussian noise attacks $\mathbf{V}$, we obtain

$$p_m = 1 - Q[Q^{-1}(p_{fa}) - \sqrt{2N}a\sigma_X^2\big/(\sigma_X^2 + \sigma_V^2)] .  \tag{3.85}$$

If $\xi \neq 2.0$, the performance of the generalized correlator (3.66) under attacks can be characterized by

$$m_0 = 0 , \ \sigma_0^2 = Var(|X + V|^{\xi})/N , \tag{3.86}$$

$$m_1 = [E(|X(1+a) + V|^{\xi}) - E(|X(1-a) + V|^{\xi})]/2 , \tag{3.87}$$

$$\sigma_1^2 = [Var(|X(1+a) + V|^{\xi}) + Var(|X(1-a) + V|^{\xi})]/(2N) . \tag{3.88}$$

The pdfs of $X(1+a)+V$, $X(1-a)+V$ and $X+V$ have to be obtained through numerical convolutions.

### 3.3.5  Performance comparisons under no attack

In this subsection, we compared MSS with ASS under no attack. Likewise, we first make performance comparisons at $c = 2.0$.

#### (1)  Gaussian hosts

Substituting $D_w = a^2$ into (3.70), we obtain

$$p_m = 1 - Q[Q^{-1}(p_{fa}) - \sqrt{ND_w}/\sigma_X] . \tag{3.89}$$

Since MVR reaches its maximum at $\xi = c$, the optimum performance for MSS is

$$p_m = 1 - Q[Q^{-1}(p_{fa}) - \sqrt{cND_w}/\sigma_X] , \tag{3.90}$$

which achieves a better performance than (3.89) at $c > 1.0$. Hence, for Gaussian hosts, MSS yields a better performance than ASS does.

#### (2)  Non-Gaussian hosts

Following the same arguments in the information hiding scenario, we concludes that MSS outperforms ASS at some $c$ larger than 1.0. In fact, at $c = 1.3$, MSS and ASS achieves almost the same performance. Fig. 3.10(a) and Fig. 3.10(b) both demonstrate that at $c > 1.3$, MSS surpass ASS in performance; otherwise, MSS is inferior to ASS in performance. We also notice from both figures that in contrast to MSS, ASS produces a nicer performance as the shape parameter $c$ decreases. These results are consistent with those obtained under data-hiding scenarios.

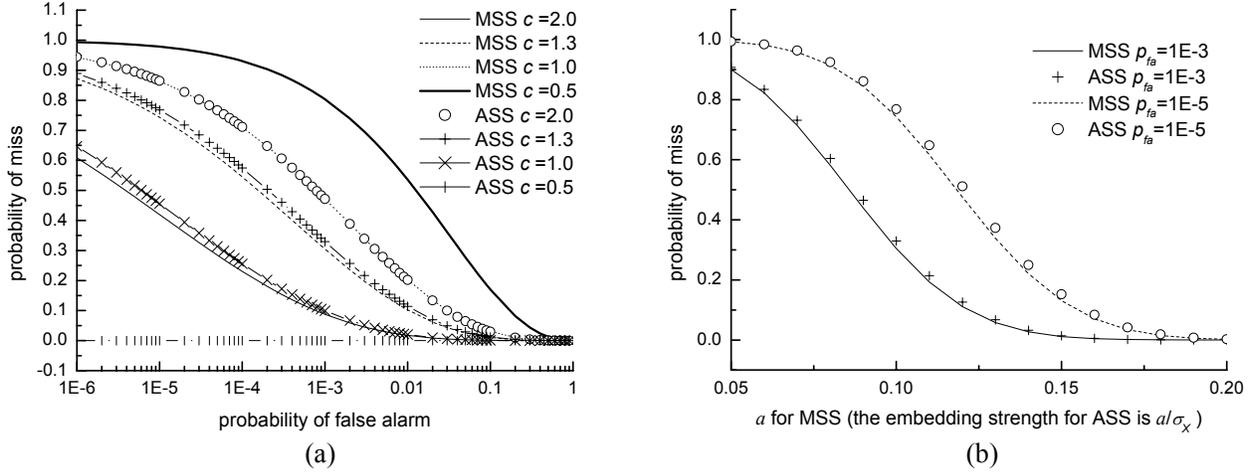

(a)                                                        (b)

Fig. 3.10. Theoretical Performance comparisons between MSS and ASS at $\xi = c$. (a) At different $c$s with $N = 1000$, DWR = 20dB and $\sigma_X = 10$. (b) Comparisons at a fixed false alarm probability with $c = 1.3$ and $\sigma_X = 10$ and $N = 1000$.

### 3.3.6  Performance comparisons under attacks

At the same distortion level $D_w$, (3.71) and (3.85) can be expressed as

$$p_m = 1 - Q\left[ Q^{-1}(p_{fa}) - \sqrt{ND_w/(\sigma_X^2 + \sigma_V^2)} \right],\tag{3.91}$$

$$p_m = 1 - Q\left[ Q^{-1}(p_{fa}) - \sqrt{2ND_w}\,\sigma_X/(\sigma_X^2 + \sigma_V^2) \right],\tag{3.92}$$

for Gaussian hosts and attacks. Comparing the above two equations, we find that MSS has a better performance if $\sigma_X > \sigma_V$. This result is clearly reflected in Fig. 3.11 where MSS intersects ASS at $\sigma_X = \sigma_V = 10$. Fig. 3.12 compares MSS with ASS at $c = 1.2$ and 1.3. It can be clearly inferred from both figures that MSS outperforms ASS at $c$ above 1.3, which coincides with the result for information hiding.

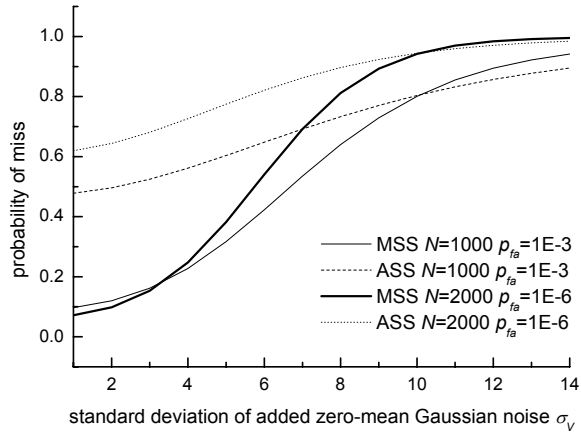

Fig. 3.11. Performance comparisons between MSS and ASS under zero-mean Gaussian noise attacks at DWR = 20dB, $\xi = c = 2.0$ and $\sigma_X = 10$.

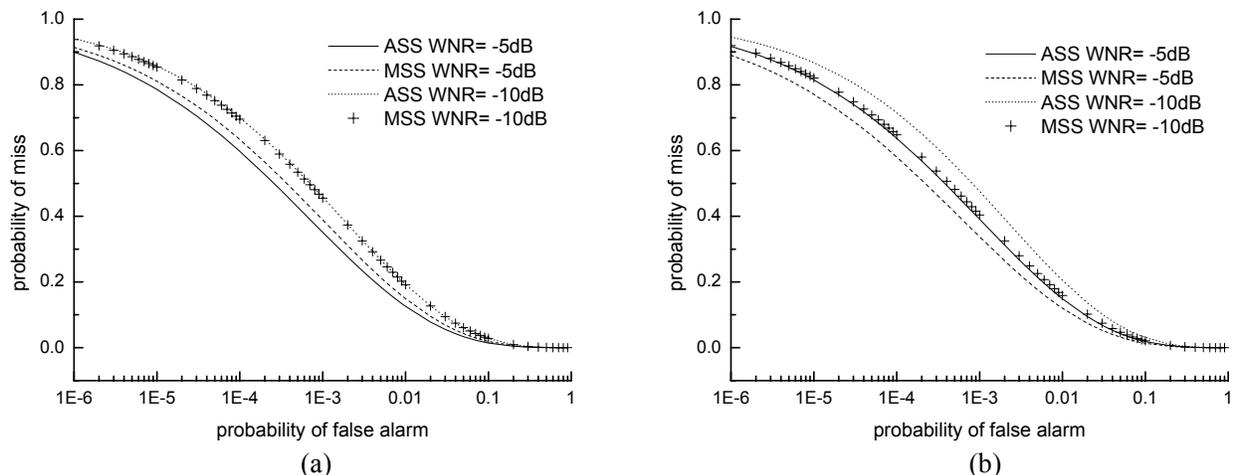

Fig. 3.12. Theoretical Performance comparisons between MSS and ASS under zero-mean Gaussian noise attacks at $N = 1000$, DWR = 20dB, $\sigma_X = 10$. (a) $\xi = c = 1.2$. (b) $\xi = c = 1.3$.

### 3.3.7 Monte-Carlo Simulations

In this section, we made Monte-Carlo simulations to verify the theoretical results obtained in the previous sections. In the legends of figures, "E" stands for empirical results, "T" for theoretical results and "OPT" for the optimum detector (3.64). Fig. 3.13(a) and Fig. 3.13(b) demonstrate the empirical results obtained for ASS under both no attack and attacks. It can be clearly observed from both figures that the theoretical and the empirical results agree well. The comparison results for MSS are displayed in Fig. 3.14(a) and Fig. 3.14(b). The same nice agreements between theoretical and empirical results are observed in these two figures. Through comparisons, we also found that the generalized correlator achieves almost the same performance with the optimum detector (3.64). This explains why we instead adopt the generalized correlator in our discussions.

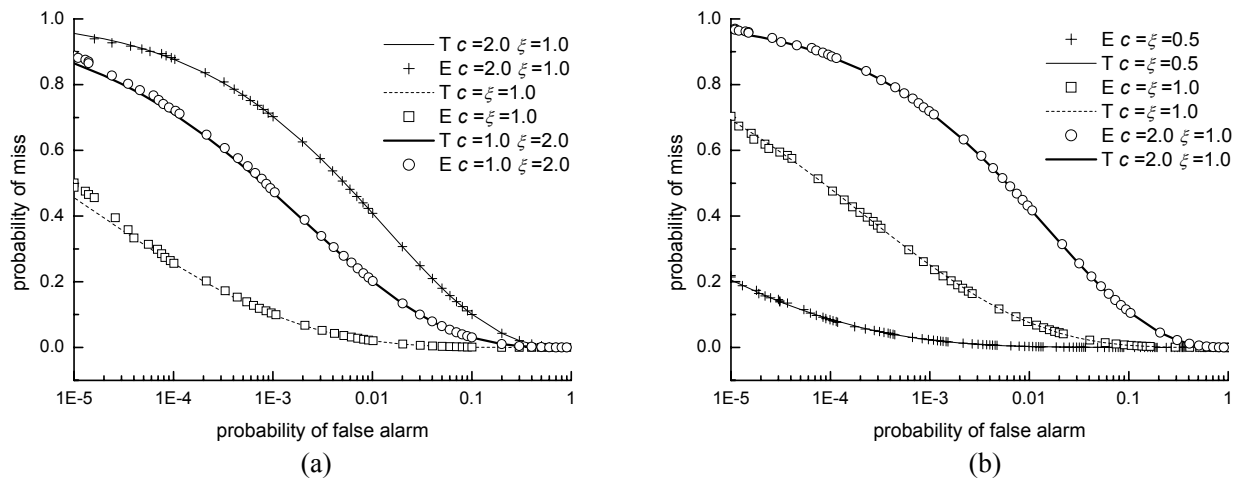

Fig. 3.13. Theoretical and empirical performances of ASS at $N = 1000$, DWR = 20dB and $\sigma_X = 10$. Empirical results are obtained on 1,000,000 groups of GGD host data. (a) Under no attack. (b) Under zero-mean Gaussian noise attacks with WNR = −5dB.

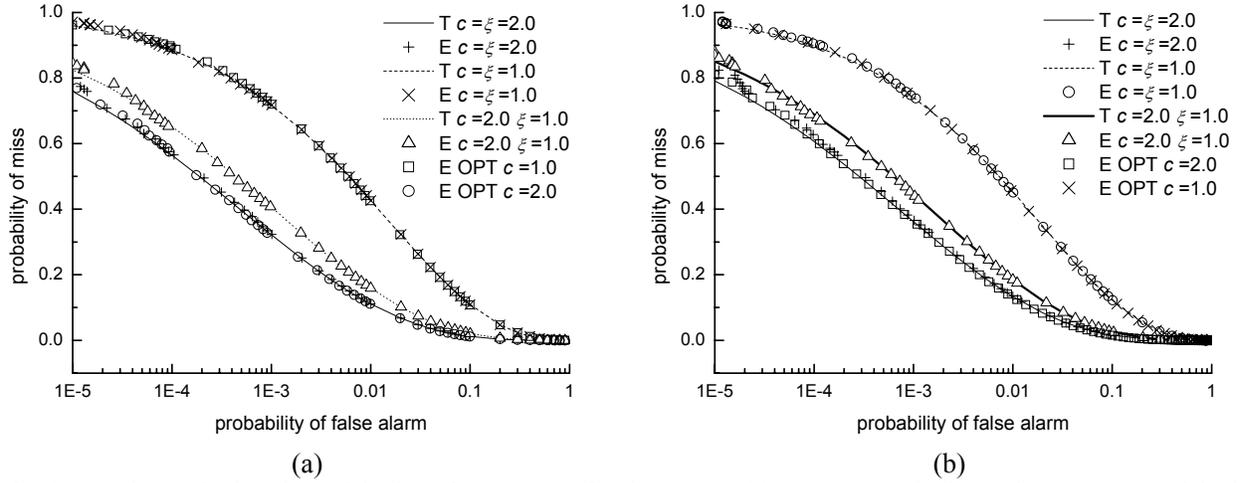

(a)                                                    (b)

Fig. 3.14. Theoretical and empirical performances of MSS at $N = 2000$, DWR = 25dB and $\sigma_X = 10$. Empirical results are obtained on 1,000,000 groups of GGD host data. (a) Under no attack. (b) Under zero-mean Gaussian noise attacks with WNR = −10dB.

## 3.4  Discussions

The authors in [77] also called $\xi$-order detectors generalized correlators. However, their analysis for the variance of the decision statistic is not correct (see (20) in [77]). In fact, we have substituted $c = 0.5$ into their variance and found that the detector with $\xi = 1$ outperforms that with $\xi = 0.5$. The mistake in their analysis lies in their assumption that both the watermark sequence and host signals are random variables, whereas their optimum test is derived under the assumption that only host signals are random variables.

Here we assume that $\mathbf{w}$ is a random vector with each element $w_i$ independent and identically distributed as $P(w_i = 1) = P(w_i = -1) = 0.5$. However, for consistency, we don't change the notation style of $\mathbf{w}$ though it should be uppercased. Under $H_0$, we have

$$L(\mathbf{S} \mid H_0) = \frac{1}{N} \sum_{i=1}^{N} |x_i|^{\xi} w_i .  \tag{3.93}$$

Therefore, we obtain

$$m_0 = 0 \text{ and } \sigma_0^2 = \frac{1}{N} E\left(|X|^{2\xi}\right) .  \tag{3.94}$$

Similarly, we have

$$L(\mathbf{S}\,|\,H_1) = \frac{1}{N}\sum_{i=1}^{N}|S_i|^{\xi}\,w_i \approx \frac{1}{N}\sum_{i=1}^{N}|x_i|^{\xi}\,w_i + \frac{\xi a}{N}\sum_{i=1}^{N}|x_i|^{\xi}\,.$$  (3.96)

Thus, it leads to

$$m_1 = \xi a E\big(|X|^{\xi}\big)\,,$$  (3.97)

$$\sigma_1^2 = Var\left(\frac{1}{N}\sum_{i=1}^{N}|X_i|^{\xi}W_i\right) + Var\left(\frac{\xi a}{N}\sum_{i=1}^{N}|X_i|^{\xi}\right) + 2Cov\left(\frac{1}{N}\sum_{i=1}^{N}|X_i|^{\xi}W_i,\ \ \frac{\xi a}{N}\sum_{i=1}^{N}|X_i|^{\xi}\right)$$

$$= \frac{E\big(|X|^{2\xi}\big)}{N} + \frac{\xi^2 a^2 Var\big(|X|^{\xi}\big)}{N} + 0$$

$$\approx \frac{E\big(|X|^{2\xi}\big)}{N}$$  (3.98)

since $a << 1$ and $Var(|X|^{\xi}) < E(|X|^{2\xi})$, where $Cov$ represents covariance. Thus, the performance is described by

$$p_m = 1 - Q\left[Q^{-1}(p_{fa}) - \xi a\sqrt{N}\,E\big(|X|^{\xi}\big)\Big/\sqrt{E\big(|X|^{2\xi}\big)}\right].$$  (3.99)

which is larger than (3.83). For instance, at $c = 0.5$, we find that $\xi E\big(|X|^{\xi}\big)\Big/\sqrt{E\big(|X|^{2\xi}\big)}$ is 0.408 and MVR($\xi$) = 0.707 if $\xi = 0.5$; $\xi E\big(|X|^{\xi}\big)\Big/\sqrt{E\big(|X|^{2\xi}\big)}$ is 0.548 and MVR($\xi$) = 0.655 if $\xi = 1.0$. In fact, $E\big(|X|^{\xi}\big)\Big/\sqrt{E\big(|X|^{2\xi}\big)}$ is an increasing function of $\xi$. Thus, there is no best performance for the case discussed in this subsection.

To verify this, we also made Monte-Carlo simulations at both $c = 0.5$ and $c = 1.0$. The experimental results are displayed in Fig. 3.15. In both Fig. 3.15(a) and Fig. 3.15(b), we found that the best performance is not achieved at $\xi = c$ and the performance increases as $\xi$ increases. Thus, we emphasize in the beginning of this work that our theoretical results are based on the assumption that the watermark sequence is given.

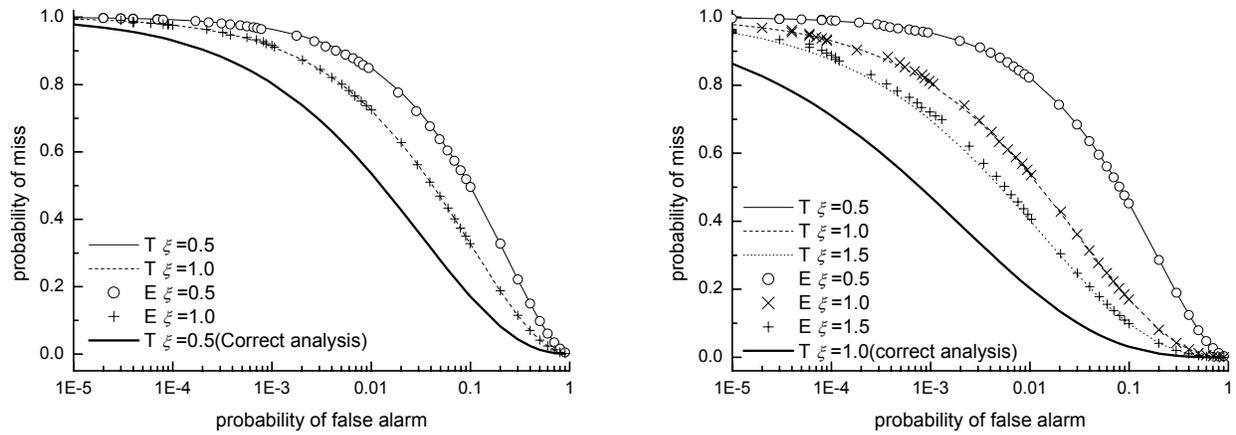

Fig. 3.15. Cheng's performance analysis and our correct analysis at $N = 1000$ and $a = 0.1$. Empirical results are obtained on 100,000 groups of host data. (a) $c = 0.5$ and $\sigma_X = 10$. (b) $c = 1.0$ and $\sigma_X = 10$.

## 3.5  Conclusions

This chapter compared the performance of additive and multiplicative spread spectrum schemes. The comparisons are carried out in the DCT domain where DCT data are assumed to follow the generalized Gaussian distributions. Through comparisons, we found that on the contrary to MSS, ASS produces a better performance as $c$ decreases. Furthermore, MSS outperforms ASS at $c$ above 1.3 (under no attack or moderate attacks). This observation also provides a criterion on selecting MSS or ASS. The second conclusion relates to the type of strongest noise attacks. We show that the Gaussian noise attacks are not the strongest additive noise attacks for watermarking schemes. For ASS, the attacking GGD noise with a shape parameter larger than 2.0 causes even more severe performance degradations.

# Chapter 4    Enhanced Multiplicative Spread Spectrum Schemes — For Watermark Verification

## 4.1 Introduction

For SS schemes, the host interference reduces the performance of the watermark detector or decoders. There are basically two approaches to improve this situation. The first approach, utilizing the host's pdf to obtain an improved detector, has been illumined in the previous chapter. Thus this chapter highlights the second approach that cancels the host interference at the embedder. Improved Spread Spectrum (ISS) scheme proposed in [43] exploits the knowledge of host content by projecting them onto the watermark, and this projected host interference is then compensated in the embedding process. A problem with ISS is that it does not take the probability distributions of the host signals into account. In fact, the implicit assumption for ISS is that the host signals are normally distributed, and thus the optimum decision statistic is the linear correlation.

The above two approaches improve the performance of SS schemes in the embedding and detection process, respectively. Can we incorporate both ideas into the same watermarking scheme? This chapter will answer this question. We will extend the ideas explored in [43] to the multiplicative watermarking. However, in our scheme, we also consider the pdf of host signals, and match the embedding with optimum detection or decoding rules. This chapter will mainly focus on watermark verification problems in the DCT domain. However, we will also extend the ideas presented in this chapter to the data hiding problems in the next chapter.

The rest of the paper is organized as follows. Section 4.2 presents the optimum embedding tailored to the optimum detection rules. Further discussion on our scheme and its performance comparisons with quantization schemes are included in Section 4.3. In section 4.4, Monte-Carlo simulations are done to verify the theoretical results derived in the previous sections. However, the experimental results on real and natural images are reported in Section 4.5. Finally, Section 4.6 summarizes this chapter.

## 4.2  Enhanced multiplicative spread spectrum scheme

### 4.2.1  New embedding rules

The watermark detection problem can be formulated by two hypotheses $H_0$: $S_i = X_i$, and $H_1$: $S_i = X_i (1 + a\, w_i)$. In the previous chapter, we have obtained an $\xi$-order detector $L(\mathbf{S}) = \sum_{1 \leq i \leq N} |S_i|^\xi \cdot w_i$. Under $H_0$, the decision statistic is

$$L(\mathbf{S} \mid H_0) = \frac{1}{N} \sum_{i=1}^{N} |X_i|^\xi w_i . \tag{4.1}$$

Its mean and variance is given by (See (3.80))

$$m_0 = 0, \ \sigma_0^2 = Var(|X|^\xi)/N . \tag{4.2}$$

Under $H_1$, we have

$$L(\mathbf{S} \mid H_1) \approx \frac{1}{N} \sum_{i=1}^{N} |X_i|^\xi w_i + \frac{\xi a}{N} \sum_{i=1}^{N} |X_i|^\xi , \tag{4.3}$$

where the approximation technique (3.12) is employed. We notice that the first term in (4.3) does not depend on the embedding strength $a$ and thus is the host interference on the decision statistic. This inference hampers the watermark detection and should be removed to improve the detector's performance. With this in mind, we propose an Enhanced Multiplicative Spread Spectrum (EMSS) scheme whose embedding rule is given by

$$s_i = x_i(1 + aw_i - \lambda \eta w_i/\gamma), \ \ i = 1, 2, \cdots, N, \tag{4.4}$$

where $\lambda$ ($0 \leq \lambda \leq 1.0$) is called the rejecting strength and $\eta$ is the distribution factor (DF) given by

$$\eta = \left( \sum_{i=1}^{N} |x_i|^\gamma w_i \right) \Big/ \sum_{i=1}^{N} |x_i|^\gamma . \tag{4.5}$$

In (4.5), the numerator comes from the host interference and spreads evenly among all host signals. Thus the third term in (4.4) cancels the host interference and $\lambda$ controls the strength of the host interference being eliminated. If $\lambda = 0$, EMSS degenerates into MSS. The above embedding rule (4.4) is called $\gamma$-order embedder and $\gamma$ is a parameter closely related with the shape parameter $c$. Our future discussion will show that the best performance for EMSS is achieved at $\gamma = \xi = c$.

## 4.2.2  Distribution Factor

The distribution factor $\eta$ is of great importance for our scheme since larger $\eta$s would incur greater embedding distortion on the host contents and invalidate the approximation technique (3.12) (which is employed in (4.9)). By invoking the central limit theorem, the pdf of $\eta$ can be approximated by a Gaussian pdf with a mean and variance given by

$$E(\eta) = 0, \ \text{var}(\eta) = \sigma_\eta^2 \approx \text{var}\big(|X|^\gamma\big) / \{N[E(|X|^\gamma)]^2\} \ . \tag{4.6}$$

The detailed proof of (4.6) can be found in Appendix A. Therefore, we see from (4.6) that $\eta$ is statistically small at a sufficiently large $N$.

## 4.2.3  Embedding distortion

Our method incurs more distortion than MSS method since it uses this extra distortion to compensate the host interference. The embedding distortion $D_w$ for EMSS is

$$D_w \approx a^2 \sigma_X^2 + (\lambda^2 / \gamma^2) \sigma_X^2 \sigma_\eta^2 \ . \tag{4.7}$$

The detailed deduction of the above equation can be found in Appendix B. Precise calculation of the expected distortion is also available in Appendix B. However, for comparison convenience, we are just satisfied with this simple approximation. Inserting (4.6) into (4.7), we obtain

$$D_w \approx a^2 \sigma_X^2 + [\lambda / \text{MVR}(\gamma)]^2 \sigma_X^2 / N \ . \tag{4.8}$$

where MVR($\cdot$) is defined in (3.33). Since MVR reaches its maximum at $\gamma = c$, the embedding distortion achieves its minimum at $\gamma = c$. Please also note that the second term in (4.8) is the extra distortion incurred by our scheme.

## 4.2.4  Performance

For EMSS, we adopt the same $\xi$-order detector (3.66) for watermark detection. It is difficult to obtain an optimum detector for EMSS since $\eta$ is a random variable. However, as $\eta$ is statistically small at a large $N$, (3.66) may be a suboptimal choice for EMSS. Plugging (4.4) into (3.66) and using the same approximation technique (3.12), we obtain

$$L(\mathbf{S} \mid H_1) \approx \frac{1}{N} \sum_{i=1}^{N} |X_i|^{\xi} (1 + \xi a w_i - \frac{\lambda \eta \xi}{\gamma} w_i) w_i = \frac{1}{N} \sum_{i=1}^{N} |X_i|^{\xi} w_i + \frac{\xi a}{N} \sum_{i=1}^{N} |X_i|^{\xi} - \frac{\lambda \xi \eta}{N \gamma} \sum_{i=1}^{N} |X_i|^{\xi} . \quad (4.9)$$

If $\gamma = \xi$, the above equation can be expressed as

$$L(\mathbf{S} \mid H_1) = \frac{(1-\lambda)}{N} \sum_{i=1}^{N} |X_i|^{\xi} w_i + \frac{\xi a}{N} \sum_{i=1}^{N} |X_i|^{\xi} . \quad (4.10)$$

Consequently, for (4.10), we have

$$m_1 \approx \xi a E(|X|^{\xi}), \quad (4.11)$$

$$\sigma_1^2 = [(1-\lambda)^2 + \xi^2 a^2] \cdot Var(|X|^{\xi}) / N . \quad (4.12)$$

Putting (4.11), (4.12) and (4.2) into (3.69), we obtain

$$p_e = 1 - Q\left[ \frac{\sqrt{var(|X|^{\xi})} Q^{-1}(p_{fa}) - \sqrt{N} \xi a E(|X|^{\xi})}{\sqrt{(1-\lambda)^2 + \xi^2 a^2} \sqrt{var(|X|^{\xi})}} \right]. \quad (4.13)$$

It is a bit more complex to derive the performance at $\gamma \neq \xi$. In Appendix C, we show that

$$m_1 = \xi a E(|X|^{\xi}), \quad (4.14)$$

$$\sigma_1^2 \approx \frac{1 + \xi^2 a^2}{N} var(|X|^{\xi}) + \frac{\lambda^2 \xi^2}{N \gamma^2} \frac{[E(|X|^{\xi})]^2 var(|X|^{\gamma})}{[E(|X|^{\gamma})]^2} - \frac{2\lambda \xi}{N \gamma} \frac{E(|X|^{\xi})[E(|X|^{\xi+\gamma}) - E(|X|^{\xi})E(|X|^{\gamma})]}{E(|X|^{\gamma})} . \quad (4.15)$$

In particular, when $\gamma = \xi$, (4.15) is identical to (4.12). Thus the detector's performance at $\gamma \neq \xi$ can be characterized by substituting (4.14), (4.15) and (4.2) into (3.69).

## 4.2.5 Optimality at $\gamma = \xi = c$

In this subsection, we prove that EMSS achieves its best performance at $\gamma = \xi = c$ for a fixed $\lambda$ and compare EMSS with MSS at the same distortion level. To continue the discussion, we define the critical false alarm probability as

$$p_{fa}^{crit} = Q[(m_1 - m_0) / \sigma_0], \quad (4.16)$$

which is the false alarm probability at which the probability of miss is 0.5. We choose 0.5 because this probability of miss is achieved at $\psi = m_1$ in Fig. 4.2 and $m_1$ is a turning point above which EMSS may outperform MSS. Substituting both (4.2) and (4.11) into (4.16) and with (4.8), we obtain

$$p_{fa}^{crit} = Q\{\text{MVR}(\xi)\sqrt{N \cdot 10^{-\text{DWR}/10} - \lambda^2/[\text{MVR}(\gamma)]^2}\} . \tag{4.17}$$

Since MVR achieves its maximum at $c$, (4.17) reaches its minimum at $\gamma = \xi = c$ for any fixed $\lambda$. In Fig. 4.1(a), $\gamma = \xi = 1.0$ has the lowest critical false alarm (shown by point 3, which is smaller than point 4 and 5) among all EMSS curves. In Fig. 4.1(b), EMSS with $\gamma = \xi = 1.0$ also has the lowest critical false alarm. Since the critical false alarm rate decides the performance at the high false alarm rates (see Fig. 4.1), EMSS with $\gamma = \xi = c$ achieves the best performance at the high false alarm rates. Substituting $\gamma = \xi = c$ into (4.17), we have

$$p_{fa}^{crit} = Q\{\sqrt{N \cdot 10^{-\text{DWR}/10} \cdot [\text{MVR}(c)]^2 - \lambda^2}\} . \tag{4.18}$$

Thus, we see that the larger the shape parameter $c$, the lower the critical false alarm and thus the better the performance.

Now we make performance comparisons between MSS and EMSS. By letting $\lambda = 0$ in (4.18), the critical false alarm for MSS is given by

$$p_{fa}^{crit} = Q\{\sqrt{N \cdot 10^{-\text{DWR}/10} \cdot [\text{MVR}(c)]^2}\} . \tag{4.19}$$

Comparing (4.18) with (4.19), we find that MSS has a lower critical false alarm (shown by point 2 in Fig. 4.1(a)). The cross false alarm probability is defined as the false alarm probability where ROC curves for MSS and EMSS intersect (see point 1 in Fig. 4.1(a)). Since MSS has a lower critical false alarm, the probability of miss at the cross false alarm probability is smaller than 0.5. However, since the critical false alarm probability for MSS (4.19) and EMSS (4.18) does not differ much if $N \cdot 10^{-\text{DWR}/10} \cdot [\text{MVR}(c)]^2 >> 1$, the probability of miss at the cross false alarm probability would be quite close to 0.5. This indicates that if MSS achieves a probability of miss smaller than 0.5 at a specified false alarm, EMSS can achieve a better performance at the same false alarm; otherwise, EMSS is inferior to MSS.

The advantage of MSS over EMSS at the low false alarm rates is further explained in Fig. 4.2. It is shown in (4.12) that the variance of decision statistic under $H_1$ for EMSS is very small when $\lambda = 1.0$. Thus, when the decision threshold $\psi$ is set between $m_0$ and $m_1$ (in Fig. 4.2), the probability of miss is about 0.0; however, when the threshold is above $m_1$, as indicated in Fig. 4.2, the probability of miss is close to 1.0. This also explains the sharp curves for EMSS at $\gamma = \xi$ in Fig. 4.1(a). However, there is no such a dramatic performance drop for MSS. Thus, if the threshold were set above $m_1$ to achieve a low false alarm, MSS would have a lower probability of miss.

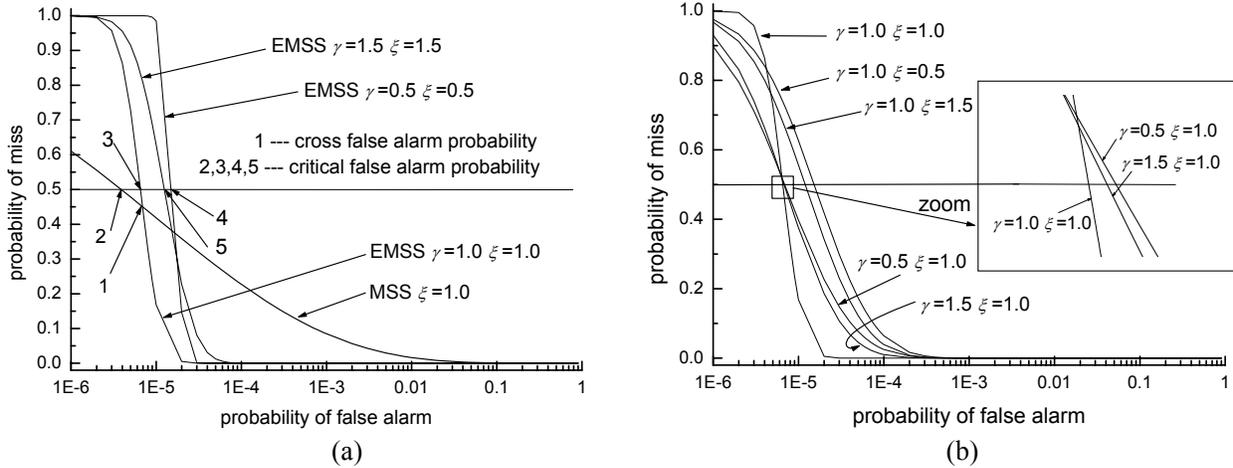

(a)                                                                                 (b)

Fig. 4.1. Optimality of EMSS at $\gamma = \xi = c$ (with DWR = 20dB, $c = 1.0$, $\sigma_X = 10.0$, $N = 2000$, $\lambda = 1.0$). (a) Critical false alarm probability at $\gamma = \xi$. (b) Critical false alarm probability at $\gamma \neq \xi$.

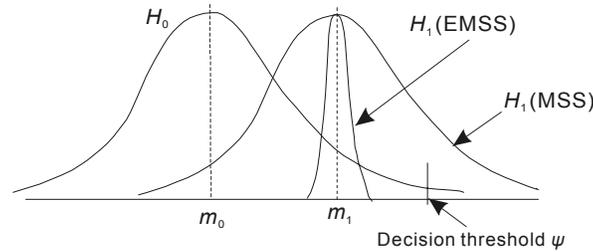

Fig. 4.2. $H_0$ and $H_1$ represent the statistic $L(\mathbf{S}\,|H_0)$ and $L(\mathbf{S}\,|H_1)$, where $m_0 = E[L(\mathbf{S}\,|H_0)]$ and $m_1 = E[L(\mathbf{S}\,|H_1)]$.

## 4.3  Discussions

### 4.3.1  Comparisons with other schemes

An interesting problem is that this chapter formulated the hypothesis where no attack is assumed. Liu [101] and Perez-Freire [102] instead considered the optimal detection for a more practical scenario where the attack is also included in the hypothesis testing. In this section, we will investigate this scenario. The scenario under attacks can be formulated as

$H_0$: $Y_i = X_i + V_i$ and $H_1$: $Y_i = S_i + V_i$ for $i = 1, 2, \ldots, N$. (4.20)

In this subsection, we consider the normally distributed host signal since the closed-form optimum detection rule for ASS is possible and the performance for EMSS can also be derived under this assumption.

*1) ASS and STDM:* For ASS, the performance is given by (3.71). Perez [102] applied the Spread Transform Dither Modulation (STDM) scheme (a special case of the general QIM scheme) to the watermark verification problems. In this case, the embedding rule is formulated as (2.31) with symmetric dithers $d = \Delta/2$. The performance is determined by [102]

$$p_{fa} = \sum_{k=-\infty}^{k=+\infty} Q\left[\frac{k\Delta + 0.5\Delta - \psi}{\sqrt{(\sigma_X^2 + \sigma_V^2)/N}}\right] - Q\left[\frac{k\Delta + 0.5\Delta + \psi}{\sqrt{(\sigma_X^2 + \sigma_V^2)/N}}\right], \tag{4.21}$$

$$p_m = 1 - \left\{\sum_{k=-\infty}^{k=+\infty} Q\left[\frac{k\Delta - \psi}{\sqrt{\sigma_V^2/N}}\right] - Q\left[\frac{k\Delta + \psi}{\sqrt{\sigma_V^2/N}}\right]\right\}. \tag{4.22}$$

where $\psi$ ($0 \le \psi \le \Delta/2$) is the decision threshold.

*2) MSS:* For MSS, the above problem (4.20) can be formulated as $H_0$: $Y_i = X_i + V_i$ and $H_1$: $Y_i = X_i(1 + a\,w_i) + V_i$. Closed-form optimum decision rules are only possible for Gaussian hosts and attacks. It is also difficult to extend the host-interference canceling technique proposed in this chapter to this scenario since the attack data cannot be known beforehand. The optimum decision rule can be easily derived from the likelihood ratio test and is given as

$$L(\mathbf{Y}) = \frac{1}{N}\sum_{i=1}^{N} Y_i^2/(\sigma_X^2 + \sigma_V^2) - Y_i^2/[\sigma_X^2(1 + aw_i)^2 + \sigma_V^2]. \tag{4.23}$$

Thus, the performance is determined by

$$m_0 = (\sigma_X^2 + \sigma_V^2)(\rho_0 + \rho_1)/2, \tag{4.24}$$

$$\sigma_0^2 = (\sigma_X^2 + \sigma_V^2)^2(\rho_0^2 + \rho_1^2)/N, \tag{4.25}$$

$$m_1 = \{[\sigma_X^2(1 + a)^2 + \sigma_V^2]\rho_0 + [\sigma_X^2(1 - a)^2 + \sigma_V^2]\rho_1\}/2, \tag{4.26}$$

$$\sigma_1^2 = \{[\sigma_X^2(1 + a)^2 + \sigma_V^2]^2\rho_0^2 + [\sigma_X^2(1 - a)^2 + \sigma_V^2]^2\rho_1^2\}/N, \tag{4.27}$$

where $\rho_0 = 1/(\sigma_X^2 + \sigma_V^2) - 1/[\sigma_X^2(1 + a)^2 + \sigma_V^2]$ and $\rho_1 = 1/(\sigma_X^2 + \sigma_V^2) - 1/[\sigma_X^2(1 - a)^2 + \sigma_V^2]$. If $a$ is small enough, (4.23) can be approximated by

$$L(\mathbf{Y}) = \frac{1}{N} \sum_{i=1}^{N} Y_i^2 \frac{\sigma_X^2 (2aw_i + a^2)}{(\sigma_X^2 + \sigma_V^2)[\sigma_X^2 (1 + 2aw_i + a^2) + \sigma_V^2]} \approx \frac{1}{N} \sum_{i=1}^{N} Y_i^2 \frac{\sigma_X^2 (2aw_i)}{(\sigma_X^2 + \sigma_V^2)(\sigma_X^2 + \sigma_V^2)} \tag{4.28}$$

We recognize that (4.28) differs from (3.66) (with $\xi = 2.0$) just by some constant and thus is equivalent to (3.66). That is, the optimum rule derived under no attack still approximately holds under attacks. Nevertheless, (4.23) still enjoys a slight performance advantage over (3.66) under the same level of attacks.

  *3) EMSS:* The empirical performance of EMSS for non-Gaussian hosts under attacks will be demonstrated through experiments in Section 4.5. The theoretical performance for Gaussian hosts can be derived by substituting (4.20) into (3.66) with $\gamma = \xi = 2.0$. Thus, under $H_0$, we have

$$m_0 = 0, \quad \sigma_0^2 = 2(\sigma_X^2 + \sigma_V^2)^2 / N . \tag{4.29}$$

Similarly, the decision statistic under $H_1$ is given by

$$L(\mathbf{Y} \mid H_1) = \frac{1}{N} \sum_{i=1}^{N} [X_i (1 + aw_i - \lambda \eta w_i / 2) + V_i]^2 w_i$$

$$\overset{(a)}{\approx} \frac{1}{N} \sum_{i=1}^{N} (1 - \lambda) X_i^2 w_i + 2aX_i^2 + V_i^2 w_i + 2V_i X_i (w_i + a - \frac{\lambda \eta}{2})$$

$$\overset{(b)}{\approx} \frac{1}{N} \sum_{i=1}^{N} (1 - \lambda) X_i^2 w_i + 2aX_i^2 + V_i^2 w_i + 2V_i X_i (w_i + a) \tag{4.30}$$

where (*a*) follows from (4.5) with $\gamma = 2.0$ and the approximation technique (3.12), and (*b*) from the fact that $\eta$ is small if $N$ is large enough (see (4.6)). Thus,

$$m_1 = 2a\sigma_X^2 , \tag{4.31}$$

$$\sigma_1^2 = \{2[(1 - \lambda)^2 + 4a^2] \sigma_X^4 + 2\sigma_V^4 + (4 + 4a^2) \sigma_X^2 \sigma_V^2\} / N . \tag{4.32}$$

Hence the performance for EMSS under attacks can be determined by the above equations.

  *4) Comparisons:* Please first note that in Fig. 4.3, the text "E" in the legend of figures stands for Empirical results, and "T" for Theoretical results. Furthermore, the step size $\Delta$ for STDM is empirically obtained through Monte-Carlo simulations. In Fig. 4.3(a), the theoretical performance for EMSS is in line with the simulation results. In the figure, we also found that MSS yields a better performance than ASS. This is in line with the results in the previous chapter (See Section 3.3.6) since $c = 2.0$ and $\sigma_V < \sigma_X$. Moreover, EMSS and STDM produce a nicer performance than ASS and MSS since both schemes cancel the host interference.

We also observed that the optimum detector derived under no attack (3.66) achieves almost the same performance as the optimum detector derived under attacks (4.23). Please note that the performance of (3.66) under attacks is given by (3.85) or obtained by setting $\lambda = 0$ in (4.32). In Fig. 4.3(b), the comparison results show that EMSS works better than STDM at the low false alarm rates. However, it is still important to note that EMSS cannot work so better since the shape parameter $c$ for GGD host signals is usually smaller than 2.0 for a real image. Nevertheless, the multiplicative spread spectrum scheme can achieve a nicer perceptual quality.

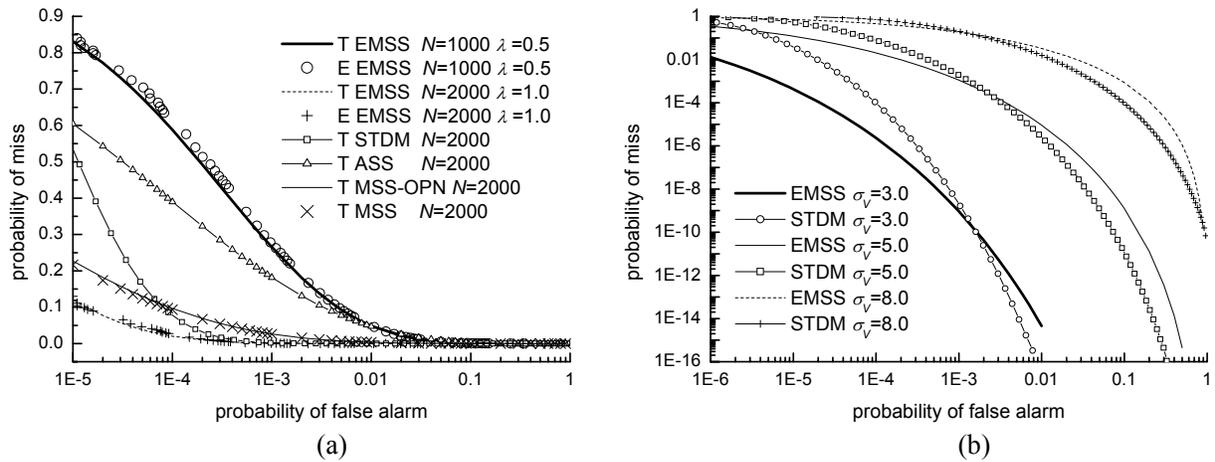

(a)                                                            (b)

Fig. 4.3. Performance comparisons between quantization and spread spectrum schemes under zero-mean Gaussian noise attacks at $c = 2.0$, $\sigma_X = 10$ and DWR = 20dB (namely, $a = 1.0$ for ASS, $\Delta = 2.34$ for STDM). (a) Performance comparisons at $\sigma_V = 5$. Empirical results for EMSS were obtained on 1,000,000 groups of data. Please note that "MSS-OPN" represents the optimum detector given by (4.23). (b) Theoretical performance comparisons between EMSS and STDM at $N = 2000$, $\lambda = 1.0$.

### 4.3.2 EMSS for multiple DCT coefficients

In this work, we assume that the host DCT signals have the same shape parameter $c$. In more practical scenarios, multiple coefficients that have different $c$s may be watermarked to achieve sufficient robustness. Here, for instance, we consider two groups of data from different frequencies. Let the first group be denoted by $\{X_1, X_2, \ldots, X_N\}$ and the second by $\{X_{N+1}, X_{N+2}, \ldots, X_{2N}\}$. Furthermore, we assume that the first group has a shape parameter $c_1$, and the second group has $c_2$. We embed watermarks into each group by EMSS, respectively. That is,

$$\begin{cases} s_i = x_i(1 + a_1 w_i - \lambda \eta_1 w_i / c_1), & i = 1, 2, \cdots, N \\ s_i = x_i(1 + a_2 w_i - \lambda \eta_2 w_i / c_2), & i = N+1, N+2, \cdots, 2N \end{cases}$$

where $a_1$ and $a_2$ are the embedding strength for the first and second group, respectively, and

$$\eta_1 = (\sum_{i=1}^{N} |x_i|^{c_1} \, w_i) \Big/ \sum_{i=1}^{N} |x_i|^{c_1} \, , \qquad \eta_2 = (\sum_{i=N+1}^{2N} |x_i|^{c_2} \, w_i) \Big/ \sum_{i=N+1}^{2N} |x_i|^{c_2} \, .$$

The decision statistic is

$$L(\mathbf{S}) = \frac{1}{2N} (\sum_{i=1}^{N} |S_i|^{c_1} \, w_i + \sum_{i=N+1}^{2N} |S_i|^{c_2} \, w_i) \, .$$

Thus, the performance is characterized by

$$m_0 = 0 \, , \quad \sigma_0^2 = [Var(|X_1|^{c_1}) + Var(|X_{N+1}|^{c_2})] / (4N) \, ,$$

$$m_1 = [a_1 c_1 E(|X_1|^{c_1}) + a_2 c_2 E(|X_{N+1}|^{c_2})] / 2 \, ,$$

$$\sigma_1^2 = [(1-\lambda)^2 + c_1^2 a_1^2] Var(|X_1|^{c_1}) / (4N) + [(1-\lambda)^2 + c_2^2 a_2^2] Var(|X_{N+1}|^{c_2}) / (4N) \, .$$

The performance analysis for this case is much complex. Thus, in this work, we use just one coefficient for simplicity. The above discussion can be easily extended to the case where more coefficients are involved.

### 4.3.3  Choice of $\lambda$

An important parameter for our scheme is $\lambda$. MSS is indeed a special case of EMSS with $\lambda = 0$. The performance of EMSS with $\lambda < 1.0$ is a trade-off between the performance at high and low false alarm rates. The choice of $\lambda$ can be made optimal (through simulations) for some attacks if the attack parameters are known. Without the knowledge of the possible attacks, an important characteristic that EMSS has a better performance at the probability of miss below 0.5 (see Section 4.2.5) may help to decide which $\lambda$ to use in a real scenario. This characteristic is also reflected in all experimental results (even under attacks) in Section 4.5. Thus, if a system requires that the probability of miss at a specified false alarm be below 0.5, then EMSS (with $\lambda = 1.0$) can be selected. However, if the probability of miss for MSS under the worst possible attacks is larger than 0.5, then MSS ($\lambda = 0.0$) can be pessimistically selected.

## 4.4  Monte-Carlo Simulations

In this section, we perform Monte Carlo simulations to verify the theoretic results derived in this paper. All the data in this section are generated by a pseudo-random number generator.  Please note that the text

"E" in the legend of the figures stands for empirical results obtained from Monte-Carlo simulations, "T" for theoretical results.

In Fig. 4.4(a), EMSS at $\gamma = \xi = 0.5$ achieves the smallest critical false alarm probability, hence validating the optimality investigated in Section 4.2.5. The performance of EMSS with $\lambda$ smaller than 1 is shown in Fig. 4.4(b). The experimental results in both figures agree with the analytical results derived in the previous sections.

Now the simulations are performed to prove the correctness of the expected distortion (4.7). Fig. 4.5(a) shows the expected distortion at different $N$s. The theoretical and simulation results match well. As $N$ increases, the distortion goes down. The simulation results at different $\lambda$s are also included in this figure. The smaller the $\lambda$, the less the distortion EMSS incurs. Fig. 4.5(b) demonstrates the correctness of the derived expected distortion for $\gamma \neq c$. The agreement between theoretical and simulation results validates the effectiveness of the approximation approach (See (A.8) in Appendix A). Moreover, as shown by (4.8), the distortion achieves its minimum at $\gamma = c$ in Fig. 4.5(b). We have done many other simulations to prove the derived theoretical distortion. However, the results are not reported here for space considerations.

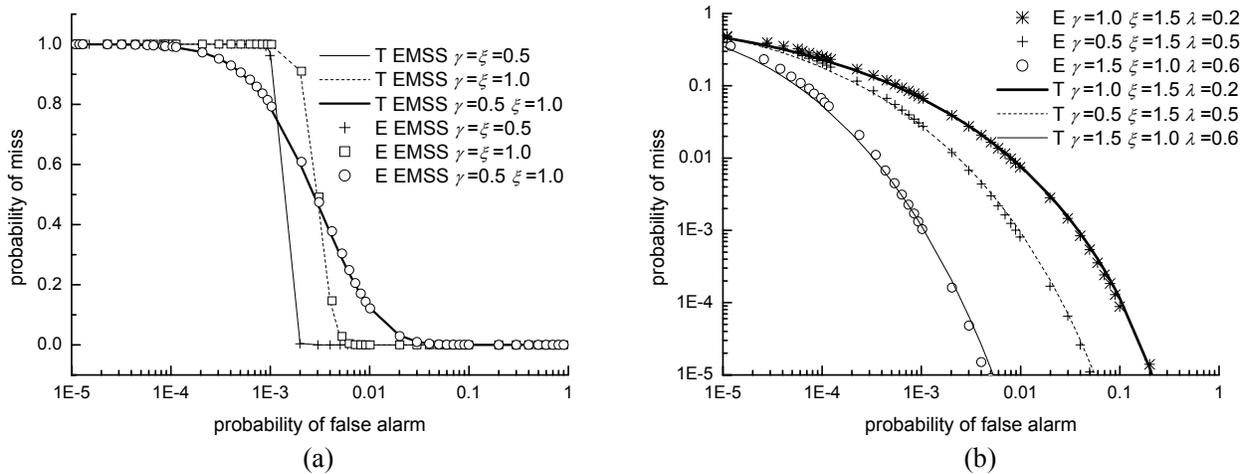

(a)                                                                                    (b)

Fig. 4.4. Theoretical and empirical ROC curves derived at DWR = 20dB and $N$ = 2000. Empirical results are obtained on 1,000,000 groups of host data. (a) Performance comparisons between EMSS and MSS at $c = 0.5$, $\sigma_X$ = 10.0 (and $\lambda = 1.0$ for EMSS). (b) Performance of EMSS at $c = 1.0$, $\sigma_X = 10.0$.

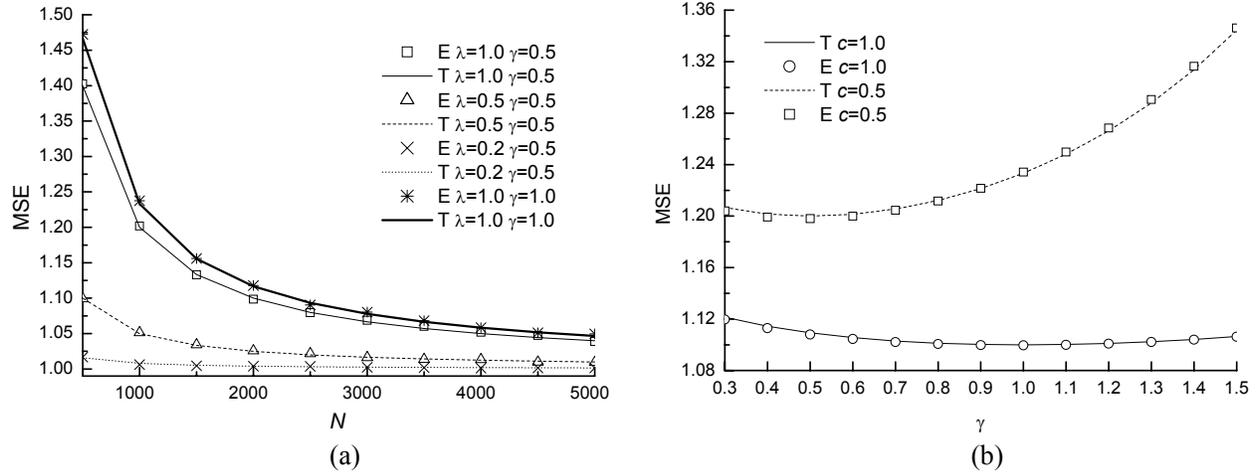

(a)                                                                          (b)

Fig. 4.5. Expected Distortion (The simulation results are obtained on 100, 000 groups of host data with $\sigma_X = 10.0$ and $a = 0.1$). (a) At different $N$ ($c = 0.5$). (b) At different $\gamma$ ($N = 1000$, $\lambda = 1.0$).

## 4.5  Experiments

First note that the text "T" in the legend of the figures stands for Theoretical results, "E" for Experimental results, "MSS" for (3.66) and "EMSS" for (4.4). In this section, the experiments are done on real natural images. For each 8×8 DCT block, we use just one coefficient for watermarking since different DCT frequency coefficients usually have different shape parameters. Thus, using one coefficient in each block can simplify the performance comparisons.

In this section, we choose $\gamma = c$ since it has been shown in Section 4.2.5 that EMSS achieves its best performance at $\gamma = \xi = c$. In the following experiments, we adopt the ML method [103] to estimate the shape parameter $c$ and standard deviation $\sigma_X$ of the host DCT coefficients. For the 5th (in Zigzag order) coefficient of Lena, its estimated $c$ is 0.69 and $\sigma_X = 19.74$; for the 5th coefficient of Peppers, $c = 1.03$ and $\sigma_X = 16.04$. Since $c$ for the watermarked coefficients does not vary much from that for the original data, we instead set $\xi$ as the estimated $c$ of the original data in detection for convenience.

Fig. 4.6 and Fig. 4.7 display the images watermarked by both MSS and EMSS methods. The averaged PSNRs in both figures are obtained over 1,000 watermarked images with different watermark sequences. We find that MSS and EMSS achieve almost the same PSNR at the same DWR. Since we do not have a large set of natural images with the same shape parameter $c$ and standard deviation $\sigma_X$, we instead permute randomly the host signals to obtain different sequences of host data $\mathbf{x}$ for the ease of performance comparisons. Each permutation produces a new **x**, and all the following experimental results are obtained on 1,000,000 such permutations. Please also refer to Appendix E for more details. Moreover, in these experiments, we also set $\xi = \gamma = c$ and $N = 2000$.

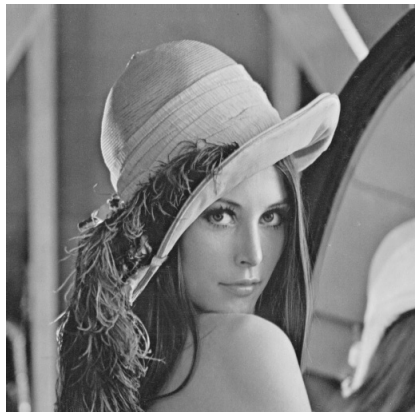
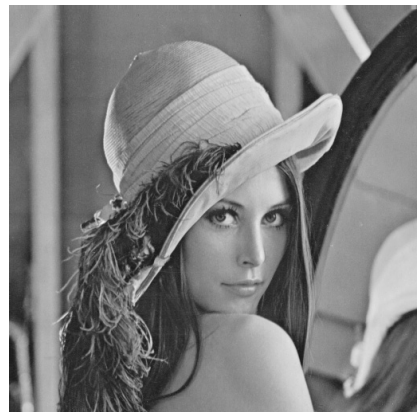

(a)                                                                                     (b)

Fig. 4.6. Watermarked Lena at DWR = 16.48dB. (a) MSS (averaged PSNR = 56.44dB). (b) EMSS (averaged PSNR = 56.36dB).

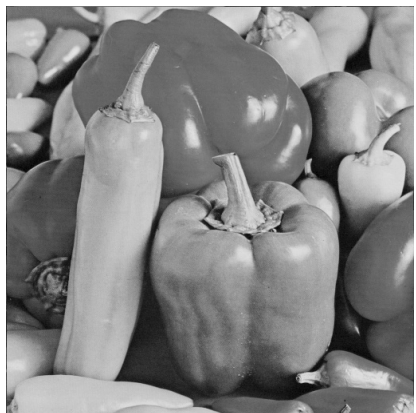
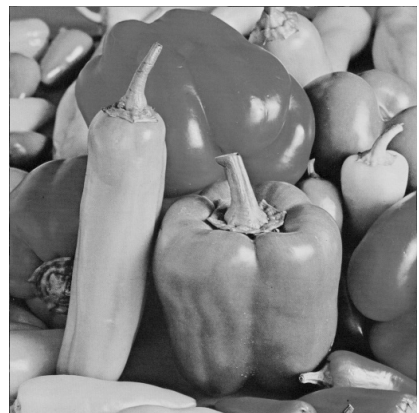

(a)                                                                                     (b)

Fig. 4.7. Watermarked Peppers at DWR = 16.48dB. (a) MSS (averaged PSNR = 58.04dB). (b) EMSS (averaged PSNR = 57.88dB).

The performance comparisons between EMSS and MSS are displayed in Fig. 4.8. The sharp curves discussed in the previous section are also observed in Fig. 4.8(a) and Fig. 4.8(b), and EMSS outperforms MSS at the false alarm rates above the cross false alarm probability. The performance for Peppers is much better since it has a larger $c$. The theoretical results for EMSS at $\lambda = 0.5$ are also included to verify the theoretical conclusion (4.13) and the validity of the estimated parameters.

All the above experiments are conducted without attacks. Now we examine the performance of our scheme under attacks. First we evaluate its performance under noise attacks. The attack noise follows a zero-mean Gaussian distribution with standard deviation 5.0. The results are displayed in Fig. 4.9. For both images, the probability of miss at the cross false alarm is about 0.45, and EMSS outperforms MSS above this miss. JPEG is a common attack the watermarked image may undergo. Fig. 4.10 displays the results under JPEG attacks with Quality Factor (QF) 50. The same observations can be made.

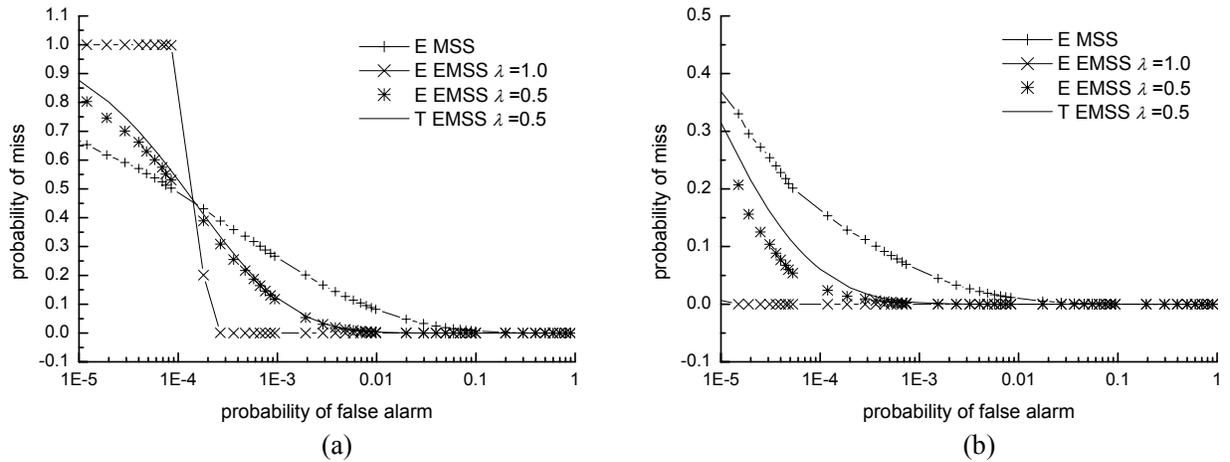

(a)                                              (b)

Fig. 4.8. Performance comparisons under no attack at DWR = 20dB. (a) For Lena. (b) For Peppers

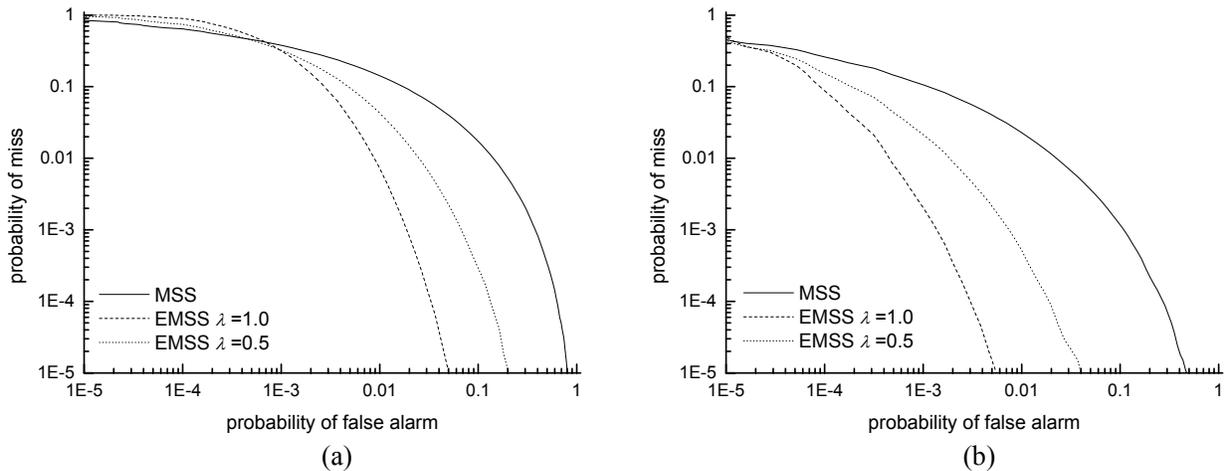

(a)                                              (b)

Fig. 4.9. Performance comparisons at DWR = 20dB under noise attacks with the noise distributed as ~ N(0, 25).
(a) For Lena (b) For Peppers

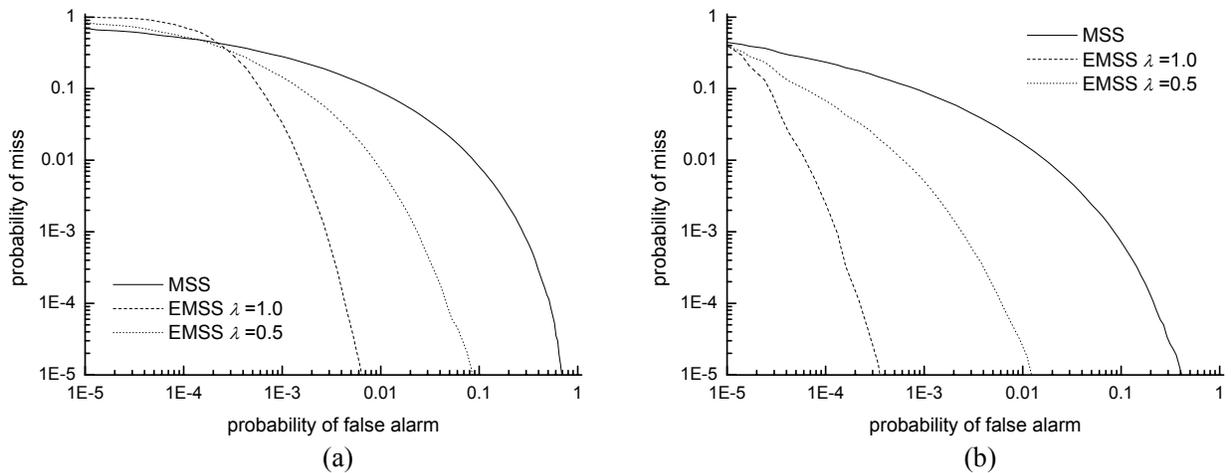

<div align="center">(a)                                                    (b)</div>

Fig. 4.10. Performance comparisons at DWR = 20dB under JPEG attacks with QF = 50. (a) For Lena (b) For Peppers

## 4.6  Conclusions

In this chapter, we proposed an Enhanced Multiplicative Spread Spectrum (EMSS) watermarking scheme that rejects the host interference at the embedder. The new embedder is designed according to the optimum decision rules for the detector. Through the matching of newly designed embedders and detectors, the performance can be improved at high false alarm rates. This is quite important since in the real scenarios, we may increase $N$ or the embedding strength to reduce the cross false alarm probability. In such a case, the performance at the high false alarm rates decides the performance of the whole watermarking systems. Furthermore, for practical applications, we usually require that the probability of miss be smaller than 0.5 at a specified false alarm probability, which thus encourages the use of EMSS.

# Chapter 5   Enhanced Multiplicative Spread Spectrum Schemes — For Data Hiding

## 5.1  Introduction

In the previous chapter, we investigated a host interference canceling technique for watermark verification problems. In this chapter, we extend this technique to the data hiding problems.

The rest of this chapter is organized as follows. In Section 5.2, we employ the host-interference rejection technique proposed in the previous chapter to design a new embedder for data-hiding problems. The simulation and experimental results are presented in Section 5.3 and 5.4, respectively, to validate the effectiveness of our proposal. The last section concludes this chapter.

## 5.2  Embedding rules

### 5.2.1  Enhanced multiplicative embedding rule

For MSS, the embedded bit can be decoded by testing the two alternative hypotheses $H_0$ ($b = -1$): $S_i = X_i(1 - a\,w_i)$ versus $H_1$ ($b = +1$): $S_i = X_i(1 + a\,w_i)$. Thus, the suboptimal generalized correlator (also called $\xi$-order decoder) discussed in Chapter 3 leads to

$$L(\mathbf{S}\,|\,b) \approx \frac{1}{N}\sum_{i=1}^{N}\left|X_i\right|^{\xi}w_i + \frac{\xi b a}{N}\sum_{i=1}^{N}\left|X_i\right|^{\xi}. \tag{5.1}$$

Please also see (3.29) for further reference. We notice that the first term in (5.1) is the host interference on the decision statistic, and therefore should be removed to reduce the probability of errors. With this in mind, we propose an Enhanced Multiplicative Spread Spectrum (EMSS) scheme whose embedding rule is given by

$$s_i = x_i(1 + baw_i - \lambda\eta w_i/\gamma),\quad i = 1,2,\cdots,N, \tag{5.2}$$

where $\lambda$ ($0 \le \lambda \le 1.0$) is called the rejecting strength (RS). In the following discussions, we shall see that $\lambda$ is used to reduce the extra distortion incurred by EMSS. Another important parameter $\eta$ is called the distribution factor (DF) and given by

$$\eta = \left(\sum_{i=1}^{N} |x_i|^{\gamma} w_i\right) \Big/ \sum_{i=1}^{N} |x_i|^{\gamma} . \tag{5.3}$$

The above embedding rule is called $\gamma$-order embedder, and $\gamma$ is closely related with the shape parameter $c$. Our future discussion will show that the best performance for EMSS is achieved at $\gamma = \xi = c$ under no attack. However, the optimum $\gamma$ under attacks, for instance, Gaussian noise attacks, is still unknown. The distribution factor is an approximately Gaussian random variable whose characteristic has been analyzed in Chapter 4. For a small $N$, it may be statistically very large. Thus, $\lambda$ often must be smaller than 1.0 to reduce the embedding distortion to an imperceptible level.

### 5.2.2  Embedding distortion

From (5.2), it is easy to see that the embedding distortion is

$$
\begin{aligned}
D_w &= \frac{1}{N} E\left[\sum_{i=1}^{N} X_i^2 (ba - \lambda \eta / \gamma)^2\right] = \frac{1}{N} E\left[\sum_{i=1}^{N} X_i^2 (a^2 - 2b\lambda \eta / \gamma + \lambda^2 \eta^2 / \gamma^2)\right] \\
&= \frac{a^2}{N} E\left[\sum_{i=1}^{N} X_i^2\right] + \frac{\lambda^2}{N\gamma^2} E\left[\eta^2 \sum_{i=1}^{N} X_i^2\right] - \frac{2}{N\gamma} E\left[b\eta \sum_{i=1}^{N} X_i^2\right] \\
&\approx a^2 \sigma_X^2 + \lambda^2 \sigma_X^2 \sigma_\eta^2 / \gamma^2
\end{aligned}
\tag{5.4}
$$

where the fourth equality follows from (B.12) in Appendix B and the fact that $p(b=1) = p(b=-1) = 0.5$. Alternatively, we have

$$D_w \approx a^2 \sigma_X^2 + [\lambda / \mathrm{MVR}(\gamma)]^2 \sigma_X^2 / N . \tag{5.5}$$

The first term in (5.5) is the distortion contributed by MSS (with $\lambda = 0$), and the second term is the extra distortion incurred by EMSS. Our method uses this extra distortion to compensate the host interference. We also notice that the extra distortion does not depend on $a$.

### 5.2.3  Performance of EMSS

We also employ the generalized correlator (3.14) for watermark decoding. Similarly as we did in the previous chapter, the decoder's performance is characterized by

$$m_1 = -m_0 \approx \xi a E(|X|^{\xi}) , \tag{5.6}$$

$$\sigma_1^2 = \sigma_0^2 \approx \frac{1 + \xi^2 a^2}{N} \text{var}(|X|^\xi) + \frac{\lambda^2 \xi^2}{N \gamma^2} \frac{[E(|X|^\xi)]^2 \text{var}(|X|^\gamma)}{[E(|X|^\gamma)]^2}$$

$$- \frac{2\lambda\xi}{N\gamma} \frac{E(|X|^\xi)[E(|X|^{\xi+\gamma}) - E(|X|^\xi)E(|X|^\gamma)]}{E(|X|^\gamma)} . \tag{5.7}$$

Please see (4.14) and (4.15) for further reference. In particular, the decoder's performance at $\gamma = \xi$ is depicted by

$$p_e = Q[a\sqrt{N}\text{MVR}(\xi) / \sqrt{(1-\lambda)^2 + \xi^2 a^2}] , \tag{5.8}$$

which is obtained by substituting both (5.6) and (5.7) (with $\gamma = \xi$) into (3.16).

### 5.2.4  Performance Comparisons at $\gamma = \xi$

In this section, we characterize the performance of EMSS at a large and small $N$ respectively. When $N$ is large enough, the second term in (5.5) is relatively small when compared with the first term in (5.5). Thus $\lambda$ can be set to a larger value, for instance, 1.0. For $\lambda = 1.0$,

$$p_e = Q[\sqrt{N}\text{MVR}(\gamma) / \gamma] . \tag{5.9}$$

Comparing (5.9) with (3.32), we find that EMSS outperforms MSS since usually $a << 1/\gamma$. We also note that (5.9) does not depend on the embedding strength $a$. However, the performance of EMSS under attacks does depend on $a$. In addition, it is easy to see that

$$\lim_{\gamma \to 0} \frac{\text{MVR}(\gamma)}{\gamma} = \lim_{\gamma \to 0} \frac{\Gamma[(\gamma+1)/c]}{\sqrt{\Gamma[(2\gamma+1)/c]\Gamma(1/c) - \{\Gamma[(\gamma+1)/c]\}^2}} = +\infty . \tag{5.10}$$

Thus, the performance improves as $\gamma$ decreases.

The above case becomes impractical when we need to embed a large amount of data. This case requires that $N$ be small, and thus we have to use $\lambda$ to reduce the extra distortion. The following discussions are based on the assumption that $(1-\lambda)^2 \gg \gamma^2 a^2$. This assumption can be easily satisfied by selecting a small $\lambda$. Thus with (5.5), (5.8) can be expressed as

$$p_e = Q\left[a\sqrt{N}\text{MVR}(\xi)/(1-\lambda)\right] \approx Q\left[\sqrt{10^{-\text{DWR}/10} \cdot N[\text{MVR}(\gamma)]^2 - \lambda^2}/(1-\lambda)\right] . \tag{5.11}$$

From (5.11), we know that the best performance for EMSS is achieved at $\gamma = \xi = c$ for a given $\lambda$. To outperform MSS, we must have

$$\sqrt{10^{-\text{DWR}/10} \cdot N[\text{MVR}(\gamma)]^2 - \lambda^2} \Big/ (1-\lambda) > \sqrt{10^{-\text{DWR}/10} \cdot N[\text{MVR}(\gamma)]^2} \; . \tag{5.12}$$

By some simple manipulations, we have

$$\lambda < \lambda_{\max} = 2 - \frac{2}{1 + 10^{-\text{DWR}/10} N[\text{MVR}(\gamma)]^2} \; . \tag{5.13}$$

It is straightforward to see that $\lambda_{\max} > 0$. Therefore, EMSS outperforms MSS if $\lambda < \lambda_{\max}$. To say it more clearly, EMSS can always have a better performance than MSS. Now we search for the optimum $\lambda$ to minimize (5.11). Let

$$k(\lambda) = \frac{10^{-\text{DWR}/10} N[\text{MVR}(\gamma)]^2 - \lambda^2}{(1-\lambda)^2} \; . \tag{5.14}$$

Therefore,

$$k'(\lambda) = \frac{dk}{d\lambda} = \frac{2 \cdot 10^{-\text{DWR}/10} N[\text{MVR}(\gamma)]^2 - 2\lambda}{(1-\lambda)^3} \; . \tag{5.15}$$

Thus, it is easy to see that (5.16) achieves its maximum at

$$\lambda_{\text{opt}} = 10^{-\text{DWR}/10} N[\text{MVR}(\gamma)]^2 \; . \tag{5.16}$$

Please also note that $\lambda_{\text{opt}}$ should be smaller than $\lambda_{\max}$.

## 5.2.5  Optimality

In the previous subsection, we found that the best performance is achieved at $\gamma = \xi = c$ when $\gamma = \xi$ and $\lambda$ is small. In this subsection, we show that the best performance is nearly achieved at $\gamma = \xi = c$ for all possible choices of $\gamma$ and $\xi$. As in the previous subsection, we consider only the scenario of small $\lambda$. Let

$$\text{MMT}(\xi, \gamma) = \frac{E(|X|^{\xi+\gamma}) - E(|X|^{\xi})E(|X|^{\gamma})}{\xi \gamma E(|X|^{\xi})E(|X|^{\gamma})} \; . \tag{5.17}$$

Since $m_1 = -m_0$ and $\sigma_0 = \sigma_1$ (see (5.6) and (5.7)), the decoder's performance is decided by

$$\frac{m_1^2}{\sigma_1^2} \approx \frac{N \cdot 10^{-\mathrm{DWR}/10} - [\lambda / \mathrm{MVR}(\gamma)]^2}{\dfrac{1}{[\mathrm{MVR}(\xi)]^2} + \dfrac{\lambda^2}{[\mathrm{MVR}(\gamma)]^2} - 2\lambda \cdot \mathrm{MMT}(\xi,\gamma)} \,. \tag{5.18}$$

It is easy to prove that if $\xi = c$,

$$\mathrm{MMT}(c,\gamma) = 1/c \,. \tag{5.19}$$

We notice that (5.19) does not depend on $\gamma$. Hence, for $\xi = c$, (5.18) achieves its maximum at $\gamma = c$. Similarly, (5.18) also reaches its maximum at $\xi = c$ when $\gamma$ is fixed at $c$. For other cases, we must resort to numerical calculations to search for the optimal $\xi$ and $\gamma$. In fact, the numerical search shows that the best performance is not always achieved at $\gamma = \xi = c$ for all $cs$. However, we compared the global optimum $p_e$ with the $p_e$ achieved at $\gamma = \xi = c$, and found that the performance at $\gamma = \xi = c$ does not deviate much from the global optimum. In Fig. 5.1 for DWR = 13.98dB, we observe that as $c$ increases, the deviation becomes larger. However, the deviation is not obvious for the small DWR = 20dB. As a matter of fact, at DWR = 13.98dB, the global optimum for $c = 2.0$ is achieved at $\gamma = 0.25$. This is because at $\lambda = 1.0$, (5.10) indicates that the smaller the $\gamma$, the better the performance. Thus, (5.10) is also partly reflected in the case of $\lambda = 0.2$. However, for practical scenarios ($c < 1.5$), it is safe to assume that the best performance is achieved at $\gamma = \xi = c$.

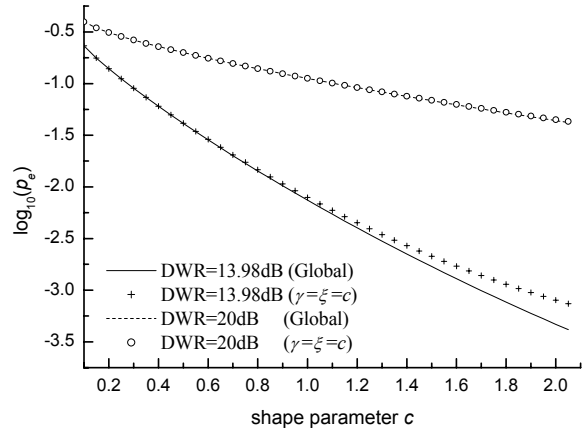

Fig. 5.1. The globally optimum performance versus the performance at $\gamma = \xi = c$ (with $N = 100$, $\lambda = 0.2$, $\sigma_X = 10$).

## 5.3 Monte-Carlo Simulations

In this section, we validate the theoretical results in the previous sections via Monte-Carlo simulations.

Please also note that in the legends of the figures, "E" means Empirical results obtained through Monte-Carlo simulations, and "T" Theoretical results. However, since the central limit theorem requires a considerably large $N$, the simulation results may not well match the theoretical results for small $Ns$, especially at large $\xi s$.

We first test the correctness of the derived expected distortion. Fig. 5.2(a) displays the distortion for $\gamma = c$. We see that the theoretical results (See (B.2) to (B.9) in Appendix B) do agree well with the simulation results. However, when $\gamma \neq c$, the case is different since the approximation (See (B.13) in Appendix B) can only provide a crude calculation of the distortion. This approximation in Fig. 5.2(b) displays a big difference from the real distortion for a small $N$. However, as $N$ increases, the approximation becomes quite accurate. Though the distortion does not well agree with the simulation results, the important point that the minimum distortion is achieved at $\gamma = c$ still holds. Moreover, for small $\lambda s$, the deviation from the real distortion becomes less significant. Fig. 5.3 demonstrates the correctness of (5.6) and (5.7). However, since the approximation by normal distributions is not accurate enough for small $Ns$, the theoretical results do not well agree with the simulation results. Nevertheless, the theoretical results do reflect the most important point that the best performance is nearly achieved at $\gamma = \xi = c$.

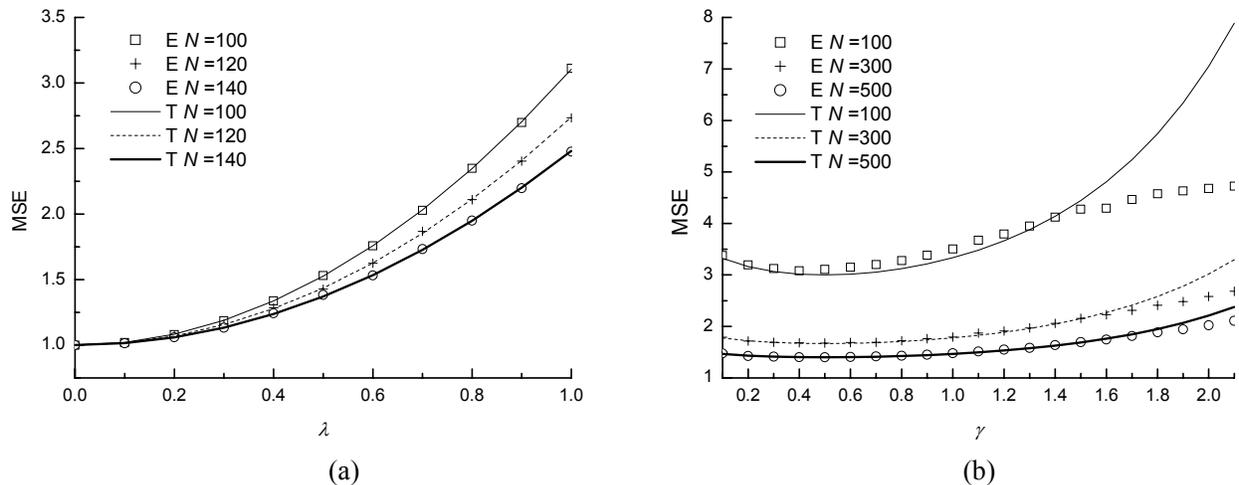

(a)                                              (b)

Fig. 5.2. Theoretical and Simulated Distortion. Results are obtained on 100, 000 groups of data with $c = 0.5$, $\sigma_X = 10$, $a = 0.1$, and $\lambda = 1.0$. (a) For $\gamma = c$. (b) For $\gamma \neq c$.

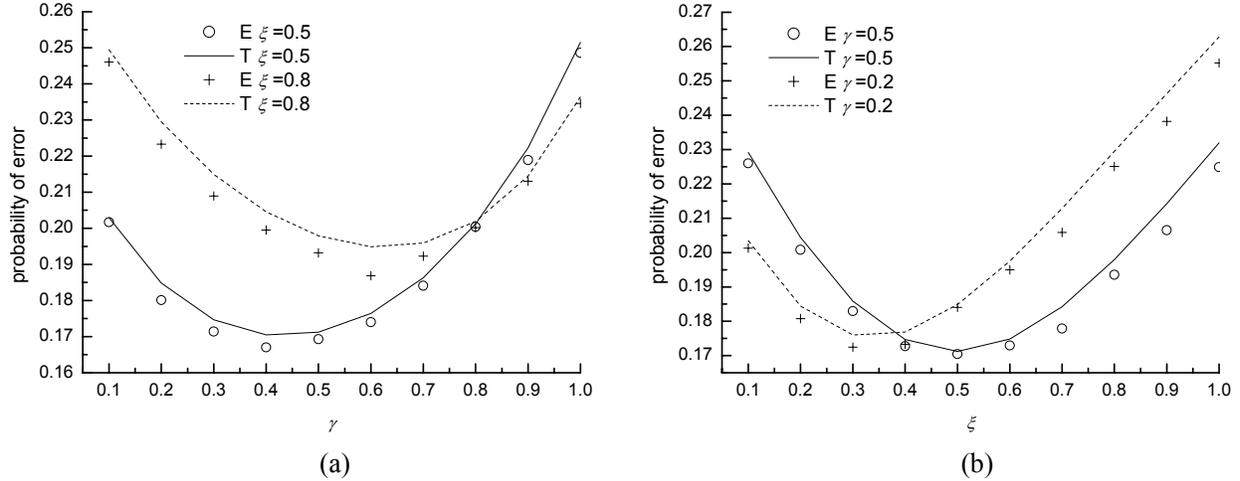

(a)                                                                              (b)

Fig. 5.3. Theoretical and simulated performance for EMSS with $\gamma \neq \xi$. Results are obtained on $100,000$ groups of data with $c = 0.5$, $\sigma_X = 10$, $N = 100$, $a = 0.1$, and $\lambda = 0.5$. (a) At different $\gamma$s. (b) At different $\xi$s.

## 5.4  Experimental results

The above section validates the theoretical results by Monte-Carlo simulations. In this section, however, experiments are carried out on real images. As we did in the previous section, we permute the original data randomly to fulfill the difficult task of finding a large database of images with the same $c$ and $\sigma_X$. The effectiveness of this permutation approach can be validated by the agreement between experimental and theoretical results. Please also refer to Appendix E for further information about the effectiveness of permutation. Also note that in this section, the text "T" in the legends of the figures stands for theoretical results, and "E" for experimental results.

The test image is Lena, and we embed data in the fifth and fifteenth (in zigzag order) AC coefficients. The shape parameter $c$ and standard deviation $\sigma_X$ are estimated by ML method [103]. For the 5th coefficient, $c = 0.69$ and $\sigma_X = 19.74$; for the 15th coefficient, $c = 0.88$ and $\sigma_X = 5.82$. Usually data embedding changes $c$ and $\sigma_X$. However, since this change is small due to the imperceptibility requirement, we instead use the parameters of the original image at the decoder for convenience.

We first compare the performance of MSS and EMSS at different $\gamma$s. The performance comparisons are shown in Fig. 5.4. For both figures, EMSS achieves its best performance at $\gamma = c$. This is in accord with the fact that MVR(0.5) = 0.823, MVR(0.69) = 0.831, MVR(1.0) = 0.814 and MVR(1.5) = 0.734 for $c = 0.69$.

For $c = 0.88$, since MVR(0.88) = 0.938 and MVR(1.0) = 0.936, the performance difference between $\gamma = 1.0$ and $\gamma = 0.88$ is very small. Theoretical results are also compared with the experimental results. Fig. 5.5 demonstrates that theoretical results do match the experimental results. Moreover, it also verifies that the estimated shape parameters do agree with the real one.

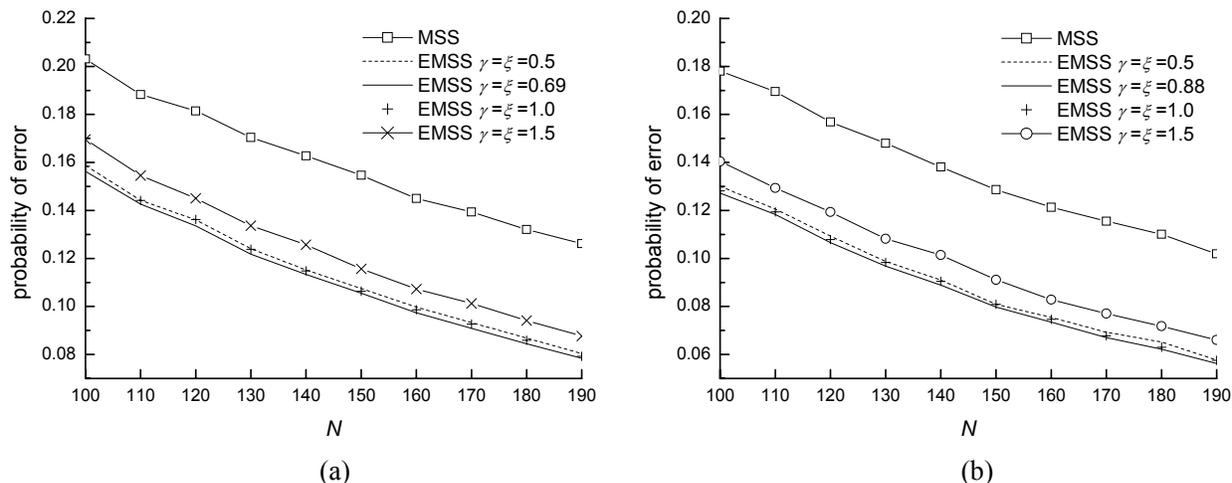

(a)                                          (b)

Fig. 5.4. MSS versus EMSS at the same distortion level (DWR = 20dB, $\lambda = 0.2$ with 100,000 embedded bits). (a) for the $5^{th}$ coefficient. (b) for the $15^{th}$ coefficient.

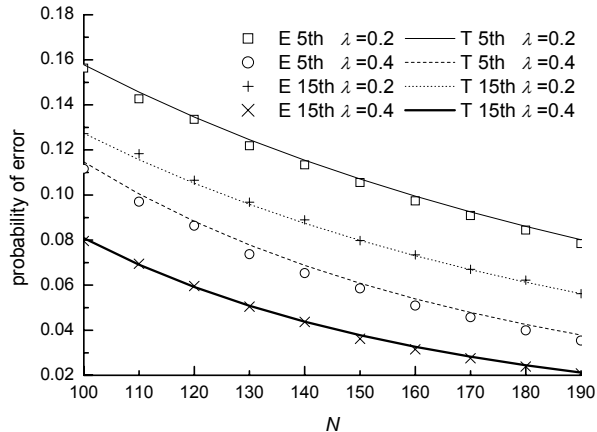

Fig. 5.5. Theoretical and experimental EMSS (DWR = 20dB with 100,000 embedded bits).

Now we examine the performance of MSS and EMSS under attacks. Fig. 5.6 shows the experimental results when the watermarked image suffers from Gaussian noise attacks. From Fig. 5.6, we see that the fifth coefficients are more robust against noise since it has a larger $\sigma_X$. As the added noise strengthens, EMSS converges to MSS since the attack noise overshadows the watermark signal. This is more saliently reflected in Fig. 5.6(b) where its $\sigma_X$ is small. Fig. 5.6 also includes the results for different $\lambda$s. In Fig. 5.6(a), the best performance under weak noise attacks is achieved at $\lambda_{opt} = 0.69$. However, it is not the case for stronger noise attacks. The performance for large $\lambda$s drops very fast as the noise increases. Thus an important point for future work is to choose a better $\lambda$ for known levels of attacks.

The second experiment reveals the performance of MSS and EMSS under JPEG compression attacks. The experimental results are displayed in Fig. 5.7. The performance curve for the 5th is completely different from that for the 15th coefficient since JPEG is a kind of quantization attack. Since the quantization step size for 5th coefficient is comparatively smaller and the 5th coefficient has a larger $\sigma_X$, the performance deteriorates gradually as quality factor (QF) decreases. However, for the 15th coefficient with a small $\sigma_X$, since the quantization step size is larger, some small values may be quantized to a larger value and thus JPEG may improve the performance at low QFs. However, as QF increases, this phenomenon becomes less protruding. The results for different $\lambda$s also show that larger $\lambda$s work better for weaker attacks in Fig. 5.7(a). In Fig. 5.7(b), EMSS at $\lambda = 0.88$ is more vulnerable to attacks due to the small standard deviation of the 15th coefficient.

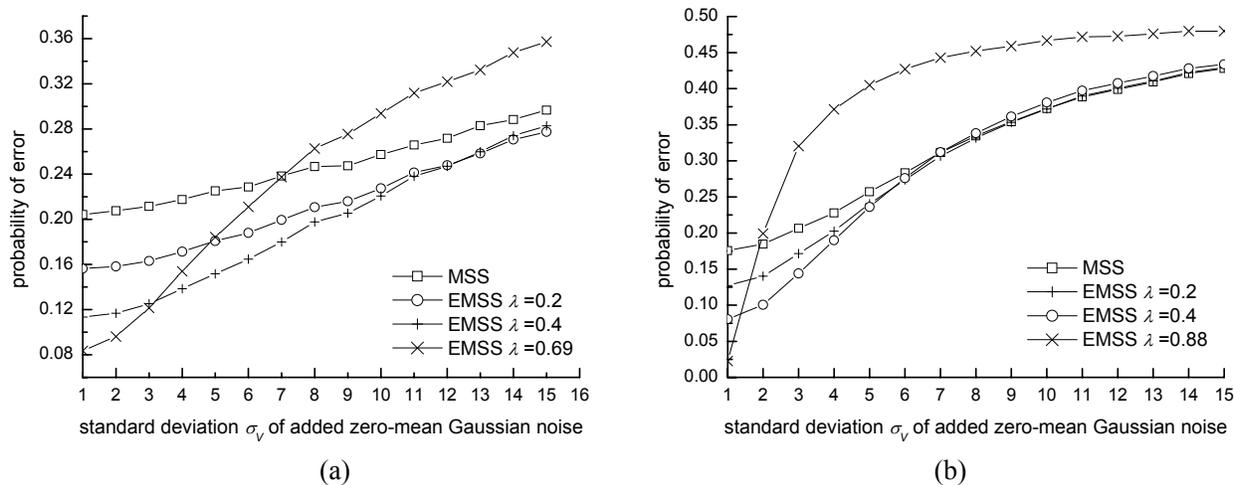

(a)                                                                (b)

Fig. 5.6. MSS versus EMSS under zero-mean Gaussian Noise attacks (DWR = 20dB, $N = 100$ with 100,000 embedded bits). (a) for the 5th coefficient. (b) for the 15th coefficient.

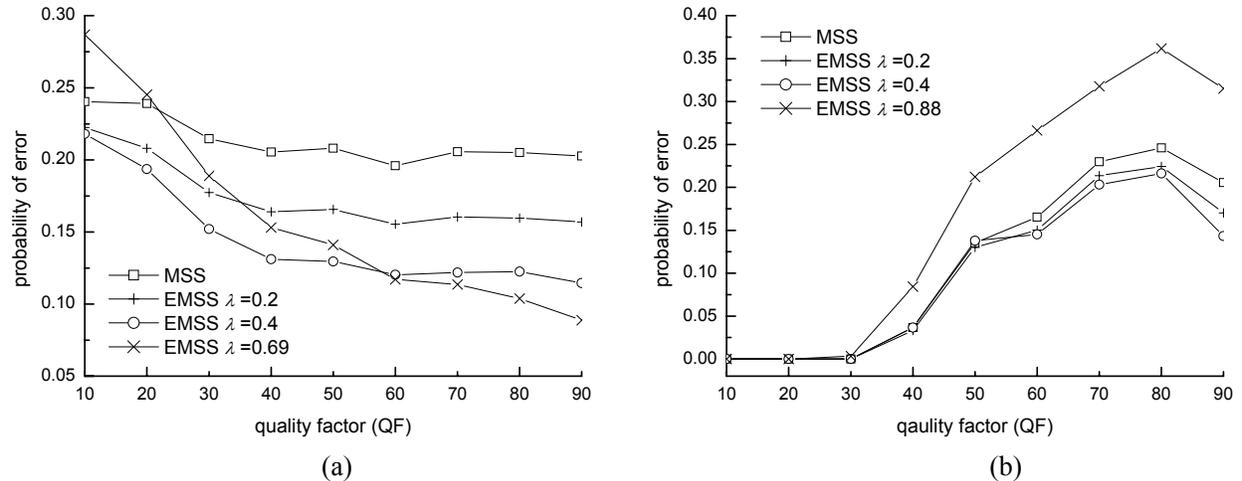

(a)                                                                  (b)

Fig. 5.7. MSS versus EMSS under JPEG attacks (DWR = 20dB, $N$ = 100 with 100,000 embedded bits). (a) for the 5[th] coefficient (b) for the 15[th] coefficient.

## 5.5 Conclusions and future directions

In this paper, we investigated how to remove the host interference in the embedding process. The idea of removing the host interference is inspired by the work [43]. However, in our scheme, the host interference is determined by the decision statistic of optimum decoding. Host signals with different shape parameters impose different interference on the decision statistic. We also proved that for any $\gamma$-order embedder and $\xi$-order decoder, the best performance is nearly achieved $\gamma = \xi = c$. $N$ is a critical factor determining the performance improvement of EMSS over MSS. Larger $N$s will demonstrate even greater improvements. However, in this chapter, we focus mainly on small $N$s since we may want to embed a large volume of data.

# Chapter 6   Double Sided Watermark Embedding and Detection

## 6.1  Introduction

In the previous chapters, the performance of the watermarking systems can be greatly improved by canceling the host interference at the embedder. However, the host interference rejection schemes, including the quantization schemes, have not taken the perceptual quality into account. Their performance advantage, in truth, is largely due to the employed MSE metric. In contrast to the SS schemes, it is difficult to implement the perceptual models in the host interference rejection schemes. This difficulty thus discourages the use of host interference rejection schemes in real scenarios. Can we instead utilize the host information without it being rejected at the embedder? This chapter shall answer this question by introducing a very simple double-sided technique. It differs from the traditional SS schemes in that it also utilizes the host information at the embedder. However, different from the host interference rejection schemes, it does not reject the host interference. Due to this nice property of not rejecting the host interference, it has a big advantage over the host interference rejection schemes in that the perceptual analysis can be easily implemented for our scheme to achieve the maximum allowable embedding level. For most of the traditional SS methods, the detector reports the existence of the embedded watermark if its response is above a given threshold. However, our double-sided detector reports the existence of the embedded watermark if the absolute value of its response exceeds a given threshold.

The rest of the work is organized as follows. Section 6.2 reviews the traditional decision rules for SS methods. Section 6.3 and 6.4 present a simple double-sided technique for both additive and multiplicative schemes. Its marriage with host interference rejection techniques will be studied in Section 6.5. However, our discussion on the host interference rejection technique is only to reveal its disadvantage as against our double-sided schemes. In the real scenarios, we strongly recommend the use of our double-sided schemes without host interference rejection. Section 6.6 makes performance comparisons to demonstrate the effectiveness of our schemes. Its performance advantages on real images are displayed in Section 6.7.

Finally, we conclude the chapter in Section 6.8.

## 6.2  Single-sided schemes

The embedded watermarks can be detected by testing the hypothesis $H_0$: the test data are not watermarked versus the hypothesis $H_1$: the test data contain the specific watermark $\mathbf{w}$. The spread spectrum schemes detect the watermark by deciding between

$$L(\mathbf{S}) > \psi \Rightarrow H_1; \ \ L(\mathbf{S}) < \psi \Rightarrow H_0, \tag{6.1}$$

where $\psi$ is the decision threshold. In this work, we call (6.1) a single-sided decision rule since $H_1$ occupies the right side of the decision space. Thus, the traditional SS schemes are also termed single-sided schemes in this work. Fig. 6.1(a) draws the pdf of the decision statistic for a typical single-sided scheme.

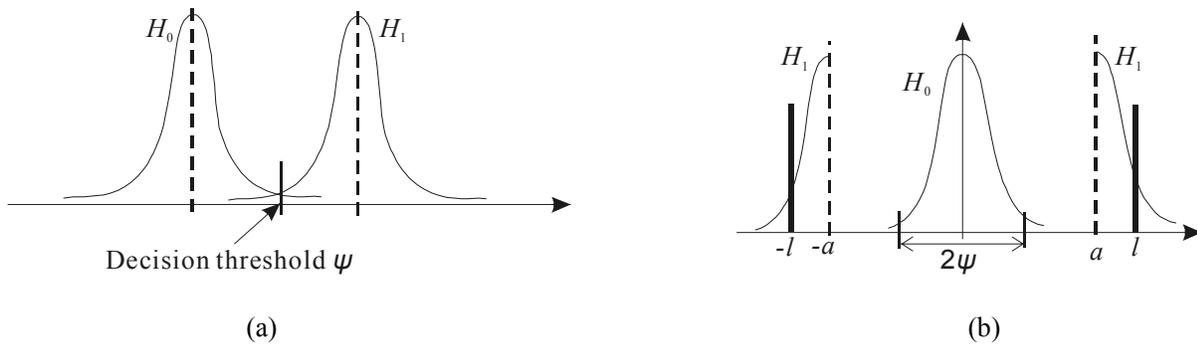

(a)                                                                                     (b)

Fig. 6.1. The illustrative pdfs of the decision statistic $L(\mathbf{S})$ under $H_0$ and $H_1$, where in the plot $H_0$ and $H_1$ represent $L(\mathbf{S}|H_0)$ and $L(\mathbf{S}|H_1)$, respectively. (a) Single-sided scheme. (b) Double-sided scheme, where two thick lines represent the pdf of $L(\mathbf{S}|H_1)$ for DS-ASS-HIR (See Section 6.5).

## 6.3  Double-sided additive spread spectrum schemes

The host interference was previously considered a nuisance for SS schemes. Host interference rejection schemes reject it at the embedder. However in this work, we introduce a double-sided technique to utilize but not reject the host interference. The basic idea comes from the following observation. For a correlation detector, the single-sided rule wastes the decision space since the statistic for $H_0$ (See Fig. 6.1(a)) with $m_0 = E(\sum_{1 \le i \le N} X_i w_i) / N = 0$ and $\sigma_0^2 = Var(\sum_{1 \le i \le N} X_i w_i) / N^2 = \sigma_X^2 / N$ has a rare chance to reach the very negative side of the axis, especially at a large $N$. For instance, the probability of it being smaller than $-5\sigma_0$ is only about $Q(5)$. However our double-sided idea can better utilize the decision space. In this section, we investigate how the new idea works for the additive spread spectrum schemes. Its extension to the multiplicative case will be examined in the next section.

### 6.3.1  Embedding rules for double-sided ASS

In this subsection, we present a new Double-Sided Additive Spread Spectrum (DS-ASS) scheme whose embedding rule is

$$s_i = x_i + aw_i, \ \text{ if } \ \bar{x} > 0; \ s_i = x_i - aw_i, \ \text{ if } \ \bar{x} \le 0 \ , \tag{6.2}$$

where $i = 1, 2, \ldots, N$ and $\bar{x}$ is the projected $\mathbf{x}$ on $\mathbf{w}$ defined as

$$\bar{x} = \frac{1}{N} \sum_{i=1}^{N} x_i w_i \ . \tag{6.3}$$

$\bar{X}, \bar{s}, \bar{S}, \bar{y}, \bar{Y}, \bar{v}, \bar{V}$ are similarly defined. Equation (6.2) is a simple informed embedder. The idea underlying (6.2) is very simple. If the correlation between the host and the watermark is positive, then the watermark is multiplied by a positive quantity when it is added to the host. However, if instead the correlation is negative, then it is multiplied by a negative quantity when it is added to the host. Thus, the host "interference" is now exploited to increase the magnitude of the output of the correlator. Compared with ISS [43] or quantization schemes [100−102], the new embedding rule utilizes but does not reject the host interference. However, as we shall see in the following text, it can also achieve a great performance improvement over ASS. For the above embedding rule (6.2), it is easy to see that we should detect the watermarks by checking whether the magnitude of the correlation is above a given threshold, that is,

$$\left| L(\mathbf{S}) \right| > \psi \Rightarrow H_1; \text{otherwise} \Rightarrow H_0 \ , \tag{6.4}$$

where $L(\mathbf{S})$ is a correlator given by (3.59). In this work, we call (6.4) a double-sided decision rule. Compared with (6.1), it reports the existence of the watermark if the absolute value of the decision statistic is above a given threshold.

Now we examine the close relation between STDM (with only two centroids) and DS-ASS. The embedding rule for STDM is $s_i = x_i + [q_\Lambda(\bar{x}) - \bar{x}]w_i$. Thus for STDM with only two centroids, if $\bar{x} > 0$, $q_\Lambda(\bar{x}) = 0.5\Delta$ and if $\bar{x} < 0$, $q_\Lambda(\bar{x}) = -0.5\Delta$. This may be taken as exploiting the host interference. However, the second term in $[q_\Lambda(\bar{x}) - \bar{x}]$ is to reject the host interference. Thus, DS-ASS is identical to a two-centroid

STDM, however without the host interference rejection.

It is also interesting to note that (6.2) gives an important hint [115] that watermark detection is much different from watermark decoding, which has been confused by many previous works (for instance, traditional SS schemes). The watermark decoder decides between +1 and −1, whereas the watermark detector answers a "yes" or "no" question. However, in traditional SS schemes, the embedder usually embeds +1 to mean "yes", and the detector thus decides between +1 and "no" since the "yes" means +1. For instance, in watermarking decoding problems, $\mathbf{s} = \mathbf{x} + a\mathbf{w}$ to embed +1 and $\mathbf{s} = \mathbf{x} - a\mathbf{w}$ to embed −1. In traditional watermarking detection problems, the embedding rule is $\mathbf{s} = \mathbf{x} + a\mathbf{w}$. Thus, the presence of watermark also means that the embedded information is +1. Here in our scheme the embedder instead embeds +1 or −1 to mean "yes" and the detector can only answer a "yes" or "no" question. For instance, the embedding rule for DS-ASS is $\mathbf{s} = \mathbf{x} \pm a\mathbf{w}$. It is not possible for the detector to decide from the "yes" answer whether +1 or −1 is embedded. That is, whether +1 or −1 is embedded does not really matter. What is important is that the watermark is present.

### 6.3.2  Advantages over host interference rejection schemes

In real scenarios, the maximum allowable embedding strength $a_{\max}$ is determined by the perceptual allowance of the image. It is easy for DS-ASS to achieve the embedding strength $a_{\max}$. However, for STDM and DS-ASS-HIR (See Section 6.5), if the embedding strength $|q_\Lambda(\bar{x}) - \bar{x}|$ and $|\pm l - \bar{x}|$ are larger than $a_{\max}$, they will incur visual distortion in the image. However, if they are smaller than $a_{\max}$, they will leave a larger perceptual allowance for the attackers to mount stronger attacks. One may argue we can select an appropriate $\lambda$ such that their embedding strength is $a_{\max}$. However, in such a case, STDM and DS-ASS-HIR instead reduce to DS-ASS. This problem thus encourages the use of our double-sided schemes without host interference rejection since the perceptual analysis can be easily implemented. However, it is difficult to implement the perceptual analysis for the host interference rejection schemes to achieve the locally bounded maximum embedding strength.

### 6.3.3 Performance under no attack

In order to evaluate the performance of DS-ASS, we project $\mathbf{S}$ on $\mathbf{w}$ and have

$$\overline{S} = \overline{X} + a, \ \text{if} \ \overline{X} > 0; \ \overline{S} = \overline{X} - a, \ \text{if} \ \overline{X} \leq 0 \,. \tag{6.5}$$

For the convenience of discussions, we define an indicator $I$ as

$$I = 1, \ \text{if} \ \overline{X} > 0; \ I = 0, \ \text{if} \ \overline{X} \leq 0 \,. \tag{6.6}$$

Since all $X_i$s are independently and identically distributed and $\sum_{1 \leq i \leq N} w_i = 0$ (See (2.2)), $\overline{X}$ has a symmetric pdf and $P(I = 0) = P(I = 1) = 0.5$. In order to obtain the pdf of $\overline{S}$, we first present a lemma whose detailed proof can be found in Appendix D.

Lemma 6.1: If $U$ is a random variable, and $Z$ is defined by $Z = U + a$ if $U > 0$ and $Z = U - a$ if $U \leq 0$, where $a$ is a const. Then $Z$ has a pdf given by

$$f_Z(z) = \begin{cases} f_U(z - a), & \text{if } z > a \\ 0, & \text{if } -a < z \leq a \\ f_U(z + a), & \text{if } z \leq -a \end{cases}$$

By invoking the central limit theorem to approximate $\overline{X}$ by a Gaussian random variable, we have from Lemma 6.1

$$f_{\overline{S}}(\overline{s}) = \begin{cases} \dfrac{1}{\sqrt{2\pi\sigma_{\overline{X}}^2}} \exp\left[ \dfrac{-(\overline{s} - a)^2}{2\sigma_{\overline{X}}^2} \right], & \text{if } \overline{s} > a \\ 0, & \text{if } -a < \overline{s} \leq a \\ \dfrac{1}{\sqrt{2\pi\sigma_{\overline{X}}^2}} \exp\left[ \dfrac{-(\overline{s} + a)^2}{2\sigma_{\overline{X}}^2} \right] & \text{if } \overline{s} \leq -a \end{cases} \tag{6.7}$$

We see that the pdf of the $\overline{S}$ can be characterized by the pdf of $\overline{X}$ with the right half shifted right by $a$ and the left half shifted left by $a$. Please refer to Fig. 6.1(b) for a better understanding of (6.7). This is why our double-sided scheme can better utilize the decision space. From Fig. 6.1(b), (6.4) and (6.7), we see that the detector's performance is delineated by

$$p_{fa} = P(|\overline{S}| > \psi \mid H_0) = 2Q[(\psi - m_0)/\sigma_0],$$ (6.8)

$$p_m = P(|\overline{S}| < \psi \mid H_1) = \begin{cases} 1 - 2Q[(\psi - m_1)/\sigma_1], & \text{if } \psi > a \\ 0.0, & \text{if } \psi \le a \end{cases},$$ (6.9)

where $\psi$ is the decision threshold and $m_0 = 0$, $m_1 = a$ and $\sigma_0^2 = \sigma_1^2 = \sigma_X^2/N$. Substituting $p_{fa}$ into $p_m$, we obtain

$$p_m = \begin{cases} 1 - 2Q[Q^{-1}(p_{fa}/2) - a\sqrt{N}/\sigma_X], & \text{if } \psi > a \\ 0.0, & \text{if } \psi \le a \end{cases}$$ (6.10)

Comparing Fig. 6.1(a) and Fig. 6.1(b), we find that our method can achieve a zero-miss if the decision threshold $\psi$ is set between 0 and $a$, whereas ASS can never achieve a zero-miss. Therefore, DS-ASS works quite as what the quantization schemes [102] do since quantization schemes can also achieve a zero miss under no attack.

Another interesting property is that DS-ASS also outperforms ASS (with correlation detector) at $\psi > a$ (see (6.10) and (3.70)), that is, $2Q[Q^{-1}(p_{fa}/2) - a\sqrt{N}/\sigma_X] > Q[Q^{-1}(p_{fa}) - a\sqrt{N}/\sigma_X]$. This performance advantage is largely due to the fact that $Q(x)$ is a fast decaying function of $x$. In order to provide a mathematically proof of it, we first give a lemma that states the fast decaying property of $Q(x)$.

Lemma 6.2: For any real $k > 0$, $Q(z+k)/Q(z)$ is a decreasing function of $z$.

The detailed proof of Lemma 6.2 can be found in Appendix D. Let $k = Q^{-1}(p_{fa}/2) - Q^{-1}(p_{fa})$ and $z = Q^{-1}(p_{fa})$. It is easy to see that $k > 0$ and $Q(z+k)/Q(z) = 0.5$. Since $a > 0$, we have $z - a\sqrt{N}/\sigma_X < z$. Thus we have from Lemma 6.2

$$\frac{Q[Q^{-1}(p_{fa}) - a\sqrt{N}/\sigma_X + Q^{-1}(p_{fa}/2) - Q^{-1}(p_{fa})]}{Q[Q^{-1}(p_{fa}) - a\sqrt{N}/\sigma_X]} = \frac{Q[Q^{-1}(p_{fa}/2) - a\sqrt{N}/\sigma_X]}{Q[Q^{-1}(p_{fa}) - a\sqrt{N}/\sigma_X]} > 0.5.$$

Thus under no attack, DS-ASS has a complete performance advantage over ASS (with correlation detector).

### 6.3.4 Performance under attacks

We now inspect the performance of DS-ASS under zero-mean noise attacks. Under attacks, we obtain $y_i = x_i \pm a w_i + v_i$. Thus by projecting $\mathbf{Y}$ on $\mathbf{w}$, we get a pair of hypotheses $H_1: \overline{Y} = \overline{X} \pm a + \overline{V}$ and $H_0: \overline{Y} = \overline{X} + \overline{V}$. Since both $\overline{X}$ and $\overline{V}$ can be approximated by Gaussian random variables, the detector's performance is given by

$$p_{fa} = P(|\overline{Y}| > \psi \mid H_0) = P(|\overline{X} + \overline{V}| > \psi) = 2Q(\psi / \sigma_0),$$
(6.11)

$$
\begin{aligned}
p_m &= P(|\overline{Y}| < \psi \mid H_1) = P(|\overline{X} + a + \overline{V}| < \psi \mid I = 1)P(I = 1) + P(|\overline{X} - a + \overline{V}| < \psi \mid I = 0)P(I = 0) \\
&= P(|\overline{X} + a + \overline{V}| < \psi, I = 1) + P(|\overline{X} - a + \overline{V}| < \psi, I = 0) \\
&= \int_0^\infty f_{\overline{X}}(\overline{x}) \cdot P(|\overline{x} + a + \overline{V}| < \psi) d\overline{x} + \int_{-\infty}^0 f_{\overline{X}}(\overline{x}) \cdot P(|\overline{x} - a + \overline{V}| < \psi) d\overline{x} \\
&= \int_0^\infty f_{\overline{X}}(\overline{x}) \cdot \left\{ Q\left[ \frac{-\psi - (\overline{x} + a)}{\sigma_1} \right] - Q\left[ \frac{\psi - (\overline{x} + a)}{\sigma_1} \right] \right\} d\overline{x} \\
&\quad + \int_{-\infty}^0 f_{\overline{X}}(\overline{x}) \cdot \left\{ Q\left[ \frac{-\psi - (\overline{x} - a)}{\sigma_1} \right] - Q\left[ \frac{\psi - (\overline{x} - a)}{\sigma_1} \right] \right\} d\overline{x} \\
&= 2\int_0^\infty f_{\overline{X}}(\overline{x}) \cdot \left\{ Q\left[ \frac{-\psi - (\overline{x} + a)}{\sigma_1} \right] - Q\left[ \frac{\psi - (\overline{x} + a)}{\sigma_1} \right] \right\} d\overline{x}
\end{aligned}
$$
(6.12)

where $\sigma_0^2 = (\sigma_X^2 + \sigma_V^2) / N$, $\sigma_1^2 = \sigma_V^2 / N$. In (6.12), the second equality follows from conditioning on $I$, and the fourth from conditioning on $\overline{X}$ and the independence of $\overline{X}$ and $\overline{V}$, and the sixth from the fact that $\overline{X}$ has a symmetric pdf.

### 6.3.5 Perceptual analysis for DS-ASS

Now we discuss how to implement the perceptual analysis in DS-ASS scheme. Its detailed discussions are delayed until Chapter 8. The global embedding strength will introduce local visual distortion on the watermarked contents. Thus, the embedding strength should be locally bounded by Human Visual Systems (HVS) for better perceptual quality. Incorporating the perceptual analysis into DS-ASS, we obtain an embedding rule

$$s_i = x_i + a_i w_i, \text{ if } \overline{x} > 0; \ s_i = x_i - a_i w_i, \text{ if } \overline{x} \leq 0,$$
(6.13)

where $i = 1, 2, \ldots, N$, $a_i > 0$ is the embedding strength locally bounded by HVS (for instance, Watson's perceptual model [55]) and $\bar{x}$ is as defined in (6.3). The same linear correlator (3.59) coupled with the double-sided rule (6.4) can be employed for watermark detection.

## 6.4 Double-sided multiplicative spread spectrum schemes

In the above section, we studied the double-sided detection for ASS. In this section, we extend this idea to the MSS schemes. We first show how the double-sided detection works for Barni's embedding rules (2.22). Then we continue to bring together the ideas of both optimum detection and double-sided detection.

### 6.4.1 Generalized Barni's rule

In this subsection, we generalize the rule (2.22) to

$$s_i = x_i + a\left|x_i\right|^{\xi} w_i \, . \tag{6.14}$$

where $i = 1, 2, \ldots, N$ and $\xi$ ($0 \leq \xi \leq 1$) is the order parameter (OP). Barni's rule can be obtained by setting $\xi = 1.0$. The order parameter $\xi$ can be chosen to match the exponent $w_{ij}$ in Watson's model (Please see [55] for a clear definition of $w_{ij}$ or $w(i, j)$ in Section 8.1.3). In particular, we may set $\xi$ to 0.7 since $w_{ij}$ has a typical empirical value of 0.7. The embedding distortion for this rule is $D_w = a^2 E(|X|^{2\xi})$.

We also employ the linear correlator (3.59) with the single-sided rule (6.1) for watermark detection. The detector decides between $H_0$: $S_i = X_i$ and $H_1$: $S_i = X_i + a|X_i|^{\xi}w_i$, and its performance under no attack can also be characterized by (3.69) but with $m_0 = 0$, $m_1 = aE(|X|^{\xi})$, $\sigma_0^2 = \sigma_X^2/N$, $\sigma_0^2 = [\sigma_X^2 + a^2 Var(|X|^{\xi})]/N \approx \sigma_X^2/N$ since $a << 1.0$ and $0 \leq \xi \leq 1$. Likewise, the performance under additive noise attacks is determined by $m_0 = 0$, $m_1 = aE(|X|^{\xi})$, $\sigma_0^2 = (\sigma_X^2 + \sigma_V^2)/N$ and $\sigma_1^2 \approx (\sigma_X^2 + \sigma_V^2)/N$ .

### 6.4.2 Double-sided detection for Barni's rule

Similarly as we did for DS-ASS, we obtain an improved embedding rule

$$s_i = x_i + a\left|x_i\right|^{\xi} w_i, \text{ if } \bar{x} > 0; \; s_i = x_i - a\left|x_i\right|^{\xi} w_i, \text{ if } \bar{x} \leq 0 \, , \tag{6.15}$$

where $i = 1, 2, \ldots, N$ and $\bar{x}$ is as defined in (6.3). $\overline{X}, \overline{s}, \overline{S}, \overline{y}, \overline{Y}, \overline{v}, \overline{V}$ are defined similarly as (6.3). This informed embedding rule is termed as Double-Sided Barni's Multiplicative Spread Spectrum (DS-BMSS) scheme. Projecting **S** on w, we obtain $L(\mathbf{S}|H_1)$ which is given by

$$\overline{S} = \overline{X} + \frac{a}{N}\sum_{i=1}^{N}|X_i|^{\xi}, \text{ if } \overline{X} > 0; \ \overline{S} = \overline{X} - \frac{a}{N}\sum_{i=1}^{N}|X_i|^{\xi}, \text{ if } \overline{X} \leq 0. \tag{6.16}$$

It is impossible to make an exact analysis for the pdf of $\overline{S}$. However, by the weak law of large numbers, $(\sum_{1 \leq i \leq N}|X_i|^{\xi})/N$ converges to its mean $E(|X|^{\xi})$ in probability as $N$ increases [107]. Thus, we may have the approximation $(\sum_{1 \leq i \leq N}|X_i|^{\xi})/N \approx E(|X|^{\xi})$. We now give Lemma 6.3 to further the discussion.

Lemma 6.3: $\sum_{1 \leq i \leq N}|X_i|^{\xi}$ is independent of the indicator $I$, where $I$ is as defined in (6.6).

It is easy to understand Lemma 6.3 since the sign of $\overline{x}$ does not influence the value of $\sum_{1 \leq i \leq N}|x_i|^{\xi}$. However, we still provide a mathematically rigorous proof for it in Appendix D. Thus Lemma 6.3 leads to $L(\mathbf{S}|H_1, I = 1) = (\overline{X}|I = 1) + a (\sum_{1 \leq i \leq N}|X_i|^{\xi})/N \approx (\overline{X}|I = 1) + aE(|X|^{\xi})$. Similarly, we have $L(\mathbf{S}|H_1, I = 0) \approx (\overline{X}|I = 0) - aE(|X|^{\xi})$. Therefore, the pdf of $L(\mathbf{S}|H_1)$ is also depicted by (6.7) (Please refer to the proof of Lemma 6.1 in Appendix D), however with $a$ replaced by $aE(|X|^{\xi})$. Thus, we have the performance

$$p_m = \begin{cases} 1 - 2Q[Q^{-1}(p_{fa}/2) - a\sqrt{N}E(|X|^{\xi})/\sigma_X], & \text{if } \psi > aE(|X|^{\xi}) \\ 0.0, & \text{if } \psi \leq aE(|X|^{\xi}) \end{cases} \tag{6.17}$$

Under attacks, we obtain $y_i = x_i \pm a|x_i|^{\xi}w_i + v_i$. Projecting **Y** on **w**, we have $\overline{Y} = \overline{X} \pm a\sum_{i=1}^{N}|X_i|^{\xi}/N + \overline{V}$. Similarly as we did for DS-ASS (See (6.11) and (6.12)), the performance of DS-BMSS under attacks is given by

$$p_{fa} = 2Q(\psi/\sigma_0), \tag{6.18}$$

$$p_m = P\left(\left|\overline{X} + a\sum_{i=1}^{N}|X_i|^{\xi}/N + \overline{V}\right| < \psi, I = 1\right) + P\left(\left|\overline{X} - a\sum_{i=1}^{N}|X_i|^{\xi}/N + \overline{V}\right| < \psi, I = 0\right)$$

$$\overset{(a)}{\approx} P\left[\left|\overline{X} + aE(|X|^{\xi}) + \overline{V}\right| < \psi, I = 1\right] + P\left[\left|\overline{X} - aE(|X|^{\xi}) + \overline{V}\right| < \psi, I = 0\right]$$

$$= 2\int_0^{\infty} f_{\overline{X}}(\overline{x}) \cdot \left\{Q\left[\frac{-\psi - (\overline{x} + m_1)}{\sigma_1}\right] - Q\left[\frac{\psi - (\overline{x} + m_1)}{\sigma_1}\right]\right\} d\overline{x}$$

$$\tag{6.19}$$

where $m_1 = aE(|X|^{\xi})$, $\sigma_0^2 = (\sigma_X^2 + \sigma_V^2)/N$, $\sigma_1^2 = \sigma_V^2/N$ and $(a)$ follows from the approximation $(\sum_{1 \le i \le N} |X_i|^{\xi})/N \approx E(|X|^{\xi})$ and Lemma 6.3.

### 6.4.3  Double-sided detection for optimum decision statistic

In this subsection, we investigate how to design an informed embedder according to the optimum decision statistic. In Chapter 3, we introduced a suboptimal generalized correlator (3.66) in no need of the embedding strength. It inspired us to propose a new Double-Sided Multiplicative Spread Spectrum (DS-MSS) scheme whose embedding rule is given by

$$s_i = x_i + ax_iw_i, \text{ if } \overline{x} > 0; \ s_i = x_i - ax_iw_i, \text{ if } \overline{x} \le 0 \ , \tag{6.20}$$

where $i = 1, 2, \ldots, N$ and

$$\overline{x} = \frac{1}{N}\sum_{i=1}^{N} |x_i|^{\xi} w_i \ . \tag{6.21}$$

$\overline{x}$ is called the projected $\mathbf{x}$ on $\mathbf{w}$ with order parameter $\xi$ ($\xi \ge 0$). In this subsection, $\overline{X}, \overline{s}, \overline{S}$ are similarly defined as (6.21). It is also interesting to note that the projected $\mathbf{X}$ shares the same form with the suboptimal detector (3.66). To detect the embedded watermarks, we also employ the double-sided decision rule (6.4), however with $L(\mathbf{S})$ given by

$$L(\mathbf{S}) = \frac{1}{N}\sum_{i=1}^{N} |S_i|^{\xi} w_i \ . \tag{6.22}$$

Thus with the approximation technique (3.12), $L(\mathbf{S}|H_1)$ is

$$L(\mathbf{S} \mid H_1) = \frac{1}{N}\sum_{i=1}^{N} |X_i(1 \pm aw_i)|^{\xi} w_i \approx \frac{1}{N}\sum_{i=1}^{N} |X_i|^{\xi} (1 \pm \xi aw_i)w_i = \frac{1}{N}\sum_{i=1}^{N} |X_i|^{\xi} w_i \pm \frac{a\xi}{N}\sum_{i=1}^{N} |X_i|^{\xi} \ . \tag{6.23}$$

Similarly, we can prove that $\sum_{1 \le i \le N} |X_i|^{\xi}$ is independent of the indicator $I$, where $I$ is as defined in (6.6), however with $\overline{X}$ replaced by (6.21). Therefore, we have $L(\mathbf{S}|H_1, I = 1) \approx (\overline{X} \mid I = 1) + a\xi E(|X|^{\xi})$ and $L(\mathbf{S}|H_1, I = 0) \approx (\overline{X} \mid I = 0) - a\xi E(|X|^{\xi})$. Thus, $L(\mathbf{S}|H_1)$ also has the pdf depicted by (6.7), however with $a$ replaced by $a\xi E(|X|^{\xi})$, and hence the performance is

$$p_m = \begin{cases} 1 - 2Q[Q^{-1}(p_{fa}/2) - a\sqrt{N}\,\mathrm{MVR}(\xi)], & \text{if } \psi > a\xi E(|X|^{\xi}) \\ 0.0, & \text{if } \psi \leq a\xi E(|X|^{\xi}) \end{cases}$$

(6.24)

which achieves its minimum at $\xi = c$. We can similarly prove that DS-MSS has a better performance than the traditional MSS schemes. However, it is difficult to analyze its performance under attacks.

## 6.5 Double-sided detection with host interference rejection

In the above sections, we have investigated the double-sided detection for SS schemes. For SS schemes, the host signals interfere with the successful detection of the embedded watermarks. There are several ways to reject this interference, for instance, ISS [43] and Spread Transform Dither Modulation (STDM) [46, 102]. In fact, the embedding strength for the host interference rejection scheme is a random variable dependent on the host signals. In this section, we investigate how to introduce the host interference rejection technique into our double-sided schemes.

### 6.5.1 Embedding rules

ISS is a technique that rejects the host interference. As noted by Pérez-Freire *et al.* [68], ISS is, in essence, a quantization scheme with two centroids. In this section, we study how to apply the same technique to our scheme. The central idea of host interference rejection in our scheme is to adjust the embedding strength $a$ so that the detector outputs only two values, for instance, $l$ and $-l$. Thus, projecting both sides of (6.2) on $\mathbf{w}$, we have

$$\overline{s} = \overline{x} + a \Rightarrow l = \overline{x} + a \Rightarrow a = l - \overline{x}, \quad \text{for } \overline{x} > 0; \quad \overline{s} = \overline{x} - a \Rightarrow -l = \overline{x} - a \Rightarrow a = l + \overline{x}, \text{ for } \overline{x} \leq 0.$$

(6.25)

where $\overline{x}$ is defined in (6.3). Furthermore, $\overline{X}, \overline{s}, \overline{S}, \overline{y}, \overline{Y}, \overline{v}, \overline{V}$ are similarly defined as (6.3). Substituting $a$ in (6.25) into (6.2) leads to a new embedding rule

$$s_i = x_i + (l - \overline{x})w_i, \text{ if } \overline{x} > 0; \ s_i = x_i - (l + \overline{x})w_i, \text{ if } \overline{x} \leq 0,$$

(6.26)

where $i = 1, 2, \ldots, N$. In this work, we call (6.26) a Double-Sided Additive Spread Spectrum with Host Interference Rejection (DS-ASS-HIR) scheme. The pdf of $L(\mathbf{S}|H_1)$ is just two pulses at $\pm l$ (See Fig. 6.1(b)). In (6.26), we see that $\overline{x}$ cancels the host $x_i$. This indeed is the key idea of ISS.

### 6.5.2  Embedding distortion

In this part, we derive the embedding distortion for DS-ASS-HIR. The embedding distortion $D_w$ is

$$
\begin{aligned}
D_w &= \frac{1}{N} E\left[ \sum_{i=1}^{N} (\pm l - \overline{X})^2 \right] = E[(l - \overline{X})^2 \mid I = 1]P(I = 1) + E[(-l - \overline{X})^2 \mid I = 0]P(I = 0) \\
&= E[(l^2 - 2l\overline{X} + \overline{X}^2) \mid I = 1]P(I = 1) + E[(l^2 + 2l\overline{X} + \overline{X}^2) \mid I = 0]P(I = 0) \\
&= l^2 + E(\overline{X}^2) + 0.5E(-2l\overline{X} \mid I = 1) + 0.5E(2l\overline{X} \mid I = 0) \\
&= l^2 + \sigma_{\overline{X}}^2 - 2l\sqrt{2\sigma_{\overline{X}}^2/\pi} \\
&= l^2 + \sigma_X^2/N - 2l\sqrt{2\sigma_X^2/(\pi N)}
\end{aligned}
\tag{6.27}
$$

where $I$ is defined in (6.6), the second equality follows from conditioning on $I$, the fourth from the fact $P(I = 1) = P(I = 0) = 0.5$, and the fifth from Lemma 6.4.

Lemma 6.4: If $Z$ is a zero-mean Gaussian random variable with standard deviation $\sigma_Z$, then

$$
E(Z \mid Z > 0) = \sqrt{2\sigma_Z^2/\pi} \text{ and } E(Z \mid Z < 0) = -\sqrt{2\sigma_Z^2/\pi} \ .
\tag{6.28}
$$

Proof: As already proved in the proof of Lemma 6.1 (See Appendix D), $f_{Z|Z>0}(z \mid Z > 0) = f_Z(z) / P(Z > 0)$ for $z > 0$ and $f_{Z|Z>0}(z \mid Z > 0) = 0$ for $z \leq 0$. Since $Z$ has a symmetric pdf, $P(Z > 0) = 0.5$ and then

$$
E(Z \mid Z > 0) = \int_{-\infty}^{\infty} z f_{Z|Z>0}(z \mid Z > 0) dz = \int_{0}^{\infty} 2z f_Z(z) dz = \int_{0}^{\infty} 2z \exp[-z^2/(2\sigma_Z^2)] \Big/ \sqrt{2\pi\sigma_Z^2} \, dz = \sqrt{2\sigma_Z^2/\pi} \ .
$$

The second equation in (6.28) can be similarly proved.                    ∎

Due to the second term in (6.27), $D_w$ may be larger than $l^2$ at small $N$s. However, it is smaller than $l^2$ at larger $N$s. Moreover, it converges to $l^2$ as $N$ tends to infinity.

### 6.5.3  Performance under attacks

DS-ASS-HIR employs the linear correlator (3.59) with the double-sided rule (6.4) to detect the embedded watermarks. Thus under no attack, we see from Fig. 6.1(b) that $p_m = 0.0$ if $\psi < l$ and $p_m = 1.0$ if $\psi \geq l$. Under attacks, the detector makes a decision between $H_0$: $\overline{Y} = \overline{X} + \overline{V}$ and $H_1$: $\overline{Y} = \pm l + \overline{V}$ . Thus, the performance is given by

$$p_{fa} = 2Q(\psi / \sigma_0), \tag{6.29}$$

$$
\begin{aligned}
p_m &= P(|\overline{Y}| < \psi \mid I = 1)P(I = 1) + P(|\overline{Y}| < \psi \mid I = 0)P(I = 0) \\
&= 0.5P(-\psi < l + \overline{V} < \psi \mid I = 1) + 0.5P(-\psi < -l + \overline{V} < \psi \mid I = 0) \\
&= 0.5P(-\psi < l + \overline{V} < \psi) + 0.5P(-\psi < -l + \overline{V} < \psi) \\
&= Q[(-\psi - l)/\sigma_{\overline{V}}] - Q[(\psi - l)/\sigma_{\overline{V}}]
\end{aligned}
\tag{6.30}
$$

where $\sigma_0^2 = (\sigma_X^2 + \sigma_V^2)/N$, $\sigma_{\overline{V}}^2 = \sigma_V^2/N$, the first equality in (6.30) follows from conditioning on $I$, the second from the fact that $P(I = 0) = P(I = 1) = 0.5$ and the third from the independence of $\overline{V}$ and $\overline{X}$.

## 6.6 Performance comparisons and discussions

In this section, we make performance comparisons at the same distortion level to demonstrate the effectiveness of our proposed schemes. In the following discussions, "ASS-OPT" stands for the optimum detector (3.58) (with $\xi = c$), and "ASS-COR" for the correlation detector (3.59), "MSS-OPT" for the optimum detector (3.66) (with $\xi = c$), "BMSS" for Barni's multiplicative schemes (6.14), "DS-ASS" for additive schemes with double-sided detection (6.2), "DS-MSS" for multiplicative schemes with double-sided detection (6.20), "DS-BMSS" for Barni's scheme with double-sided detection (6.15), "E" for Empirical results and "T" for Theoretical results. Please also note that the empirical results in this section are obtained through Monte-Carlo simulations.

### 6.6.1 DS-ASS versus ASS

Fig. 6.2 compares the performance of ASS and DS-ASS at DWR = 20dB. Please note that the performance of ASS-COR and DS-ASS is invariant to the shape parameter $c$. It can be clearly observed from the comparison results that DS-ASS has a much better performance than ASS-COR (which is optimum at $c = 2.0$). At the high false alarm rates, above $10^{-3}$ in Fig. 6.2(a), DS-ASS can achieve a zero miss. Since the performance of ASS-OPT increases as $c$ decreases, DS-ASS, essentially designed for correlation detectors, is inferior to ASS-OPT in performance at the low false alarm rates in Fig. 6.2(a) and Fig. 6.2(b). However, DS-ASS is also applicable to the data that are difficult to model by a mathematically tractable pdf, for which case it is hard to find an optimum detector. Another disadvantage of ASS-OPT over DS-ASS is that ASS-OPT has to be informed of the embedding strength $a$, which is image-dependent for perceptual constraints.

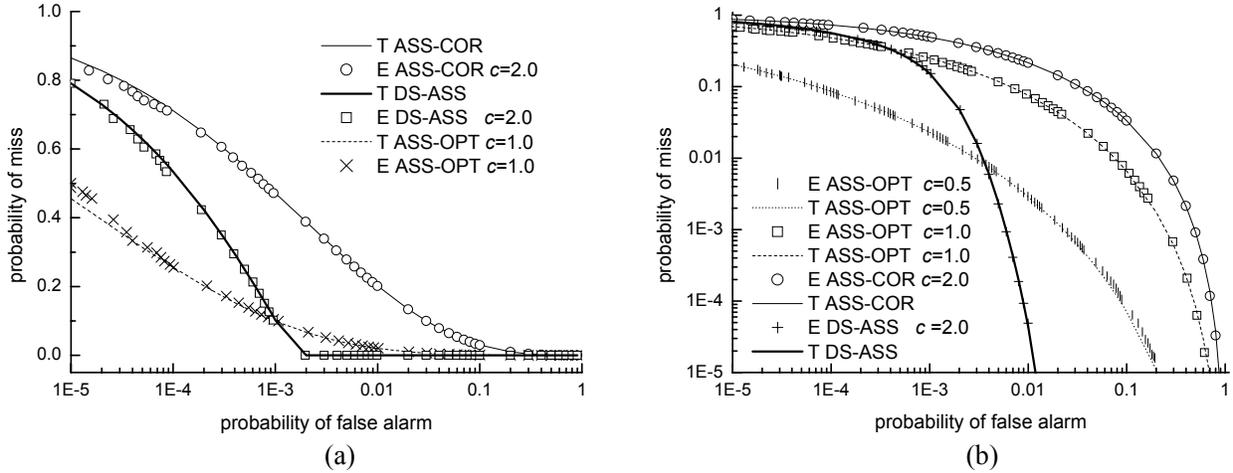

<div align="center">(a)                                                      (b)</div>

Fig. 6.2. ASS-OPT (with $\xi = c$) versus DS-ASS at DWR = 20dB, $N = 1000$ and $\sigma_X = 10.0$. Experimental results are obtained on 1,000,000 groups of GGD data with shape parameter $c$ and $\sigma_X = 10.0$. (a) Performance comparisons under no attack. (b) Performance comparisons under zero-mean Gaussian noise attacks at WNR = $-5$dB.

### 6.6.2    DS-ASS, DS-ASS-HIR and STDM

Pérez-Freire *et al.* [102] applied STDM scheme [46] to the watermark verification scenarios. Its performance is given by (4.22) and (4.23). In Fig. 6.3(a) and Fig. 6.3(b), we notice that DS-ASS-HIR achieves almost the same performance with STDM. In fact, as $N \to \infty$, $\bar{x} \to 0$ and $a = q_\Lambda(\bar{x}) - \bar{x} \to \pm\Delta/2$. Thus STDM (2.31) converges to DS-ASS-HIR (6.26) as $N$ increases. It can also be observed from Fig. 6.3(a) that both host-interference canceling schemes outperform DS-ASS at the high false alarm rates. However, DS-ASS can achieve a better performance at the low false alarm rates in Fig. 6.3(a).

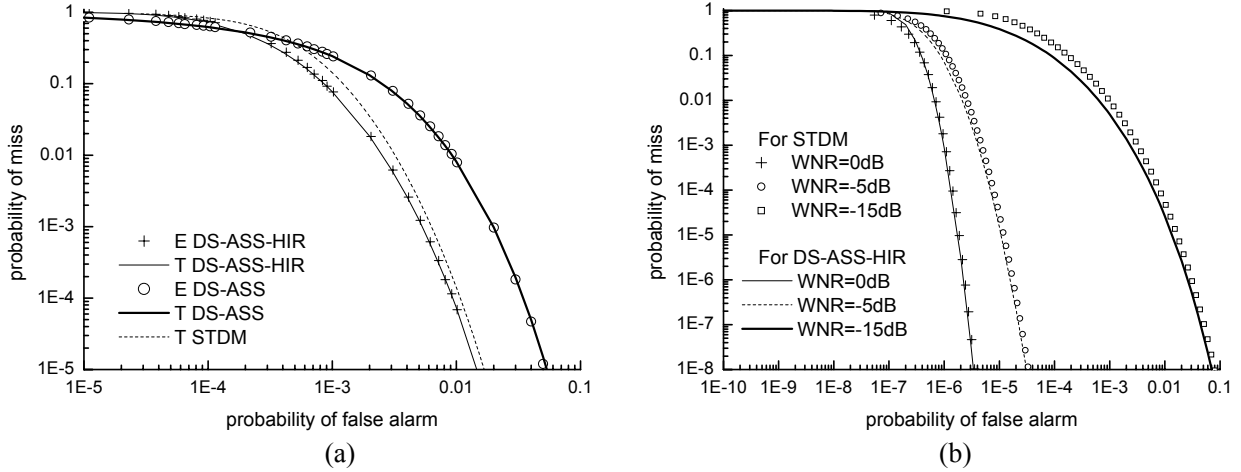

(a)                                                    (b)

Fig. 6.3. (a) DS-ASS, DS-ASS-HIR and STDM at $N = 1000$, DWR = 20dB, WNR = $-10$dB, and $\sigma_X = 10.0$. Empirical results are obtained on 1,000,000 groups of GGD data with $c = 1.0$ and $\sigma_X = 10$. (b) Theoretical performance comparisons between STDM and DS-ASS-HIR at DWR = 20dB and $N = 2000$.

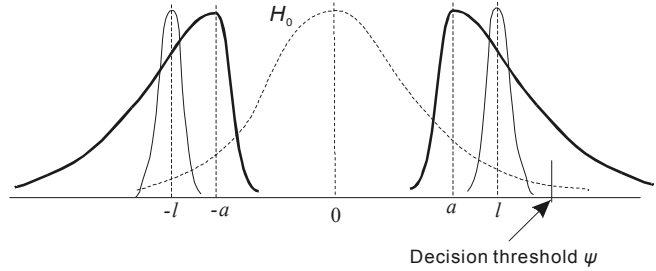

Fig. 6.4. The illustrative pdf of the decision statistics $L(\mathbf{Y}|H_0)$ and $L(\mathbf{Y}|H_1)$ under noise attacks. The thick and solid curves represent $L(\mathbf{Y}|H_1)$ for DS-ASS and DS-ASS-HIR, respectively.

Since the embedding distortion for DS-ASS is $a^2$, $l$ in (6.27) must be larger than $a$ to achieve the same distortion for DS-ASS-HIR. Under noise attacks, since $L(\mathbf{Y}|H_1) = \bar{X} \pm a + \bar{V}$ for DS-ASS and $L(\mathbf{Y}|H_1) = \pm a + \bar{V}$ for DS-ASS-HIR, $Var(L(\mathbf{Y}|H_1))$ for DS-ASS is larger than that for DS-ASS-HIR (see Fig. 6.4). Thus, if $\psi < l$, DS-ASS-HIR has a smaller $p_m$. However, if $\psi > l$ (as indicated in Fig. 6.4), then $p_m$ for DS-ASS-HIR is larger than that for DS-ASS. This explains why DS-ASS-HIR can achieve a better performance at the high false alarm rates and a poorer performance at the low false rates. The distortion $D_w$ for STDM is [102]

$$D_w = \sum_{i=-\infty}^{\infty} \int_{i\Delta}^{i\Delta+\Delta} (x - 0.5\Delta - i\Delta)^2 f_{\bar{X}}(x)dx < \sum_{i=-\infty}^{\infty} \int_{i\Delta}^{i\Delta+\Delta} (0.5\Delta)^2 f_{\bar{X}}(x)dx = (0.5\Delta)^2 \qquad (6.31)$$

since $-0.5\Delta < x - 0.5\Delta - i\Delta < 0.5\Delta$ for $i\Delta < x < i\Delta+\Delta$. Thus to achieve the same distortion $a^2$, $\Delta/2$ must be larger than $a$, which leads to the same advantage of STDM over DS-ASS at the high false alarm rates. Though STDM and DS-ASS-HIR achieve a better performance at high false alarm rates, they are not encouraged in real scenarios due to the reason outlined in Section 6.3.2.

### 6.6.3 BMSS and DS-BMSS

In Fig. 6.5, we compared the performance of BMSS and DS-BMSS. It is clearly shown that DS-BMSS yields a better performance than BMSS does. Moreover, the effectiveness of the approximation ($\sum_{1 \le i \le N} |X_i|^\xi$ )$/N \approx E(|X|^\xi)$ in Section 6.4.2 is also verified by the nice agreement between theoretical and experimental results.

### 6.6.4 MSS and DS-MSS

The same improvement over MSS-OPT can be observed from both Fig. 6.6(a) and Fig. 6.6(b). Moreover in Fig. 6.6(a), the empirical performance of DS-MSS agrees with the theoretical one predicted in (6.24). The performance decreases as the shape parameter $c$ decreases. The disadvantage of DS-MSS is that it relies on the knowledge of the shape parameter $c$ to choose an optimum $\xi$. This knowledge can be obtained from the estimation of the shape parameter $c$ [69, 103]. If $\xi$ is not known beforehand at the detector, it has to be estimated from the attacked image and the estimated $c$ would differ from that estimated from the original image. However, this difference is quite small [77] due to the imperceptibility requirement. An alternative is to adopt the same $\xi$ at both embedders and detectors. However, it is at the cost of performance.

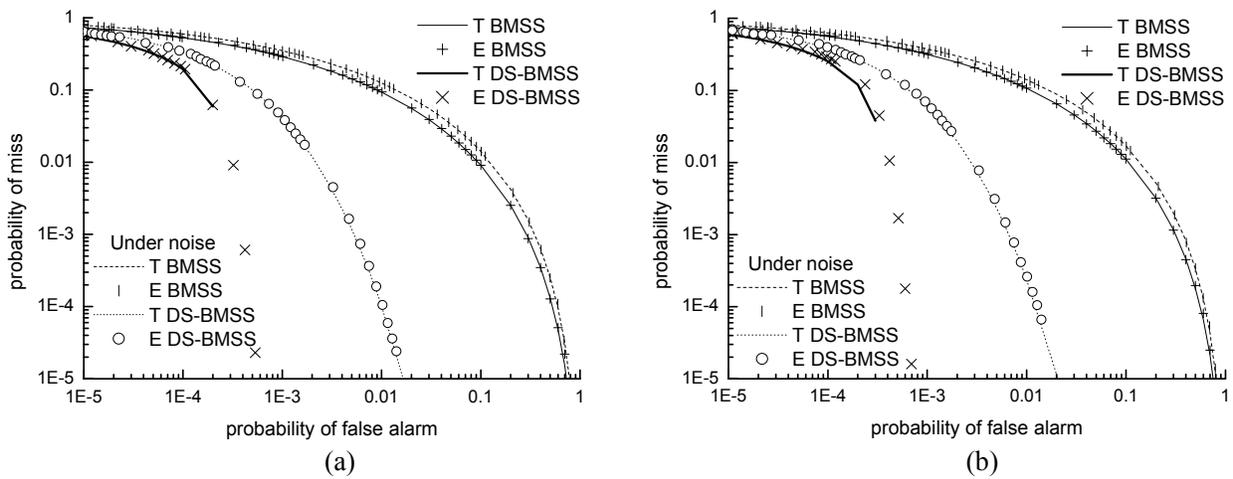

Fig. 6.5. BMSS versus DS-BMSS at DWR = 20dB, WNR = −10dB, $\sigma_X$ = 10.0 and $N$ = 2000. Empirical results are obtained on 1,000,000 groups of data with the specified $c$ and $\sigma_X$ = 10.0. In the legends of this figure, "noise" represents the results obtained under zero-mean Gaussian noise attacks. (a) $c$ = 1.0 and $\xi$ = 0.7. (b) $c$ = 2.0 and $\xi$ = 1.0.

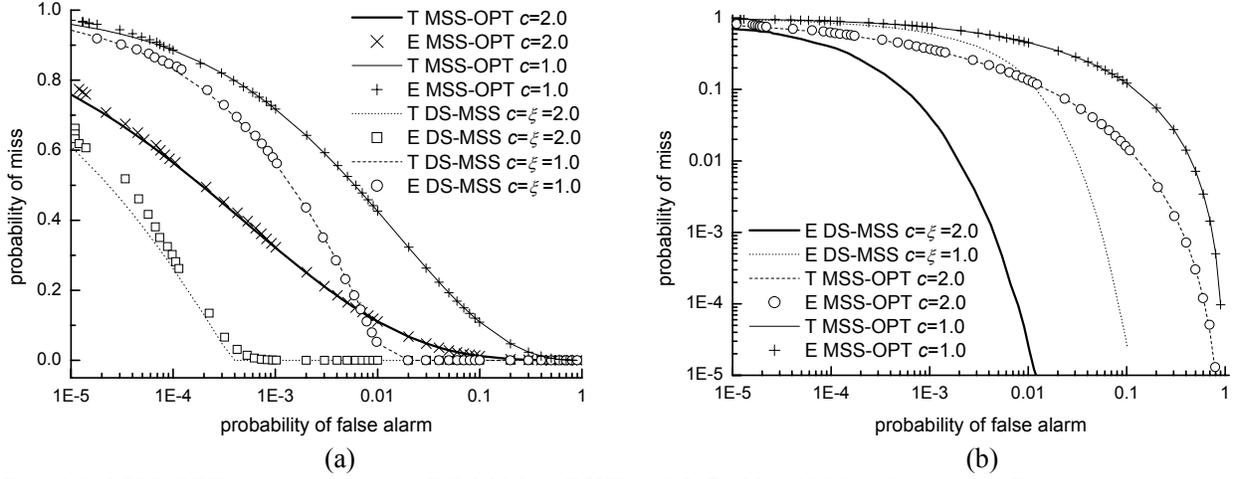

Fig. 6.6. MSS-OPT (with $\xi = c$) and DS-MSS at DWR = 25dB, $N = 2000$ and $\sigma_X = 10$. Empirical results are obtained on 1,000,000 groups of data with shape parameter $c$ and $\sigma_X = 10$. (a) Under no attack. (b) Under zero-mean Gaussian noise attacks at WNR = −10dB.

### 6.6.5 Comparisons among double-sided schemes

In this subsection, we make performance comparisons among all double-sided detection schemes. Since $D_w = a^2$ for DS-ASS, plugging DWR into (6.10) and substituting $\psi$ with $p_{fa}$ (see (6.8)) leads to

$$p_m = \begin{cases} 1 - 2Q[Q^{-1}(p_{fa}/2) - \rho], & \text{if } p_{fa} < 2Q(\rho) \\ 0.0, & \text{if } p_{fa} \geq 2Q(\rho) \end{cases}. \tag{6.31}$$

where $\rho = 10^{-\text{DWR}/20}\sqrt{N}$ . \hfill (6.32)

Similarly, the performance of DS-BMSS and DS-MSS can also expressed by (6.31), but with $\rho$ replaced by

$$\rho = 10^{-\text{DWR}/20}\sqrt{N}\, E(|X|^{\xi}) \Big/ \sqrt{E(|X|^{2\xi})} , \tag{6.33}$$

$$\rho = 10^{-\text{DWR}/20}\sqrt{N}\,\text{MVR}(\xi) . \tag{6.34}$$

for DS-BMSS and DS-MSS respectively. It is easy to see from (6.33) and (6.34) that DS-BMSS has a worse performance than DS-ASS since $E(|X|^{\xi})\Big/\sqrt{E(|X|^{2\xi})} < 1.0$. However, DS-BMSS can achieve a better perceptual quality since it automatically implements a simple contrast masking of Watson's perceptual model [55]. Moreover, DS-BMSS produces a better performance as $\xi$ decreases. In fact, DS-BMSS degenerates into DS-ASS at $\xi = 0$ (See (6.15) and (6.2)). Since MVR(1.0) = 1.0 at $c = 1.0$, DS-MSS (at $\xi = c$) outperforms DS-ASS at $c$ above 1.0. Compared with DS-BMSS, DS-MSS achieves a better performance if $c > 0.5$. These conclusions are also reflected in Fig. 6.7(a). In Fig. 6.7(b), we compared their performances under attacks. It can be seen from Fig. 6.7(b) that at $c = 2.0$, DS-MSS outperforms DS-ASS-HIR and DS-ASS at the low false alarm rates, and however, is inferior to DS-ASS-HIR and DS-ASS at $c < 1.0$. Moreover, DS-BMSS also yields a poorer performance than DS-ASS does under attacks.

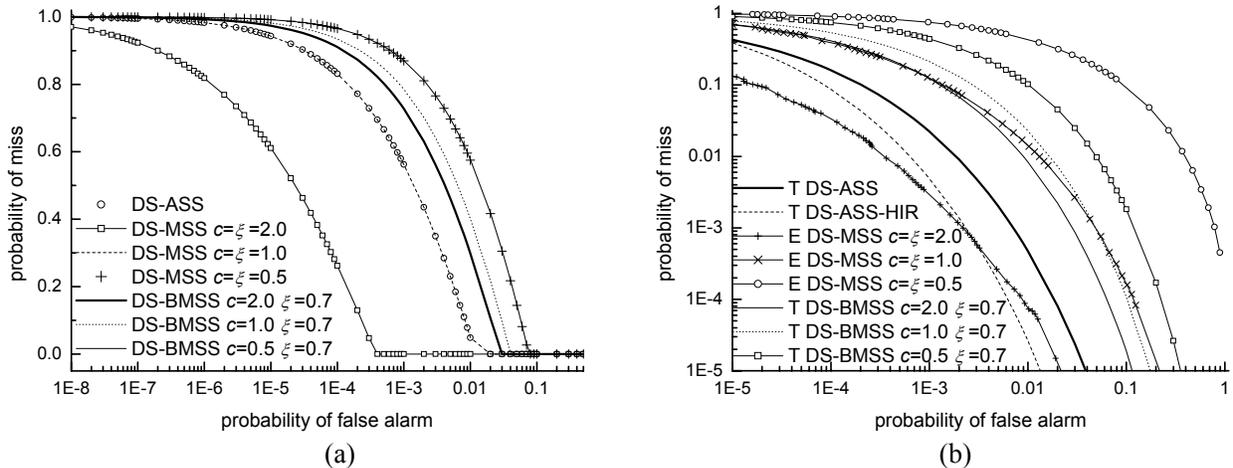

Fig. 6.7. Performance comparisons among double-sided schemes at $N = 2000$ and $\sigma_X = 10$. (a) Theoretical performance comparisons under no attack at DWR = 25dB. (b) Under zero-mean Gaussian noise attacks at DWR = 20dB and WNR = −15dB. Empirical results are obtained on 1,000,000 groups of data with shape parameter $c$ and $\sigma_X = 10$.

## 6.7 Experimental results

In this section, the experiments are conducted on real natural images, namely, Lena, Peppers, Boat and Baboon. For the ease of performance comparisons, we don't consider the perceptual analysis in this chapter. Performance comparisons with perceptual analysis on real images are listed in Chapter 8. For each 8×8 DCT block, we use just one coefficient for watermarking since different DCT frequency coefficients usually have different shape parameters. Thus, using one coefficient in each block can simplify the performance comparisons.

The shape parameter $c$ and standard deviation $\sigma_X$ of the host DCT coefficients are estimated by the ML method [103]. For the 5[th] (in Zigzag order) coefficient of Lena, its estimated $c$ is 0.69 and $\sigma_X = 19.74$; for the 5[th] coefficient of Peppers, $c = 1.03$ and $\sigma_X = 16.04$; for the 5[th] coefficient of Boat, $c = 0.73$ and $\sigma_X = 24.32$; for the 5[th] coefficient of Baboon, $c = 1.11$ and $\sigma_X = 32.12$. Since $c$ for the attacked data does not vary much from that for the original data due to the imperceptibility requirement, we instead use the estimated $c$ of the original data in detection for convenience.

Since we do not have a large set of natural images with the same $c$ and $\sigma_X$, we instead permute the host DCT coefficients to obtain different $\mathbf{x}$s. Each permutation obtains a new test data $\mathbf{x}$, and our experimental results are obtained on 1,000,000 such permutations. Please also refer to Appendix E for further reference.

The comparison results under zero-mean Gaussian noise attacks are displayed in Fig. 6.8, Fig. 6.9 and Fig. 6.10. It can be clearly observed that our double-schemes dramatically improve the corresponding single-sided schemes. As also stated in Section 6.6.2, STDM and DS-ASS-HIR achieves almost the same performance. Despite their performance advantages over DS-ASS in Fig. 6.8 and Fig. 6.11, the host inference rejection schemes are not encouraged in real scenarios for the reason outlined in Section 6.3.2. The comparison results under JPEG attacks are displayed in Fig. 6.11, Fig. 6.12 and Fig. 6.13. At Quality Factor (QF) = 50, the same performance advantage of double-sided schemes over single-sided schemes can be observed. Though ASS-OPT achieves a better performance than DS-ASS does at the low false alarm rates under Gaussian noise attacks (See Fig. 6.8), it is inferior to DS-ASS under JPEG attacks.

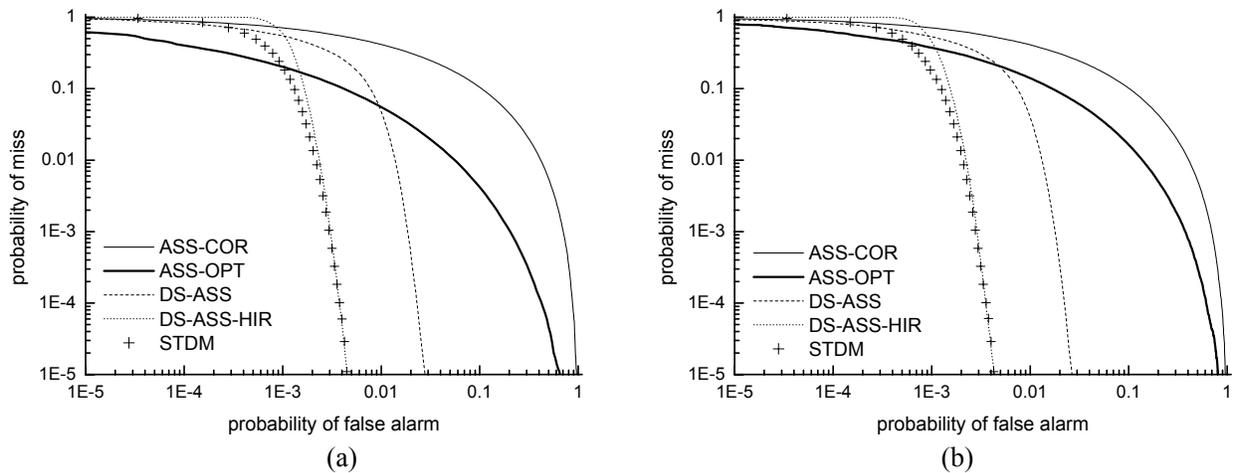

(a)                                                      (b)

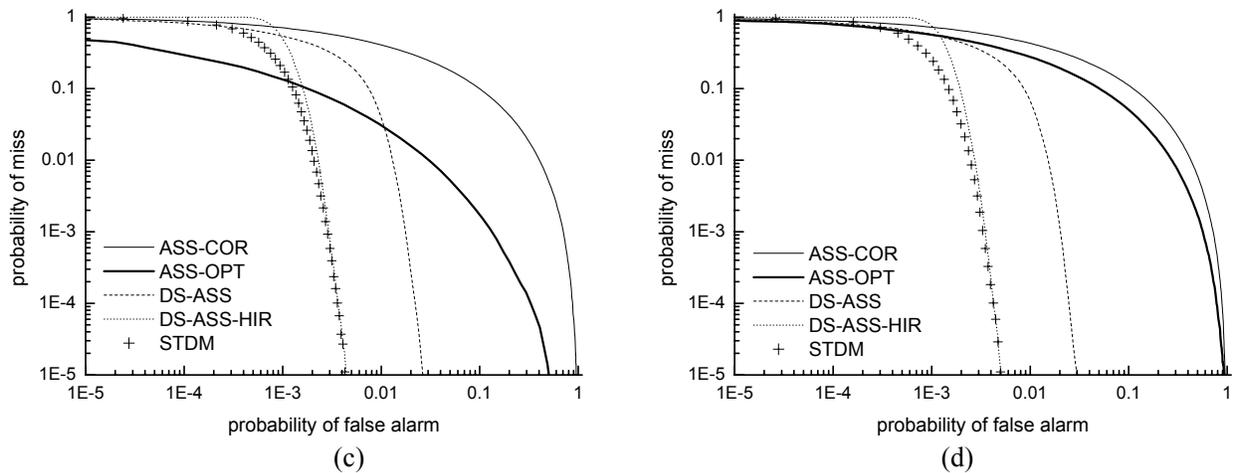

(c)                                                                    (d)

Fig. 6.8. Performance comparisons among additive schemes under zero-mean Gaussian noise attacks at DWR = 25dB, WNR = −5dB and $N$ = 2000. Please also note that $\xi = c$ for ASS-OPT. (a) For the 5th coefficient of Lena. (b) For the 5th coefficient of Peppers. (c) For the 5th coefficient of Boat. (d) For the 5th coefficient of Baboon.

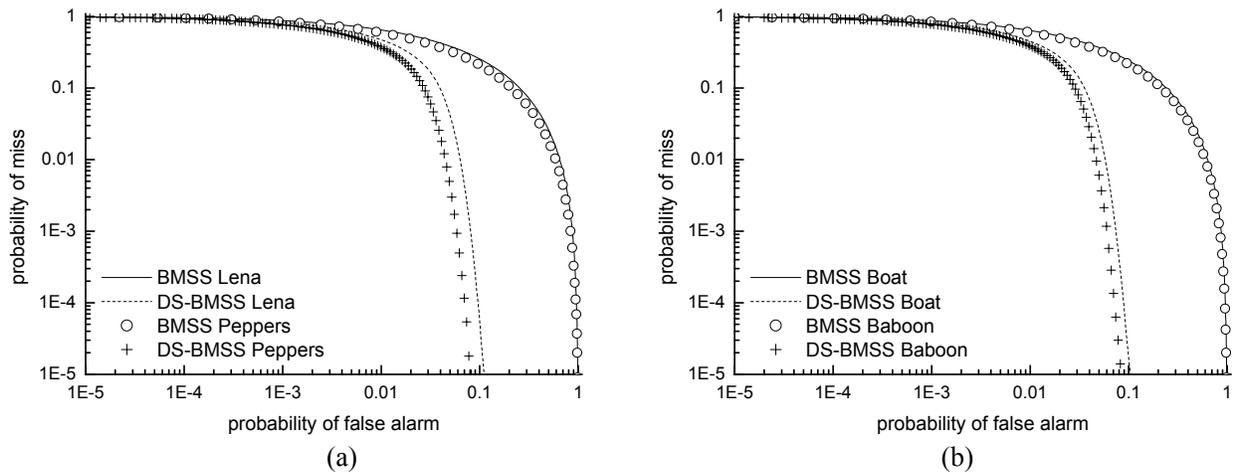

(a)                                                                    (b)

Fig. 6.9. Performance comparisons between BMSS ($\xi = 0.7$) versus DS-BMSS ($\xi = 0.7$) under zero-mean Gaussian noise attacks at DWR = 25dB, WNR = −5dB and $N$ = 2000. (a) For the 5th coefficients of Lena and Peppers. (b) For the 5th coefficients of Boat and Baboon.

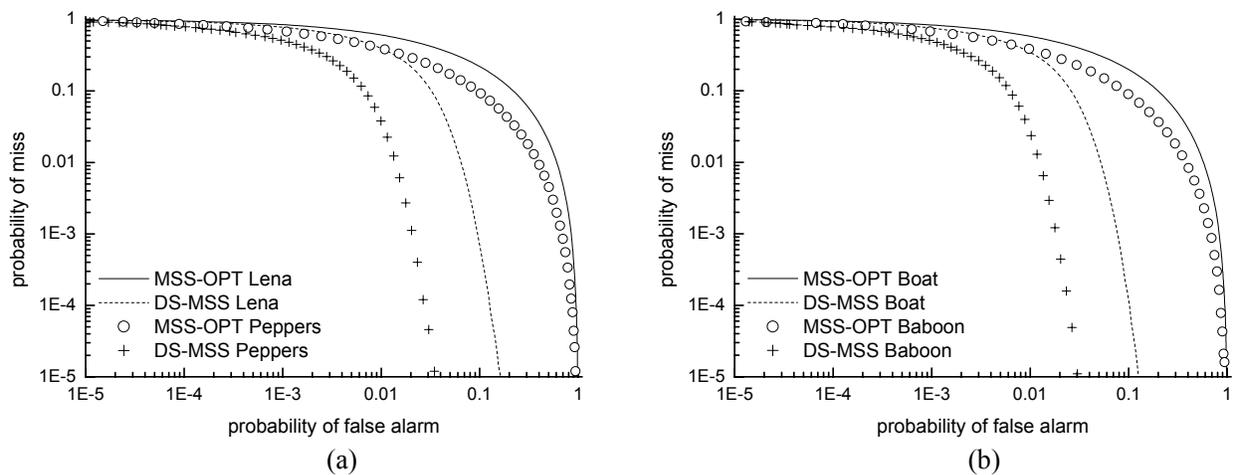

(a)                                                                    (b)

Fig. 6.10. Performance comparisons between MSS-OPT ($\xi = c$) and DS-MSS ($\xi = c$) under zero-mean Gaussian noise attacks at DWR = 25dB, WNR = −5dB and $N = 2000$. (a) For the $5^{\text{th}}$ coefficients of Lena and Peppers. (b) For the $5^{\text{th}}$ coefficients of Boat and Baboon.

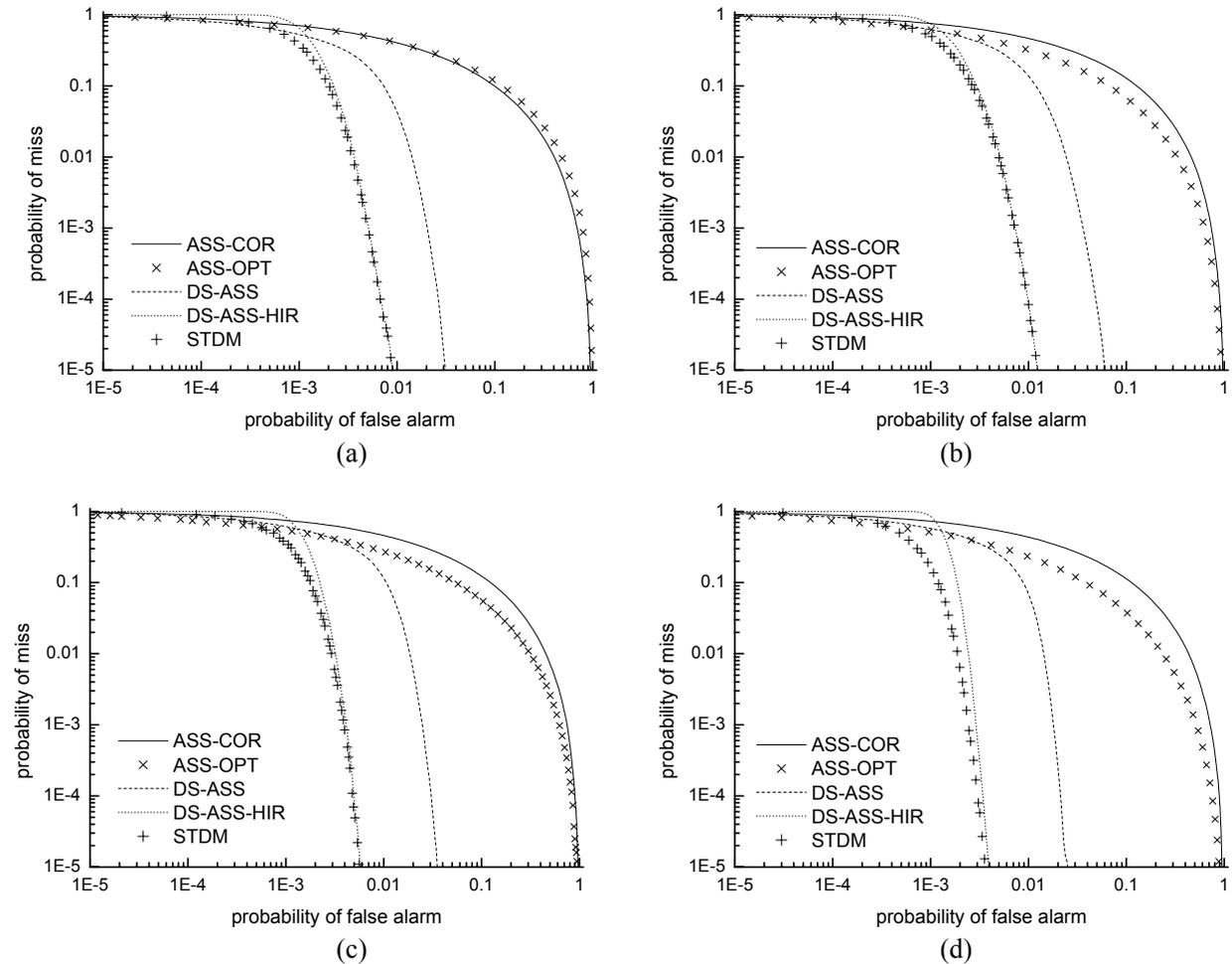

(a)                                      (b)

(c)                                      (d)

Fig. 6.11. Performance comparisons between additive schemes under JPEG (QF = 50) attacks at DWR = 25dB and $N = 2000$. Please also note that $\xi = c$ for ASS-OPT. (a) For the $5^{\text{th}}$ coefficient of Lena. (b) For the $5^{\text{th}}$ coefficient of Peppers. (c) For the $5^{\text{th}}$ coefficient of Boat. (d) For the $5^{\text{th}}$ coefficient of Baboon.

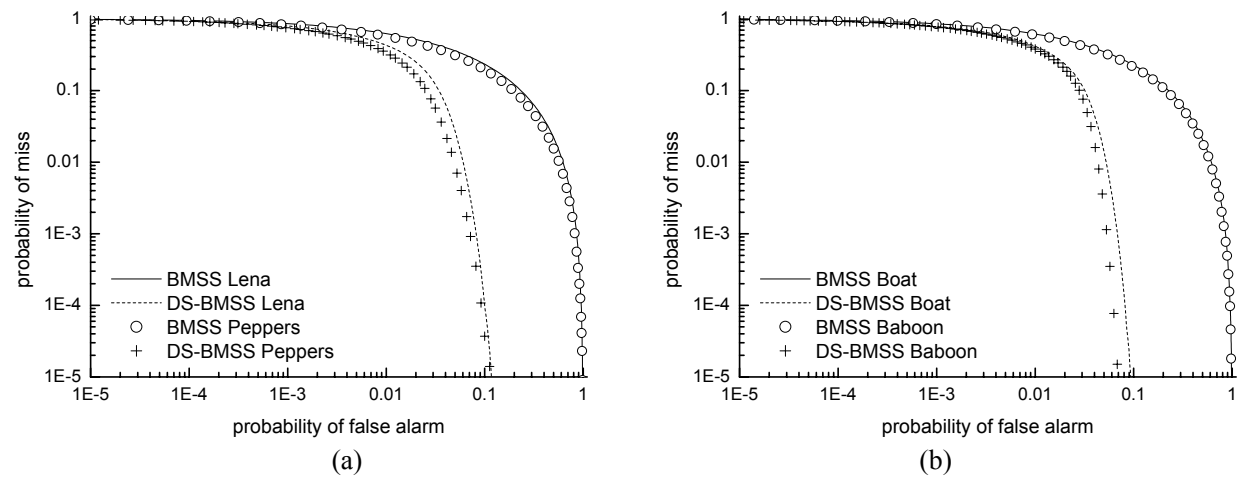

(a)                                      (b)

Fig. 6.12. Performance comparisons between BMSS ($\xi = 0.7$) and DS-BMSS ($\xi = 0.7$) under JPEG (QF = 50) attacks at DWR = 25dB and $N = 2000$. (a) For the $5^{th}$ coefficients of Lena and Peppers. (b) For the $5^{th}$ coefficients of Lena and Peppers.

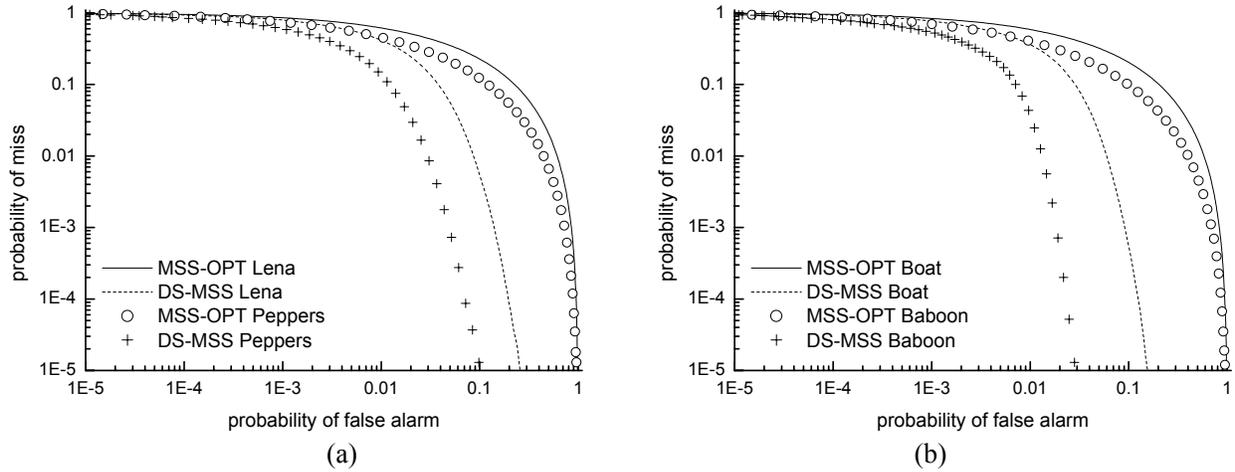

(a)                                         (b)

Fig. 6.13. Performance comparisons between MSS-OPT ($\xi = c$) and DS-MSS ($\xi = c$) under JPEG (QF = 50) attacks at DWR = 25dB and $N = 2000$. (a) For the $5^{th}$ coefficients of Lena and Peppers. (b) For the $5^{th}$ coefficients of Boat and Baboon.

## 6.8  Conclusions

In this chapter, we presented a double-sided technique for SS schemes. Different from the traditional SS and host interference rejection schemes, it utilizes but does not reject the host interference. Through both theoretical and empirical comparisons, we found that the proposed double-sided schemes could achieve a great performance advantage over the traditional SS schemes and overcome the drawbacks of the host interference rejection schemes. Though DS-ASS and DS-BMSS are not the best in performance, they are the two most appealing schemes that offer advantages

*1) Over its single-sided counterparts (such as ASS-COR and BMSS) in performance*.

*2) Over host interference rejection schemes (such as STDM and DS-ASS-HIR)*: The perceptual analysis can be easily implemented to achieve the locally bounded maximum embedding strength, whereas it is difficult for host interference rejection schemes.

*3) Over optimal detectors (such as ASS-OPT, MSS-OPT and DS-MSS)*: They have an advantage over ASS-OPT in that they require no knowledge of the embedding strength, especially in the case where the embedding strength should be locally bounded by the perceptual analysis. The second advantage is that they do not need to estimate the shape parameter for watermark detection. The third is that they both employ a simple linear correlator in no need of expensive computations. The fourth is that they are universally applicable to all kinds of data whatever their probabilities of distributions are.

The idea of DS-ASS is also reflected in a recently submitted paper [116] where the watermark embedding and detection are investigated from an information-theoretic approach. This work is quite interesting and may be coupled with our work to gain a deeper understanding of the double-sided technique. Moreover, it would be much interesting if [116] could produce a better embedding rule without host interference rejection. However, ours offers a different treatment of the idea.

Our embedding rule is also designed according to a given detection function. However, different from the works [115] and [116], we did not choose to optimize the embedding function. We follow a very intuitive approach to utilize side information to increase the magnitude of the detector's output. For a simple correlator, the correlation between the host signals and the watermark determines whether the embedding should take an additive or a subtractive form. Of course, owning to LMP tests, the simple linear correlation is further generalized to a generalized form in our work. For instance, DS-MSS and DS-Cauchy (See Chapter 8) adopt the generalized correlation in their embedding rules. In fact, these correlators have a nice property that their decision statistics for the unwatermarked content are symmetrical about zero. Thus given any possible correlator, we can propose a corresponding double-sided embedding function. Therefore with our approach, we investigated the double-sided idea in both additive and multiplicative cases.

Second, we also investigated how to implement the perceptual analysis for DS-ASS (See Chapter 8 for details). In fact, in our work, we emphasized on the nice property of our double-sided scheme that it utilizes but not reject the host interference. Due to this nice property, double-sided schemes offer a big advantage over the host interference rejection schemes that the perceptual analysis can be easily implemented to achieve a maximum possible robustness.

Third, we analyzed its performance under both no attack and attacks on a more practical model of host signals, namely, GGD model. Specifically, we provided a simpler proof of its performance advantage over ASS-COR for any given $N$. However, they gave a much involved proof of its performance advantage in the asymptotic sense.

Fourth, ours offered extensive performance comparisons with other schemes. For instance, in our work, performance comparisons under both Gaussian noise attacks and JPEG attacks were conducted to verify the performance advantage of our double-sided schemes. However, no empirical performance comparisons were provided in their work.

# Chapter 7   Double Sided Schemes in the DFT Domain

In the previous chapter, we presented a new double-sided technique and inspected its performance in the DCT domain. In this chapter, however, we investigate the performances of the double-sided schemes in the DFT domain. DS-ASS is not applicable to the DFT magnitudes since the watermarked data should be nonnegative. Thus, we only examine the performance gain of the double-sided multiplicative schemes over the traditional multiplicative schemes.

In the first section of this chapter, we review the single-sided schemes and provide a mathematically rigorous analysis of these schemes. As of this writing, no solid performance analysis has been given for the optimal detectors in the DFT domain. In the second section, the double-side schemes in the DFT domain are briefly outlined with special attention paid to the difference from those in the DCT domain. The following section then conduct performance comparisons to justify the advantages of our double-sided schemes. In the last section, we perform Monte-Carlo simulations to validate the conclusions drawn in the previous sections.

## 7.1  Single-sided schemes

### 7.1.1  Correlator

Since the DFT magnitudes are nonnegative, the Barni's multiplicative rule reduces to the multiplicative rule (2.21) which for reference is listed as

$$s_i = x_i + a \cdot x_i \cdot w_i \tag{7.1}$$

The performance of the correlator is determined by [42]

$$m_0 = 0, \ \sigma_0^2 = Var(X), \ m_1 = aE(X), \ \sigma_1^2 = Var(X)(1 + a^2) \approx Var(X) \tag{7.2}$$

since $a$ is rather small to keep the embedding distortion imperceptible. In order to evaluate its performance under attacks, we consider a kind of absolute Gaussian noise attacks to avoid the negative values of Gaussian noise. The attack can be expressed as $\{|V_1|, |V_2|, \ldots, |V_N|\}$, where $V_i$s are i.i.d. Gaussian random variables with a zero mean and standard deviation $\sigma_V$. Thus under attacks, the detector decides between $H_0$: $Y_i = X_i + |V_i|$ and $H_1$: $Y_i = X_i + a \cdot X_i \cdot w_i + |V_i|$. It is also easy to see that its performance under attacks is depicted by

$$
\begin{aligned}
m_0 &= 0, & \sigma_0^2 &= Var(X) + Var(|V|) \\
m_1 &= aE(X), & \sigma_1^2 &= Var(X)(1+a^2) + Var(|V|) \approx Var(X) + Var(|V|)
\end{aligned}
\tag{7.3}
$$

### 7.1.2  Optimum decision statistic

In [75, 76], DFT magnitudes are assumed to follow a Weibull distribution. The estimation of the shape and scale parameters has been outlined in Chapter 2. This assumption thus leads to the likelihood test

$$
l(\mathbf{s}) = \frac{f_{\mathbf{S}|H_1}(\mathbf{s})}{f_{\mathbf{S}|H_0}(\mathbf{s})} = \frac{\prod_{i=1}^{N} \frac{1}{1+aw_i} f_{X_i}\left(\frac{s_i}{1+aw_i}\right)}{\prod_{i=1}^{N} f_{X_i}(s_i)} = \frac{\exp\left[-\sum_{i=1}^{N} \frac{s_i^{\delta}}{(1+aw_i)^{\delta}}\right] \prod_{i=1}^{N} \frac{1}{(1+aw_i)^{\delta}}}{\exp\left[-\sum_{i=1}^{N} s_i^{\delta}\right]}
\tag{7.4}
$$

It thus follows that

$$
L(\mathbf{s}) = \ln[l(\mathbf{s})] = \delta \ln \frac{1}{1+aw_i} + \sum_{i=1}^{N} s_i^{\delta}\left[1 - \frac{1}{(1+aw_i)^{\delta}}\right]
\tag{7.5}
$$

Since the first term in the above equation is a fixed value for a given $\mathbf{w}$, the above optimum decision statistic is thus

$$
L(\mathbf{S}) = \sum_{i=1}^{N} S_i^{\delta}\left[1 - \frac{1}{(1+aw_i)^{\delta}}\right]
\tag{7.6}
$$

It is identical to (3.64) in the DCT domain. Since $a << 1$, this optimum decision rule can also be replaced by a suboptimal decision statistic [74, 77]

$$
L(\mathbf{S}) = \frac{1}{N} \sum_{i=1}^{N} S_i^{\delta} w_i
\tag{7.7}
$$

Although it is suboptimal, it offers an advantage over (7.6) in that it entails no knowledge of the embedding strength. This advantage is vital to the corresponding double-sided schemes.

### 7.1.3  Performance of optimal detectors under no attack

The performance analysis for (7.7) bears a close resemblance to that of the generalized correlators in the DCT domain. Therefore similar results can be obtained as

$$p_m = 1 - Q[Q^{-1}(p_{fa}) - a\sqrt{N}\text{MVR}(\delta)] \tag{7.8}$$

However, for Weibull pdf,

$$
\begin{aligned}
E(X^\xi) &= \int_0^\infty \frac{\delta}{\theta}(x/\theta)^{\delta-1}\exp[-(x/\theta)^\delta]x^\xi dx \\
&= \int_0^\infty \frac{\delta}{\theta}(x/\theta)^{\delta+\xi-1}\theta^\xi \exp[-(x/\theta)^\delta]dx \\
&= \int_0^\infty y^{(\delta+\xi-1)/\delta}\theta^\xi \exp(-y)y^{(1/\delta)-1}dy \quad (\text{Let } x = \theta y^{1/\delta}) \\
&= \int_0^\infty y^{\xi/\delta}\theta^\xi \exp(-y)dy \\
&= \Gamma(\xi/\delta+1)\int_0^\infty y^{(\xi/\delta+1)-1}\theta^\xi \exp(-y)\big/\Gamma(\xi/\delta+1)\,dy \\
&= \Gamma(\xi/\delta+1)\theta^\xi
\end{aligned}
\tag{7.9}
$$

and thus

$$Var(X^\xi) = E(X^{2\xi}) - [E(X^\xi)]^2 = \Gamma(2\xi/\delta+1)\theta^{2\xi} - [\Gamma(\xi/\delta+1)]^2\theta^{2\xi} \tag{7.10}$$

Hence, MVR is

$$\text{MVR}(\xi) = \frac{\xi E(|X|^\xi)}{\sqrt{Var(|X|^\xi)}} = \frac{\xi E(X^\xi)}{\sqrt{Var(X^\xi)}} = \frac{\xi\Gamma(\xi/\delta+1)}{\sqrt{\Gamma(2\xi/\delta+1) - [\Gamma(\xi/\delta+1)]^2}} \tag{7.11}$$

We see from the above equation that MVR($\cdot$) does not depend on $\theta$. It is difficult to maximize (7.11) analytically. However, we found through the numerical search that MVR achieves its maximum at $\xi = \delta$ for all $\delta$ in [0.1, 4.0]. Moreover, substituting $\xi = \delta$ into the above equation, we have

$$\text{MVR}(\delta) = \delta . \tag{7.12}$$

### 7.1.4  Performance of generalized correlators under attacks

The decision statistics for $H_0$ and $H_1$ under attacks are

$$L(\mathbf{Y}\,|\,H_0) = \frac{1}{N}\sum_{i=1}^N [X_i + |V_i|]^\delta w_i \text{ and } L(\mathbf{Y}\,|\,H_1) = \frac{1}{N}\sum_{i=1}^N [X_i(1+aw_i) + |V_i|]^\delta w_i , \tag{7.13}$$

respectively. Therefore, the performance for the generalized correlator (7.7) is determined by

$$m_0 = 0, \quad \sigma_0^2 = Var([X + |V|]^\delta) / N \,,$$
(7.14)

$$m_1 = \frac{E\{[X(1+a) + |V|]^\delta\} + E\{[X(1-a) + |V|]^\delta\}}{2},$$
(7.15)

$$\sigma_1^2 = \frac{Var\{[X(1+a) + |V|]^\delta\} + Var\{[X(1-a) + |V|]^\delta\}}{2N},$$
(7.16)

where numerical convolutions are employed to obtain the pdf of $X + |V|$, $X(1+a) + |V|$ and $X(1-a) + |V|$.

## 7.2  Double-sided schemes in the DFT domain

### 7.2.1  Double-sided scheme for Barni's rules

In this subsection, we investigate the performance of DS-BMSS schemes in the DFT domain. The embedding rule for DS-BMSS (with $\xi = 1.0$) is given by

$$s_i = x_i + ax_iw_i, \quad \text{if } \bar{x} > 0; \; s_i = x_i - ax_iw_i, \quad \text{if } \bar{x} \le 0 \,,$$
(7.17)

where

$$\bar{x} = \frac{1}{N} \sum_{i=1}^{N} x_iw_i \,.$$
(7.18)

The performance is thus given by (6.17), however with $\xi = 1.0$. Please also note that $E(|X|) = E(X)$ since $X$ is nonnegative. The performance under attacks is similarly characterized by (6.18) and (6.19), but with

$$m_1 = aE(X), \; \sigma_0^2 = [Var(X) + Var(|V|)] / N, \; \sigma_1^2 = Var(|V|) / N$$
(7.19)

### 7.2.2  Double-sided scheme for optimum decision rules

In this section, we examine the performance of DS-MSS in the DFT domain. The embedding rule is described in (6.20) and (6.21). Its performance under no attack is also depicted in (6.24), however with MVR replaced by (7.11). It also achieves its best performance at $\xi = \delta$.

## 7.3  Performance comparisons

In this section, we make performance comparisons between the double and single sided schemes in the DFT domain. Please first note that in the legends of the following figures, "MSS-OPT" stands for the suboptimal detector (7.7), "E" for Empirical results, "T" for Theoretical results and other notations follow the definitions in the previous chapter.

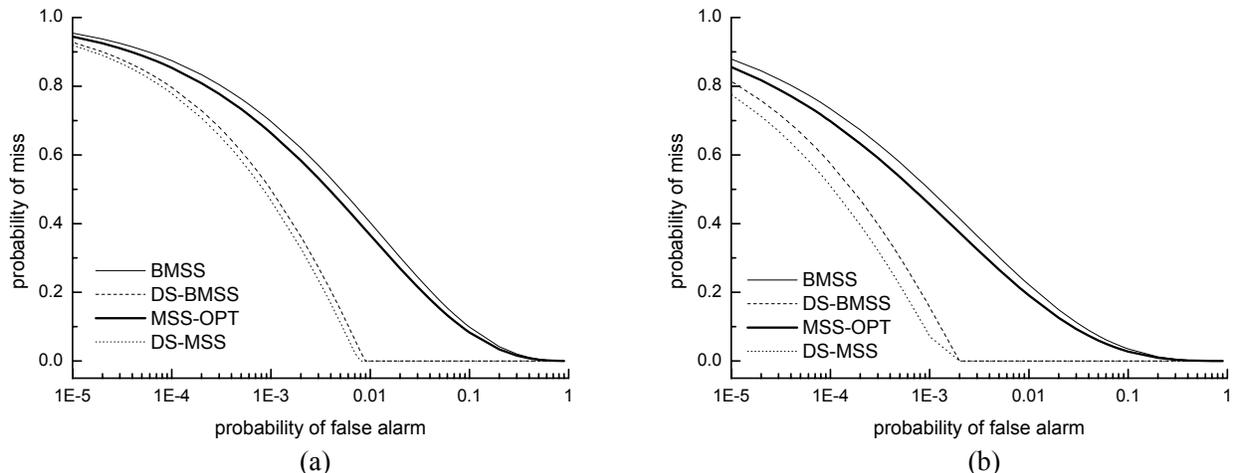

Fig. 7.1. Theoretical performance comparisons between single and double-sided schemes at DWR = 25dB and $N$ = 1000 (a) $\theta$ = 0.05, $\delta$ = 1.5. (b) $\theta$ = 0.1, $\delta$ = 1.8.

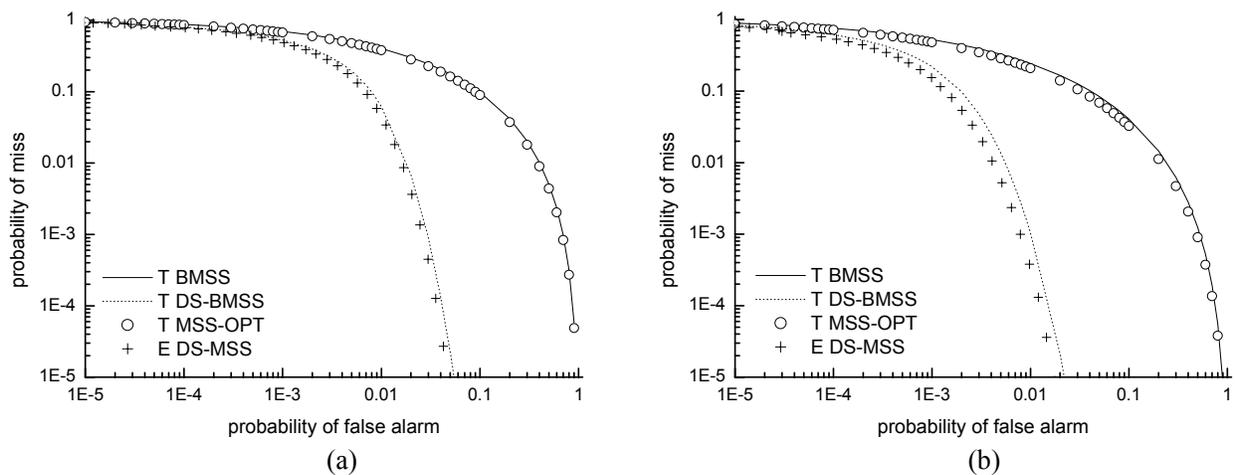

Fig. 7.2. Performance comparisons between single and double sided schemes under absolute zero-mean Gaussian noise attacks at DWR = 25dB, WNR = −10dB and $N$ = 1000. Empirical results are obtained on 1,000,000 Weibull data with the designated scale and shape parameter. (a) $\theta$ = 0.05, $\delta$ = 1.5. (b) $\theta$ = 0.1, $\delta$ = 1.8.

We see from Fig. 7.1 and Fig. 7.2 that the double-sided schemes have a great performance advantage over the corresponding single-sided schemes, especially at large false alarm rates. It is also interesting to note that the suboptimal detector only enjoys a slight performance advantage over BMSS. This observation also holds for the corresponding double-sided schemes, that is, DS-MSS is only slightly better than DS-BMSS in performance. Now we see why the performance gain of MSS-OPT over BMSS is so small. Since at $\xi$ =

1.0, MSS-OPT and DS-MSS reduce to BMSS and DS-BMSS, respectively, the performance of MSS-OPT can be obtained by substituting $\xi$ with 1.0 in (6.24). Thus, the performance of BMSS can also be characterized by

$$p_m = 1 - Q[Q^{-1}(p_{fa}) - a\sqrt{N}\text{MVR}(1.0)]$$

Thus the performance of BMSS or DS-BMSS is determined by MVR(1.0), whereas MSS-OPT's or DS-BMSS's performance is decided by MVR($\delta$). Fig. 7.3 plots the different MVR values at different $\delta$s. The figure clearly indicates that MVR(1.0) does not differ much from MVR($\delta$) = $\delta$ and this difference increases as $\delta$ grows. Thus, we conclude that using BMSS and DS-BMSS instead does not result in a big performance loss.

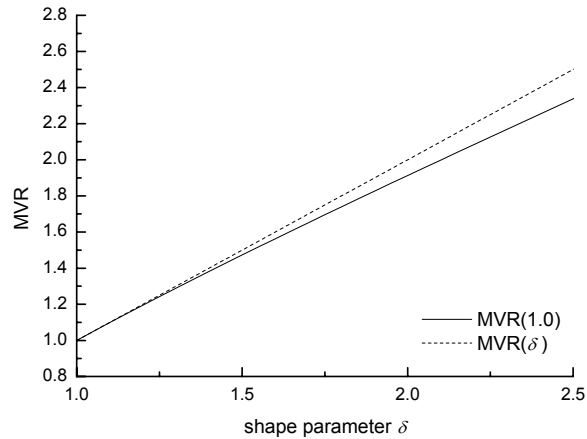

Fig. 7.3. MVR(1.0) and MVR($\delta$)

## 7.4 Monte-Carlo Simulations

In this section, we verify the theoretical results obtained in the previous sections through Monte-Carlo simulations. The comparison results are displayed in Fig. 7.4, Fig. 7.5 and Fig. 7.6. All these figures demonstrate the nice agreements between theoretical and empirical results. Moreover, the advantages of double-sided schemes over single-sided ones are well observed in Fig. 7.4 and Fig. 7.5. Please also note the sharp turns for double-sided schemes, above which the double-sided schemes can achieve a zero probability of miss. Another important observation is that BMSS and DS-BMSS produce a better performance at a larger shape parameter since MVR(1.0) increases as $\delta$ increases.

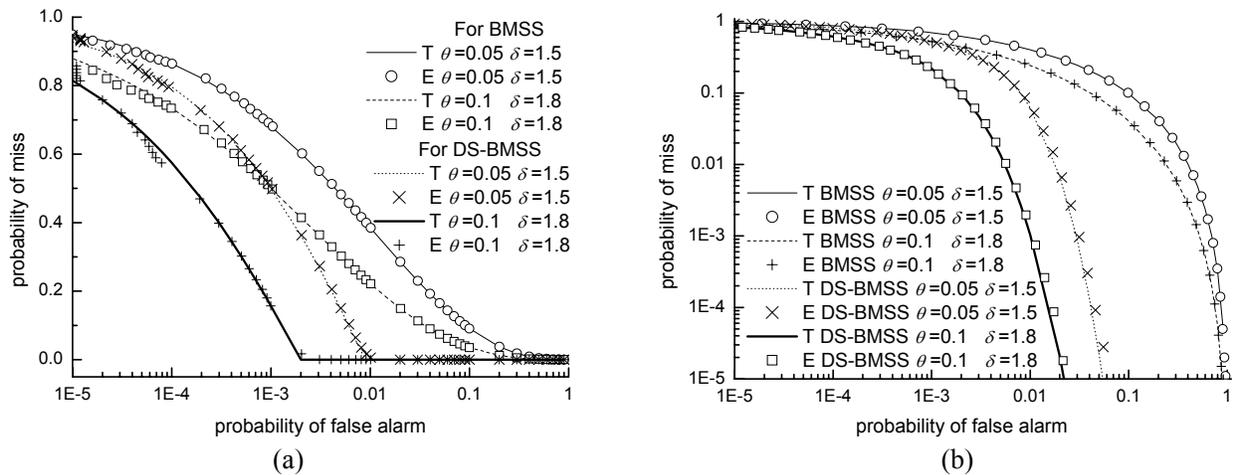

(a)                                          (b)

Fig. 7.4. Theoretical and empirical performances for BMSS and DS-BMSS at DWR = 25dB and N = 1000. Empirical results are obtained on 1,000,000 Weibull data with the designated scale and shape parameter. (a) Under no attack. (b) Under absolute zero-mean Gaussian noise at WNR = −10dB.

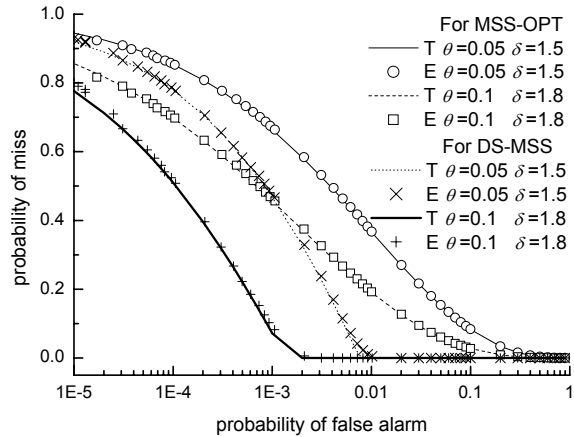

Fig. 7.5. Theoretical and empirical performances for MSS-OPT and DS-MSS at DWR = 25dB and N = 1000. Empirical results are obtained on 1,000,000 Weibull data with the designated scale and shape parameter.

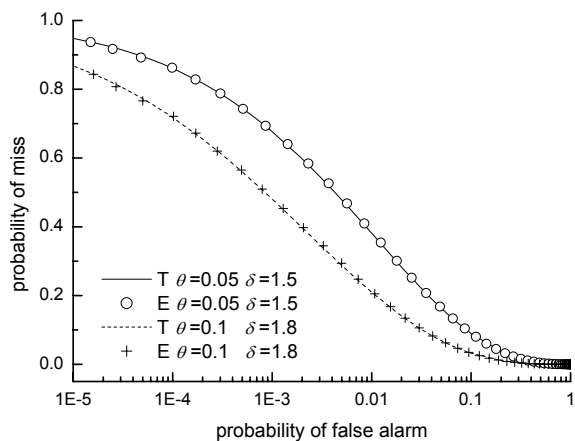

Fig. 7.6. Theoretical and empirical performances under absolute zero-mean Gaussian noise attacks for MSS-OPT at DWR = 25dB, WNR = −10dB and N = 1000. Empirical results are obtained on 1,000,000 Weibull data with the designated scale and shape parameter.

# Chapter 8   Double-Sided Schemes with Perceptual Analysis

## 8.1  Watson's perceptual model

In Chapter 6, we have outlined how to implement the perceptual analysis for the double-sided additive spread spectrum schemes. In this chapter, we detail its implementations in the DCT (8×8) domain and evaluate its performance by experiments.

Perceptual analyses have been widely employed in watermarking systems to improve the image fidelity at a maximum allowable embedding level [56−63, 117, 118]. Watson [55] originally proposed a perceptual model to design a custom quantization matrix tailored to a particular image. The model consists of frequency, luminance and contrast masking. In this section, we first review this model.

Let the $(i, j)$th DCT coefficient at block $k$ of the image be denoted by $x(i, j, k)$ with $0 \le i, j \le 7$. The masked threshold $m(i, j, k)$ estimates the amounts by which $x(i, j, k)$ may be changed before resulting in any perceptible distortions.

### 8.1.1  Frequency masking

Peterson *et al.* [109, 110, 111] have provided measurements of thresholds $m(i, j, k)$ for DCT basis functions. For each frequency $(i, j)$ they measured psychophysically the smallest coefficient that yielded a visible signal. By Peterson's method, Cox [3] computed the masks $m(i, j, k)$ and tabled them in a DCT frequency sensitivity table (which for reference is given as Table 8.1.)

Table 8.1. DCT frequency sensitivity table (See Table 7.2 in Chapter 7 of [3]).

| 1.4 | 1.01 | 1.16 | 1.66 | 2.4 | 3.43 | 4.79 | 6.56 |
|------|------|------|------|------|------|------|------|
| 1.01 | 1.45 | 1.32 | 1.52 | 2 | 2.71 | 3.67 | 4.93 |
| 1.16 | 1.32 | 2.24 | 2.59 | 2.98 | 3.64 | 4.6 | 5.88 |
| 1.66 | 1.52 | 2.59 | 3.77 | 4.55 | 5.3 | 6.28 | 7.6 |
| 2.4 | 2 | 2.98 | 4.55 | 6.15 | 7.46 | 8.71 | 10.17 |
| 3.43 | 2.71 | 3.64 | 5.3 | 7.46 | 9.62 | 11.58 | 13.51 |
| 4.79 | 3.67 | 4.6 | 6.28 | 8.71 | 11.58 | 14.5 | 17.29 |
| 6.56 | 4.93 | 5.88 | 7.6 | 10.17 | 13.51 | 17.29 | 21.15 |

### 8.1.2  Luminance masking

The frequency sensitivity measures the thresholds for DCT basis functions as a function of the mean luminance of the display. However, the local mean luminance in the image also has a great influence on the DCT thresholds. Watson called this luminance masking and formulated it as

$$m(i, j, k) = m(i, j, k) \cdot [x(0,0,k) / \overline{x}(0,0)]^{a_T} , \tag{8.1}$$

where $m(i, j, k)$ in the right side of (8.1) is the frequency masking discussed in the previous subsection, $x(0, 0, k)$ is the mean luminance (or DC coefficient) of the block $k$ of the original image, $\overline{x}(0, 0)$ is the mean luminance of the original image (or the mean of DC coefficients), and $a_T$ is a constant with a suggested value of 0.649.

### 8.1.3  Contrast masking

The threshold for a visual pattern is typically reduced in the presence of other patterns, particularly those of similar spatial frequency and orientation. Watson called this contrast masking and described the threshold for a coefficient as a function of its magnitude, that is,

$$m(i, j, k) = \max\{m(i, j, k), \quad |x(i, j, k)|^{w(i,j)} \cdot m(i, j, k)^{1-w(i,j)}\} , \tag{8.2}$$

where $m(i, j, k)$ in the right side of (8.2) is obtained in the previous subsection and $w(i, j)$ is an exponent that lies between 0 and 1 with a typical empirical value of 0.7. Since the threshold $m(i, j, k)$ roughly scales proportionally to the host coefficient $x(i, j, k)$, the multiplicative rule can thus automatically achieve a nicer perceptual quality. In fact, if the perceptual model is employed, an additive embedding rule is sort of a multiplicative rule.

## 8.2  Previous single-sided works

Podilchuk and Zeng [56] are the first to take the perceptual quality into account. The embedding rule for such a case becomes

$$s_i = x_i + am_i w_i , \tag{8.3}$$

where $m_i$ is the mask threshold determined in the previous section. The early methods detect the embedded watermark by a simple linear correlator. Later, Hernandez [69] designed an optimum detector by assuming that the host DCT coefficients follow a GGD distribution. The pdf of a GGD distribution is given by (2.7). Based on this assumption, an optimum decision statistic can be formulated as [69]

$$L(\mathbf{S}) = \frac{1}{N} \sum_{i=1}^{N} |S_i|^c - |S_i - am_i w_i|^c \, ,$$  (8.4)

which is in line with the optimum detector (3.57) in Chapter 3 where the perceptual analysis is not considered. In [72, 73], the author describes the host signals by a Cauchy distribution whose pdf is given by

$$f_X(x) = \frac{1}{\pi} \frac{\gamma}{\gamma^2 + (x - \delta)^2} \, ,$$  (8.5)

where $\gamma$ is the scale parameter and $\delta$ is the location parameter. For a symmetric pdf, $\delta = 0$. In this work, we focus primary on symmetrical data. Thus, the above pdf is simplified as

$$f_X(x) = \frac{1}{\pi} \frac{\gamma}{\gamma^2 + x^2} \, .$$  (8.6)

The scale parameter can be estimated by [112, 113] or by solving numerically the equation

$$\frac{N}{\gamma} = \sum_{i=1}^{N} \frac{2\gamma}{\gamma^2 + x_i^2} \, .$$  (8.7)

In [72], the authors give a non-linear Cauchy detector with a decision statistic

$$L(S) = \sum_{i=1}^{N} \frac{2(S_i - \delta) w_i}{\gamma^2 + (S_i - \delta)^2} \, .$$  (8.8)

Since $\delta = 0$ in this work, the above statistics can be rewritten as

$$L(S) = \frac{1}{N} \sum_{i=1}^{N} \frac{S_i w_i}{\gamma^2 + S_i^2} \, .$$  (8.9)

It is also important to note in [69, 72], only frequency and luminance masking are considered. Though they call their detectors optimal, they are never optimal since $m_i$ is taken as a fixed value in their derivation of the optimum detectors. In real scenarios, $m_i$ is image-dependent and not fixed. However, their detectors do achieve a performance improvement over linear correlators if only frequency and luminance masking are considered. This is because $m_i$ does not vary much and thus can be roughly taken as fixed. Nevertheless, if the contrasting masking is implemented, $m_i$ would be dependent on $x_i$ (See (8.2)) and their detector should never be optimal at all. Instead, we will see in the future experiments that their detectors yield a very poor performance in such a case.

## 8.3  Double-sided schemes

### 8.3.1  Double-sided Additive Spread Schemes

In Section 6.3.5, we have briefed how to implement the perceptual analysis in the double-sided additive spread spectrum schemes (DS-ASS). The embedding rule is given by

$$s_i = x_i + am_iw_i, \ \text{if} \ \bar{x} > 0; \ s_i = x_i - am_iw_i, \ \text{if} \ \bar{x} \le 0 \,, \tag{8.10}$$

where $\bar{x}$ is as defined in (6.3). We employ the same linear correlator with the double-sided rule for watermark detection.

### 8.3.2  Double-sided Cauchy schemes

It is difficult to propose a double-sided scheme for the optimum detector designed by Hernandez [69] since their detector requires the knowledge of embedding strength. However, it is possible to design a Double-Sided Cauchy (DS-Cauchy) scheme for Briassouli's scheme with an embedding rule

$$s_i = x_i + am_iw_i, \ \text{if} \ \bar{x} > 0; \ s_i = x_i - am_iw_i, \ \text{if} \ \bar{x} \le 0 \tag{8.11}$$

where

$$\bar{x} = L(x) = \frac{1}{N}\sum_{i=1}^{N}\frac{x_iw_i}{\gamma^2 + x_i^2} \,. \tag{8.12}$$

We apply the same decision statistic (8.9) with the double-sided rule (6.4) to the watermark detection. To see how this rule works, substituting (8.11) into (8.9), we obtain

$$L(\mathbf{s}\,|\,H_1) = \frac{1}{N}\sum_{i=1}^{N}\frac{(x_i \pm am_iw_i)w_i}{\gamma^2 + (x_i \pm am_iw_i)^2} = \frac{1}{N}\sum_{i=1}^{N}\frac{x_iw_i}{\gamma^2 + (x_i \pm am_iw_i)^2} \pm a\frac{1}{N}\sum_{i=1}^{N}\frac{m_i}{\gamma^2 + (x_i \pm am_iw_i)^2} \,. \tag{8.13}$$

Since $a \cdot m_i$ is usually much smaller than $x_i$, thus

$$L(\mathbf{s} \mid H_1) \approx \frac{1}{N} \sum_{i=1}^{N} \frac{x_i w_i}{\gamma^2 + x_i^2} \pm \frac{1}{N} \sum_{i=1}^{N} \frac{a m_i}{\gamma^2 + (x_i \pm a m_i w_i)^2} = \overline{x} \pm \frac{1}{N} \sum_{i=1}^{N} \frac{a m_i}{\gamma^2 + (x_i \pm a m_i w_i)^2} \qquad (8.14)$$

Thus, we see from (8.11) and (8.14) that the second term of (8.14) is of the same sign with $\overline{x}$, which is the essence of double-sided schemes.

## 8.4  Discussions

### 8.4.1  Performance in the sense of Mean Squared Errors (MSE) metric

In real scenarios, $m_i$ depends on the host signal $x_i$. However, to simplify the discussion, we assume that $m_i$ is fixed in this subsection. Thus, the performance of DS-ASS can be described as

$$p_m = \begin{cases} 1 - 2Q[Q^{-1}(p_{fa}/2) - k\sqrt{N}/\sigma_X], & \text{if } \psi > k \\ 0.0, & \text{if } \psi \leq k \end{cases},$$

where

$$k = \frac{1}{N} \sum_{i=1}^{N} a \cdot m_i$$

Please refer to Section 6.6.3 for the detailed derivation of the above equation. In the sense of MSE metric, we may assume that the embedding distortion $D_w$ is

$$D_w = \frac{1}{N} \sum_{i=1}^{N} a^2 m_i^2 .$$

By Cauchy's inequality, we have

$$k^2 = \left( \frac{1}{N} \sum_{i=1}^{N} a m_i \right)^2 \leq \frac{1}{N} \sum_{i=1}^{N} a^2 m_i^2 = D_w ,$$

where the equality holds at $m_1 = m_2 = \ldots = m_N$. That is, the best performance of DS-ASS is achieved at $m_1 = m_2 = \ldots = m_N$. Hence, the perceptual shaping reduces the performance of DS-ASS at the same distortion level (in the sense of MSE metric). It also implies that performance analyses without taking perceptual quality into account may be quite misleading. Of course, the above rationale holds similarly for ASS schemes.

### 8.4.2  Disadvantages of host interference rejection schemes

The host interference rejection schemes have been adapted to watermark detection problems and proven to be superior in performance. However, they haven't taken the perceptual analysis into account. Indeed, it is difficult to implement the perceptual analysis for them. In this subsection, we conjecture a possible way to implement the perceptual analysis for STDM and then discuss its disadvantages.

Suppose that the maximum allowable embedding strength for the host data is $a_{max}$. The perceptual analysis for our schemes can be simply implemented as $s_i = x_i \pm a_{max} m_i w_i$ to achieve the maximum possible embedding strength. With the help of the spread transform technique, we can also obtain the embedding rule for STDM with perceptual analysis. Projecting both sides of (8.3) on $\mathbf{w}$, we have

$$\overline{s} = \overline{x} + a \frac{1}{N} \sum_{i=1}^{N} m_i \tag{8.15}$$

where both $\overline{x}$ and $\overline{s}$ are defined as (6.3). Let $\overline{s} = q_\Lambda(\overline{x})$ and we immediately have

$$a = \frac{q_\Lambda(\overline{x}) - \overline{x}}{\left( \sum_{i=1}^{N} m_i \right) \big/ N} . \tag{8.16}$$

Substituting (8.16) into (8.3) leads to an embedding rule for STDM scheme with perceptual analysis. A recent paper [119] proposed to implement the perceptual analysis in STDM scheme for watermark decoding problems. The underlying idea is much similar to our conjecture. Now, we examine the disadvantages of the above embedding rule. In the above equation, $a$ should be smaller than $a_{max}$ to keep the distortion imperceptible. However, if it is smaller than $a_{max}$, it leaves a larger perceptual allowance for the attacker. One may argue that we can select an appropriate step size to make the embedding strength $a = a_{max}$. However, it is not workable since in such a case the detector has no chance of knowing what the step size is used. In fact, the major inherent problem with the host interference rejection schemes is that it is hard to control the embedding strength since it depends on $\overline{x}$ over which we have no control.

## 8.5  Experimental results

In this section, we make performance comparisons between the above single-sided and double-sided schemes. The host data come from the fifth coefficient (in Zigzag order) of Lena, Peppers, Boat and Baboon. As also explained in the previous chapters, we permute the host signals to obtain different **x** for tests. Each such permutation generates one **x**. In the following experiments, the empirical results are obtained over 1,000,000 such permutations. Finally, in the legends of the figures, "Briassouli" stands for the scheme [72] or (8.9), "Hernandez" for the scheme [69] or (8.4), "ASS-COR" for the widely used linear correlator, "DS-ASS" for the scheme (8.10) and "DS-Cauchy" for the scheme (8.11).

### 8.5.1  Perceptual analysis with frequency and luminance masking

As done in [69, 72], only frequency and luminance masking are considered in this subsection. In the first experiment, the $\gamma$ and $c$ at the detector are the same with those of the original images. The experimental results with this setup are displayed in Fig. 8.1 to Fig. 8.3. In the second experiment (only for Lena and Peppers), the $\gamma$ and $c$ at the detector are instead estimated from the test images. Table 8.2 and Table 8.3 compare the parameters estimated from the original images and watermarked images under both no attack and attacks. These results are obtained over an average of the estimated parameters on 1,000 watermarked images (with 1,000 different watermark sequences). In real scenarios, each run of test should be fed with a different parameter estimated from the test data; however, it is time-consuming to perform the parameter estimations. Thus, we instead use the average of the estimated parameters (shown in Table 8.2 and Table 8.3) to save the simulation time. Moreover, even under attacks, the masks estimated from the test image are almost the same with those estimated from the original image. Thus, for Hernandez's detector, we use $m_i$s of the original image for convenience. The experimental results obtained on the parameters displayed in Table 8.2 and Table 8.3 are listed in Fig. 8.4, Fig. 8.5 and Fig. 8.6.

From Fig. 8.1 to Fig. 8.6, we conclude that

First, Hernandez's and Briassouli's schemes enjoy a better performance than ASS-COR. This is because $m_i$ does not vary much and can be roughly taken as fixed. It is also quite interesting to point out that the theoretical analysis in their works [69, 72] are unreasonable since the optimum tests are derived on the assumption that $w_i$ is not a random variable whereas their analyses take $w_i$ as random. The correct analysis should take $x_i$ as random and $w_i$ as fixed. Please refer to Appendix E for further information.

Second, it is hard to conclude that a Cauchy non-linear detector produces a better performance than a GGD optimum detector. In [72], its detector's performance is compared with a GGD model with the shape parameter $c$ fixed at $c = 0.5$ (Note: Cheng [77] believes that $c = 0.5$ is very typical of the shape parameters of most images.) However, if $c$ is not fixed and instead estimated from the host image, then Cauchy non-linear detectors do not necessary yield a better performance. This is also reflected in another paper by Briassouli [73].

Third, the double-sided detection can be implemented for Briassouli's scheme since its detector does not require the knowledge of the embedding strength and the perceptual masks. In these figures, we observe that DS-Cauchy does achieve a better performance than its single-sided counterpart.

Fourth, in accordance with conclusions drawn in the previous chapter, DS-ASS improves ASS-COR.

Finally, it is really hard to decide whether DS-ASS is better than DS-Cauchy. DS-Cauchy can achieve a better false alarm since it is based on a suboptimal detector. Nevertheless, as explained in the previous chapter, DS-ASS is much simpler and entails no parameter estimation.

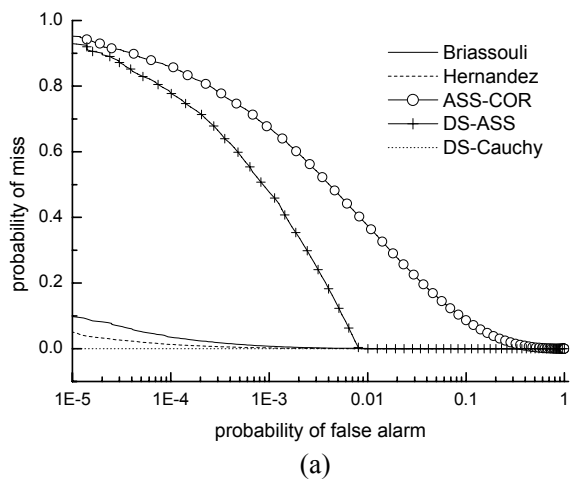 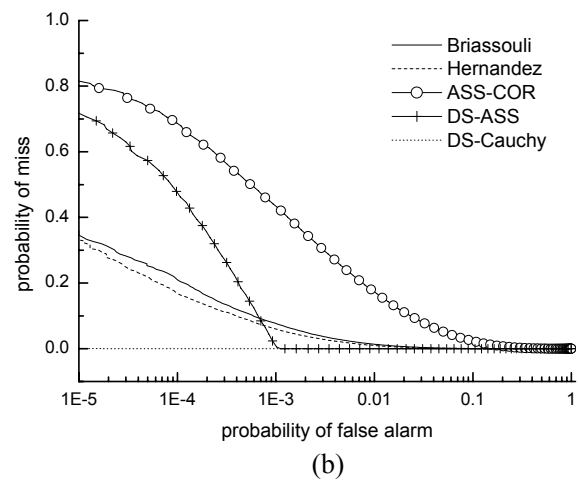

(a)                                        (b)

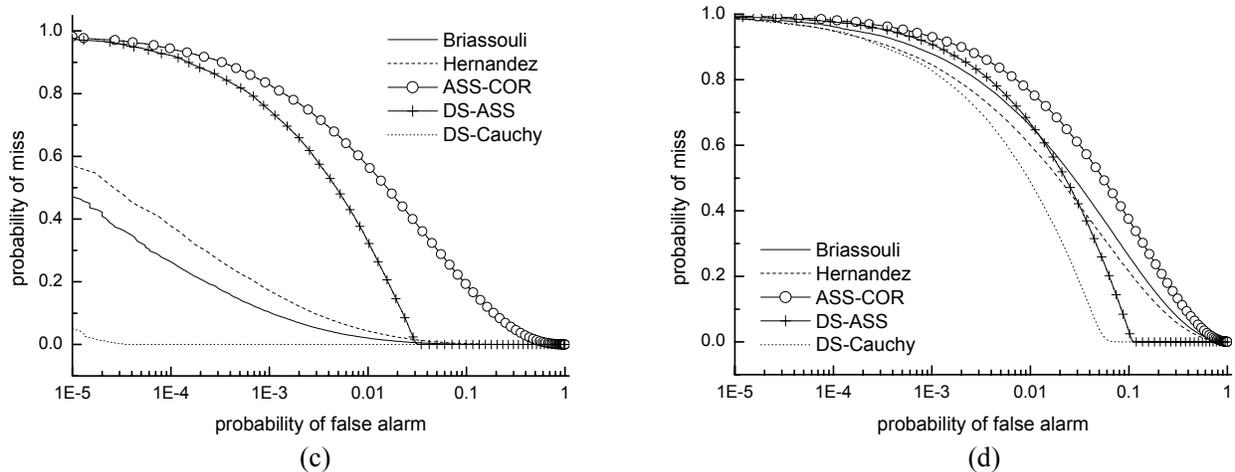

(c)                                                          (d)

Fig. 8.1. Performance comparisons under no attack (with $a = 1.0$, $N = 2000$). (a) Lena. At embedder, $\gamma = 6.69$; at detector, $\gamma = 6.69$ and $c = 0.69$. (b) Peppers. At embedder, $\gamma = 7.35$; at detector, $\gamma = 7.35$ and $c = 1.03$. (c) Boat. At embedder, $\gamma = 8.65$; at detector, $\gamma = 8.65$ and $c = 0.73$. (d) Baboon. At embedder, $\gamma = 15.59$; at detector, $\gamma = 15.59$ and $c = 1.11$.

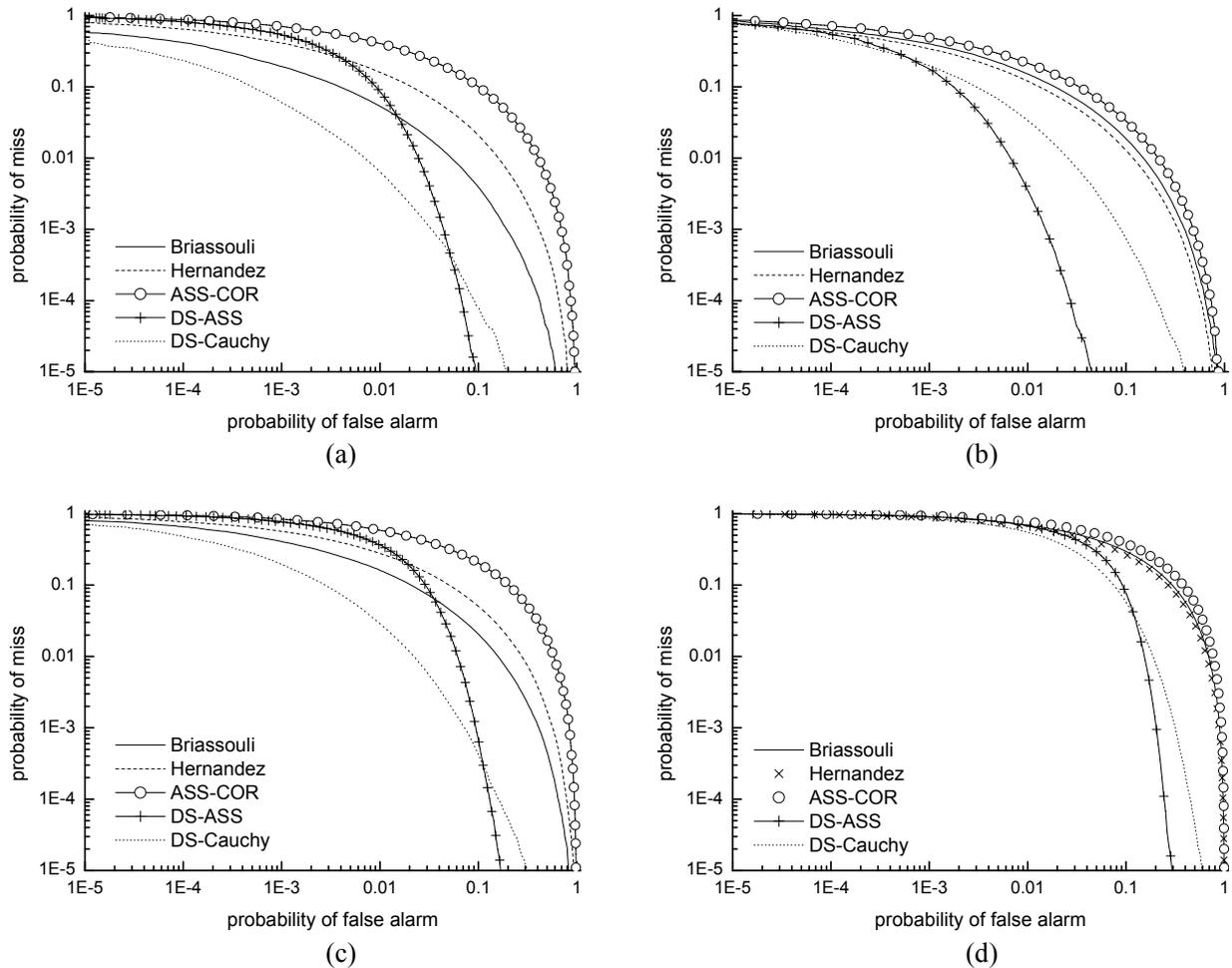

Fig. 8.2. Performance comparisons under zero-mean Gaussian noise attacks N(0, 25) (with $a = 1.0$, $N = 2000$). (a) Lena. At embedder, $\gamma = 6.69$; at detector, $\gamma = 6.69$ and $c = 0.69$. (b) Peppers. At embedder, $\gamma = 7.35$; at detector,

$\gamma = 7.35$ and $c = 1.03$. (c) Boat. At embedder, $\gamma = 8.65$; at detector, $\gamma = 8.65$ and $c = 0.73$. (d) Baboon. At embedder, $\gamma = 15.59$; at detector, $\gamma = 15.59$ and $c = 1.11$.

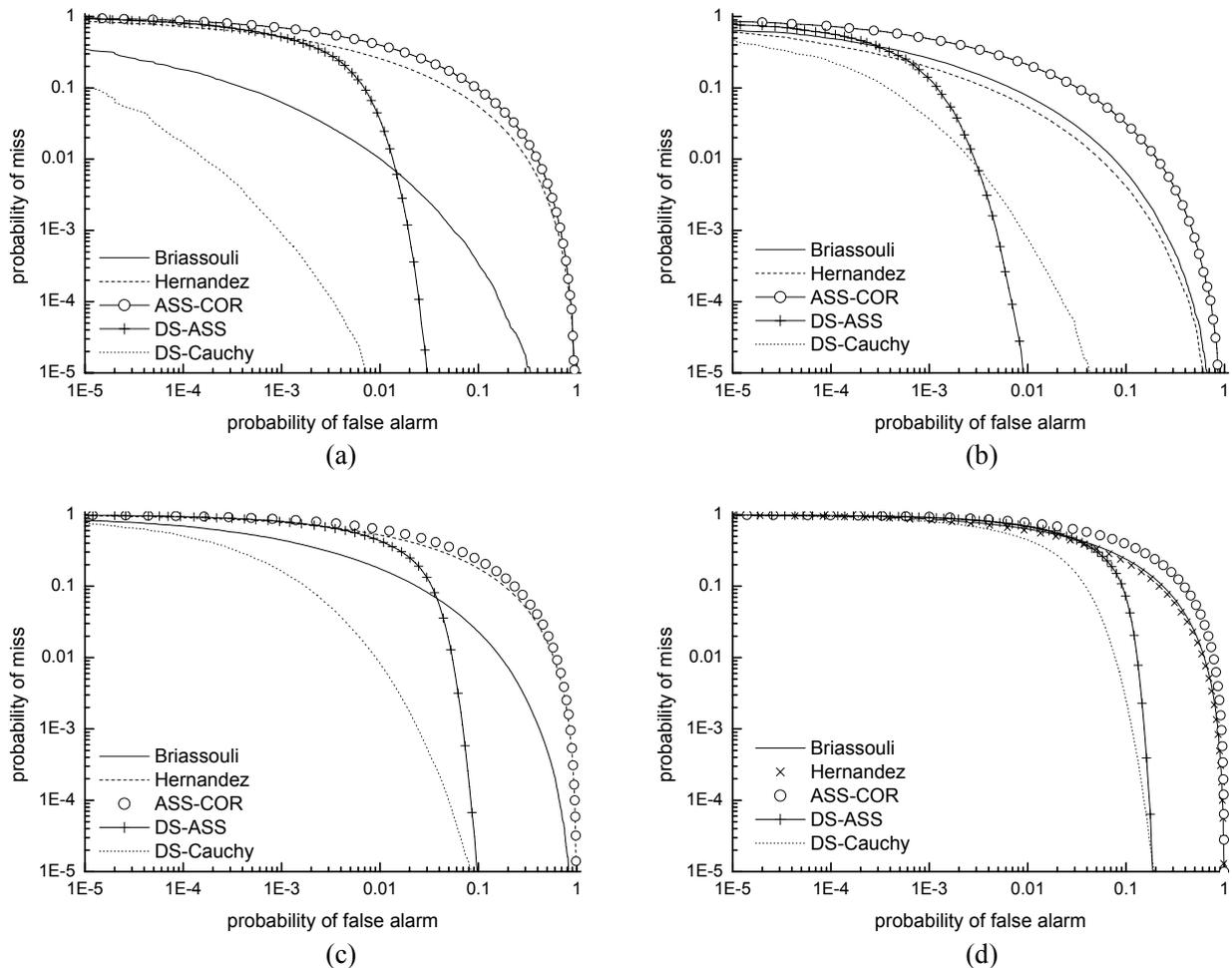

(a)                                                    (b)

(c)                                                    (d)

Fig. 8.3. Performance comparisons under JPEG (QF = 50) attacks (with $a = 1.0$, $N = 2000$). (a) Lena. At embedder, $\gamma = 6.69$; at detector, $\gamma = 6.69$ and $c = 0.69$. (b) Peppers. At embedder, $\gamma = 7.35$; at detector, $\gamma = 7.35$ and $c = 1.03$. (c) Boat. At embedder, $\gamma = 8.65$; at detector, $\gamma = 8.65$ and $c = 0.73$. (d) Baboon. At embedder, $\gamma = 15.59$; at detector, $\gamma = 15.59$ and $c = 1.11$.

Table 8.2. Estimated scale parameter $\gamma$

| Image | Scheme | Hypothesis | Unattacked | JPEG QF=50 | Noise N(0, 25) |
|-------|--------|------------|------------|------------|----------------|
| Lena | Briassouli | unwatermarked($H_0$) | 6.69 | 5.9 | 8.04734 |
| Lena | Briassouli | watermarked($H_1$) | 6.79513 | 6.00041 | 8.11382 |
| Lena | DS-Cauchy | unwatermarked($H_0$) | 6.69 | 5.9 | 8.04734 |
| Lena | DS-Cauchy | watermarked($H_1$) | 6.79943 | 6.00866 | 8.1178 |
| Peppers | Briassouli | unwatermarked($H_0$) | 7.35 | 6.86 | 8.47844 |
| Peppers | Briassouli | watermarked($H_1$) | 7.43631 | 7.03595 | 8.52656 |
| Peppers | DS-Cauchy | unwatermarked($H_0$) | 7.35 | 6.86 | 8.47844 |
| Peppers | DS-Cauchy | watermarked($H_1$) | 7.44517 | 7.05036 | 8.53412 |

Table 8.3. Estimated shape parameter $c$

| Image | Scheme | Hypothesis | Unattacked | JPEG QF=50 | Noise N(0, 25) |
|-------|--------|------------|------------|------------|----------------|
| Lena | Hernandez | unwatermarked(H0) | 0.69 | 0.79 | 0.84393 |
| Lena | Hernandez | watermarked(H1) | 0.71259 | 0.7897 | 0.85095 |
| Peppers | Hernandez | unwatermarked(H0) | 1.03 | 1.11 | 1.22832 |
| Peppers | Hernandez | watermarked(H1) | 1.04191 | 1.09775 | 1.23613 |

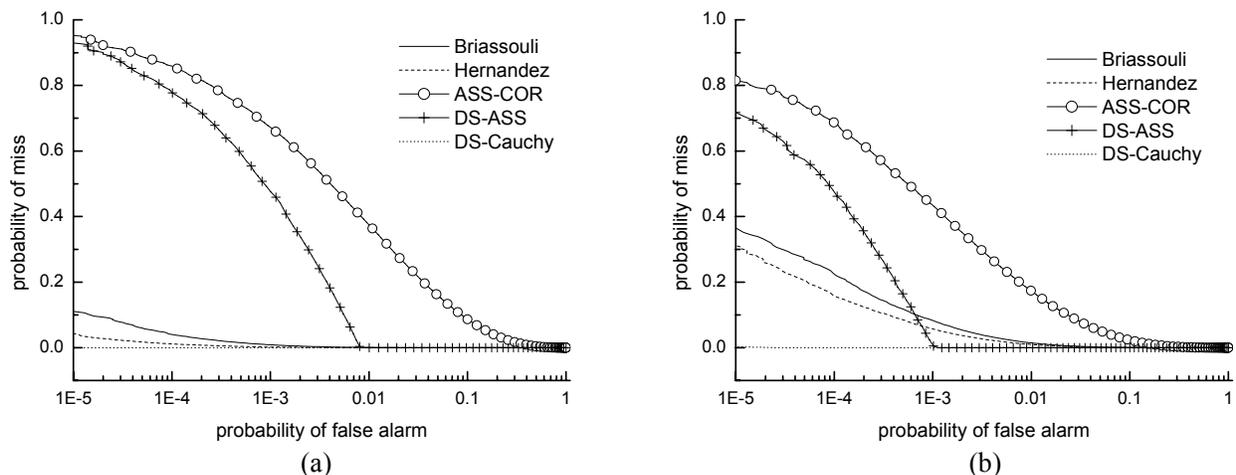

(a)                                                    (b)

Fig. 8.4. Performance comparisons under no attack (with $a = 1.0$, $N = 2000$). The parameters used at the detector are specified in Table 8−2 and Table 8−3. (a) Lena. At embedder, $\gamma = 6.69$. (b) Peppers. At embedder, $\gamma = 7.35$.

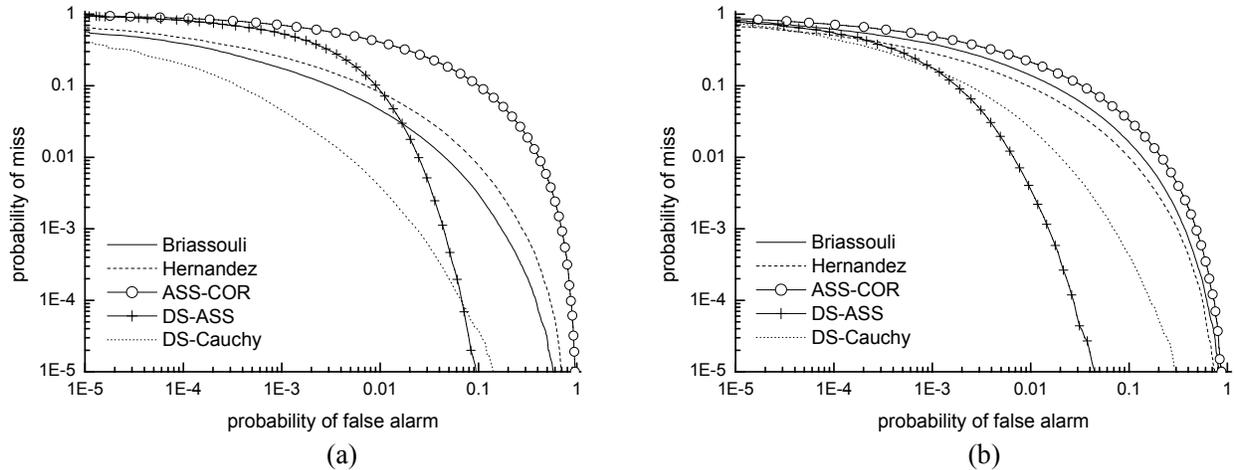

Fig. 8.5. Performance comparisons under zero-mean Gaussian noise attacks N(0,25) (with $a = 1.0$, $N = 2000$). The parameters used at the detector are specified in Table 8−2 and Table 8−3. (a) Lena. At embedder, $\gamma = 6.69$. (b) Peppers. At embedder, $\gamma = 7.35$.

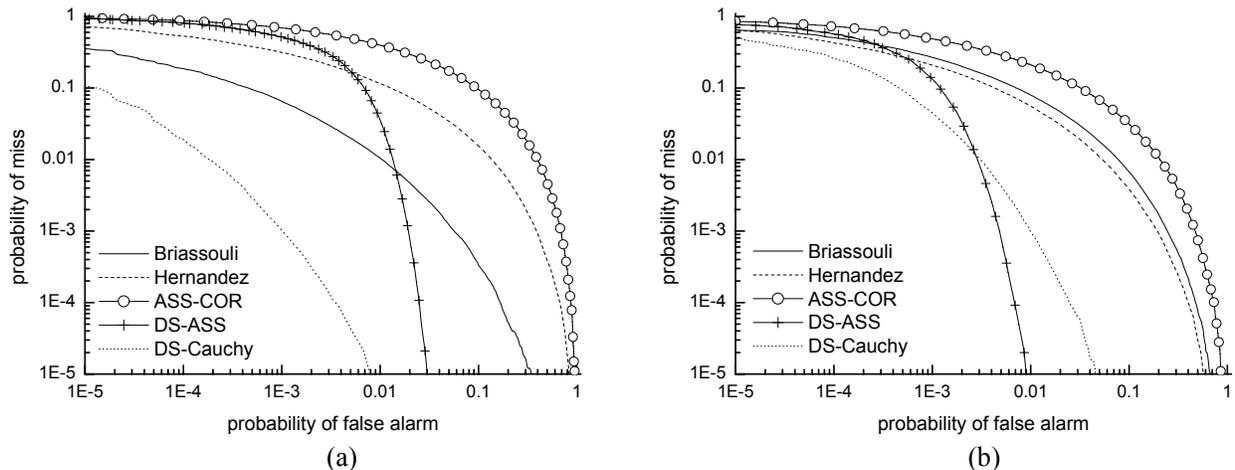

Fig. 8.6. Performance comparisons under JPEG (QF = 50) attacks (with $a = 1.0$, $N = 2000$). The parameters used at the detector are specified in Table 8−2 and Table 8−3. (a) Lena. At embedder, $\gamma = 6.69$. (b) Peppers. At embedder, $\gamma = 7.35$.

From Table 8.2 and Table 8.3, we see that the parameters estimated from the test images do not differ much from those of the original images. This is the reason why in almost all experiments, we just use the parameters of the host images for the ease of performance comparisons. In fact, using the parameters of the host images does not make much influence on the detector's performance. This we show in Fig. 8.7 to Fig. 8.9. In these figures, E1 denotes the scenario where the detector uses the parameters of the original images and E2 the scenario where the detector uses the parameters estimated from the test images (shown in Table 8.2 and Table 8.3).

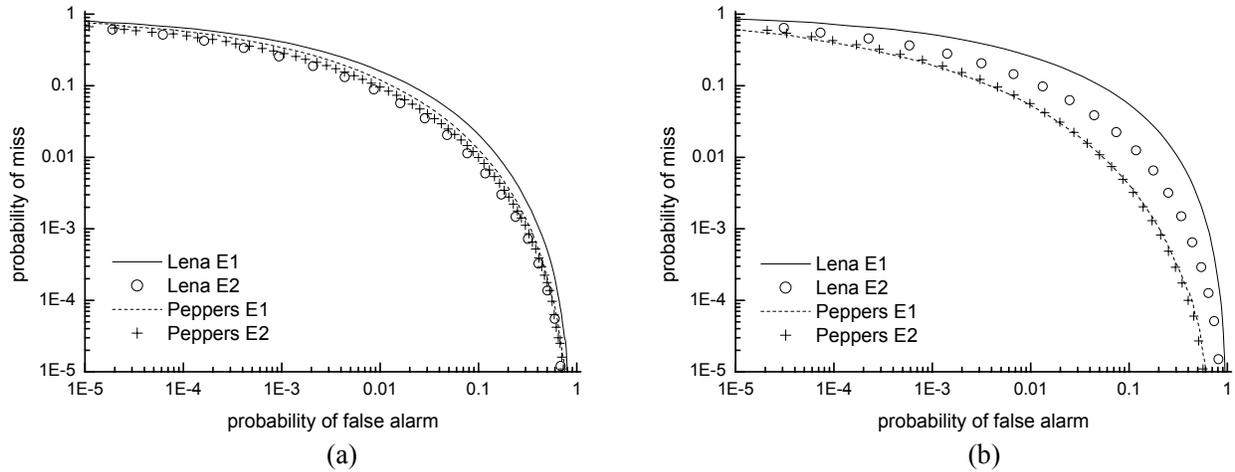

(a)                                                    (b)

Fig. 8.7. Performance comparisons between E1 and E2 for Hernandez's scheme. (a) Under zero-mean Gaussian noise attacks N(0, 25). (b) Under JPEG (QF = 50) attacks.

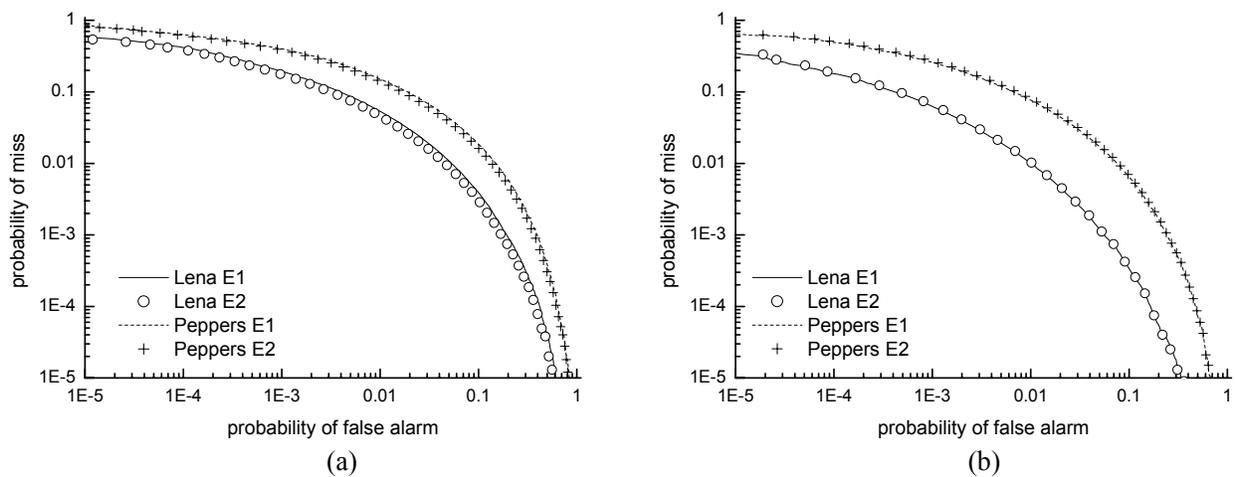

(a)                                                    (b)

Fig. 8.8. Performance comparisons between E1 and E2 for Briassouli's scheme. (a) Under zero-mean Gaussian noise attacks N(0, 25). (b) Under JPEG (QF = 50) attacks.

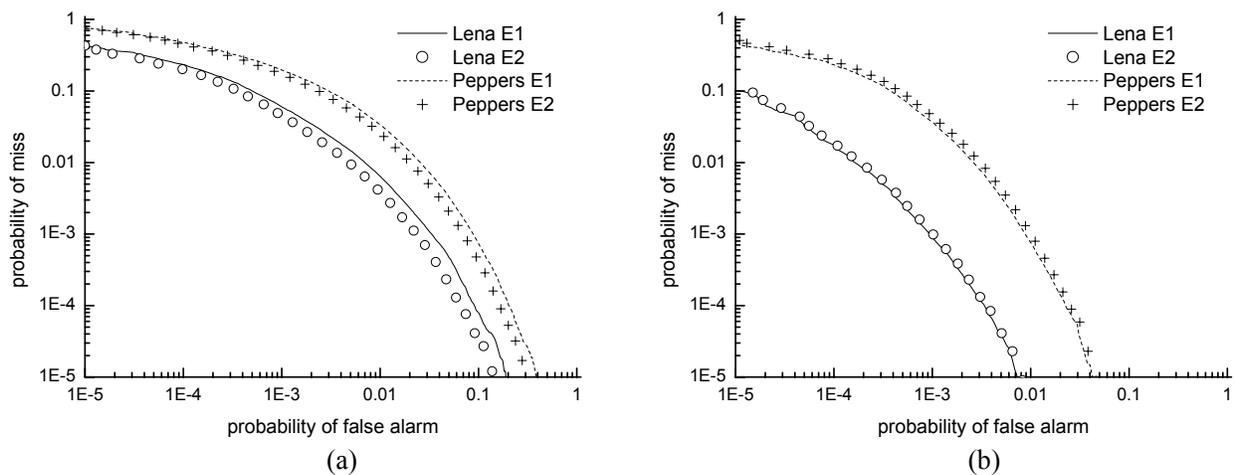

(a)                                                    (b)

Fig. 8.9. Performance comparisons between E1 and E2 for DS-Cauchy. (a) Under zero-mean Gaussian noise attacks N(0, 25). (b) Under JPEG (QF = 50) attacks.

## 8.6  Perceptual analysis with frequency, luminance and contrast masking

In this subsection, the contrast masking is also included in the perceptual analysis. For the ease of performance evaluations, the scale and shape parameters adopted at the detector are the same with those estimated from the original images. Moreover, the perceptual masks $m_i$ and the embedding strength $a$ are also presumed to be known at the Hernandez's detector. In real scenarios, $m_i$ has to be computed independently from the test images. The comparison results are demonstrated in Fig. 8.10 to Fig. 8.12.

Hernandez's scheme is very poor in performance since the perceptual masks $m_i$ (with contrast masking) depends on host signals $x_i$ and thus the proposed detector is not optimal at all. Briassouli's scheme achieves a slight better performance, but is also not optimal. Their performance is even much worse than that of the correlator. This thus gives rise to an interesting problem whether we should implement contrast masking in real scenarios. However, if we do not implement contrast masking at the embedder, it will leave a larger allowance for the attackers to mount stronger attacks on the watermarked contents.

Second, as expected, DS-Cauchy still achieves a better performance than Briassouli's scheme.

Third, most important of all, DS-ASS yields a much better performance than ASS-COR does. This thus encourages the use of DS-ASS in real scenarios.

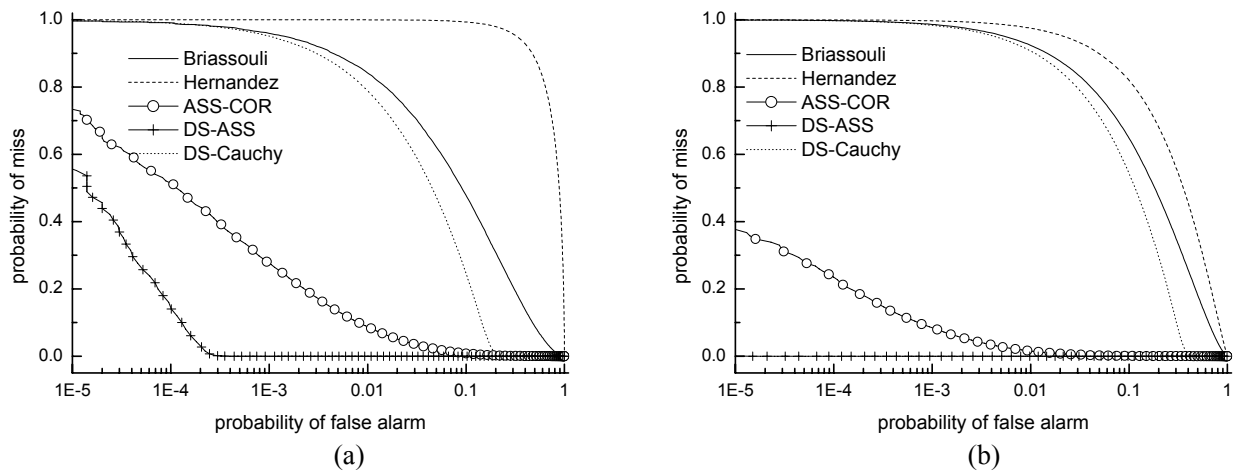

(a)                                    (b)

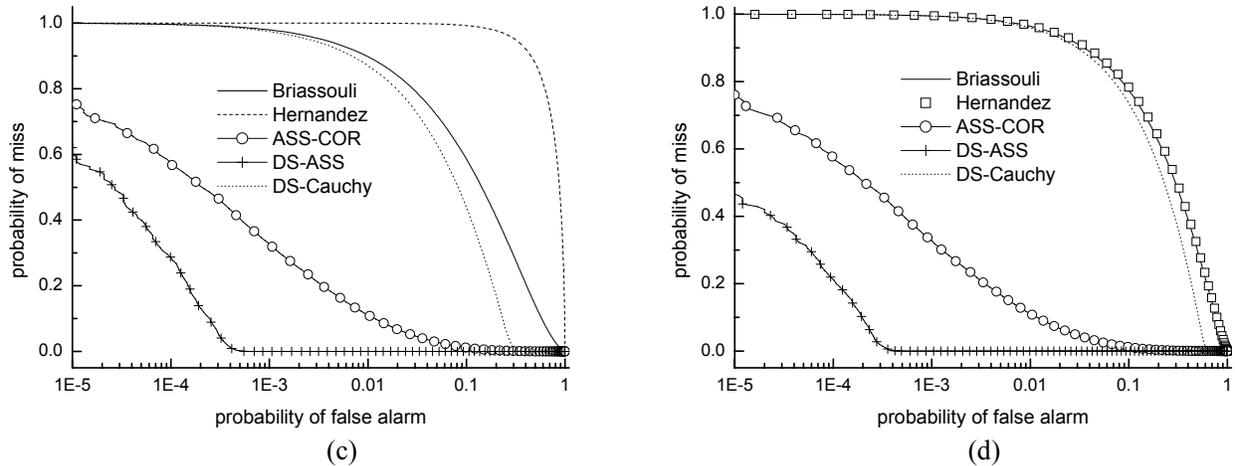

(c)                                                           (d)

Fig. 8.10. Performance comparisons under no attack (with $a = 0.3$, $N = 2000$). (a) Lena. At embedder, $\gamma = 6.69$; at detector, $\gamma = 6.69$ and $c = 0.69$. (b) Peppers. At embedder, $\gamma = 7.35$; at detector, $\gamma = 7.35$ and $c = 1.03$. (c) Boat. At embedder, $\gamma = 8.65$; at detector, $\gamma = 8.65$ and $c = 0.73$. (d) Baboon. At embedder, $\gamma = 15.59$; at detector, $\gamma = 15.59$ and $c = 1.11$.

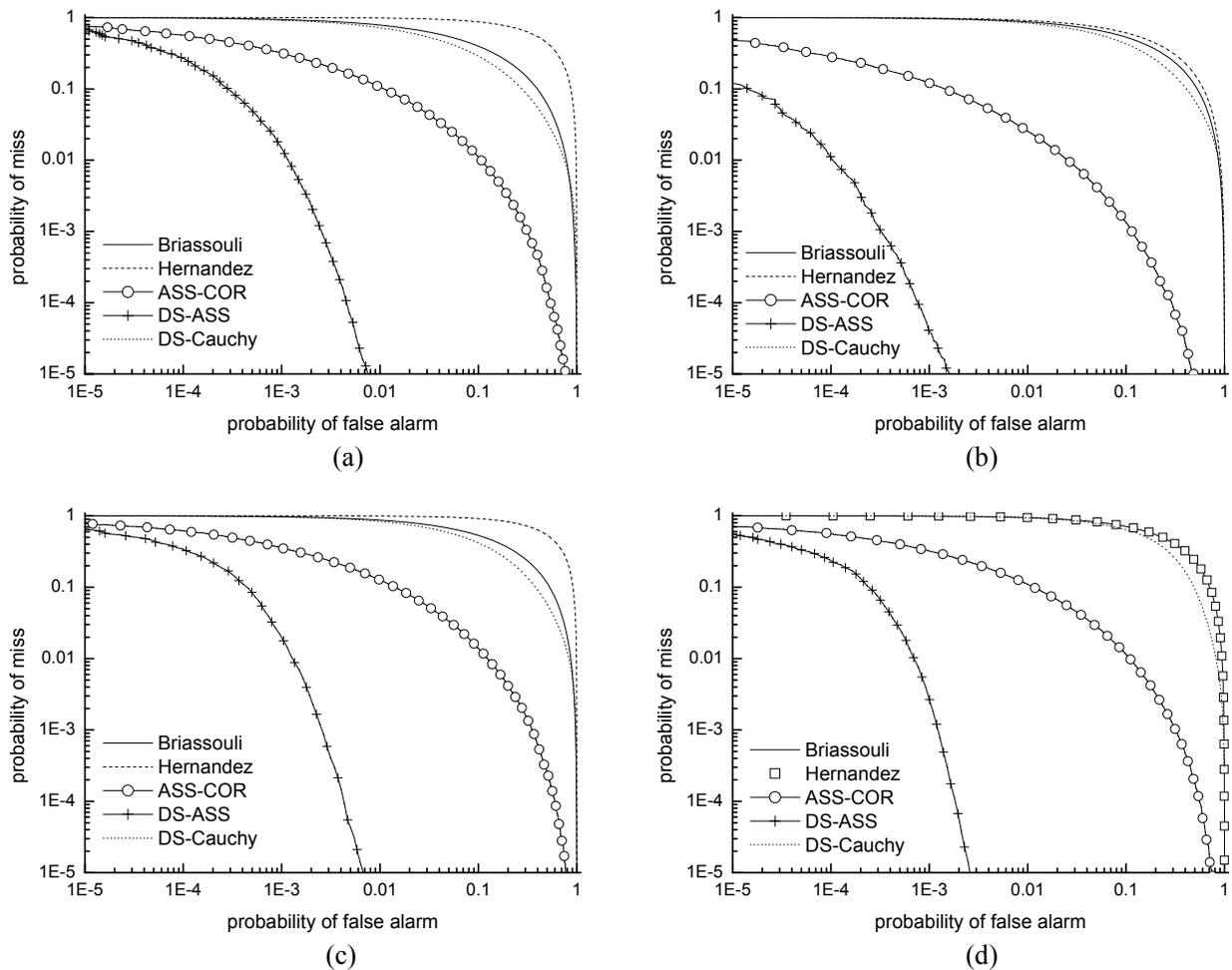

(a)                                                           (b)

(c)                                                           (d)

Fig. 8.11. Performance comparisons under zero-mean Gaussian noise attacks N(0, 25) (with $a = 0.3$, $N = 2000$). (a) Lena. At embedder, $\gamma = 6.69$; at detector, $\gamma = 6.69$ and $c = 0.69$. (b) Peppers. At embedder, $\gamma = 7.35$; at detector,

$\gamma = 7.35$ and $c = 1.03$. (c) Boat. At embedder, $\gamma = 8.65$; at detector, $\gamma = 8.65$ and $c = 0.73$. (d) Baboon. At embedder, $\gamma = 15.59$; at detector, $\gamma = 15.59$ and $c = 1.11$.

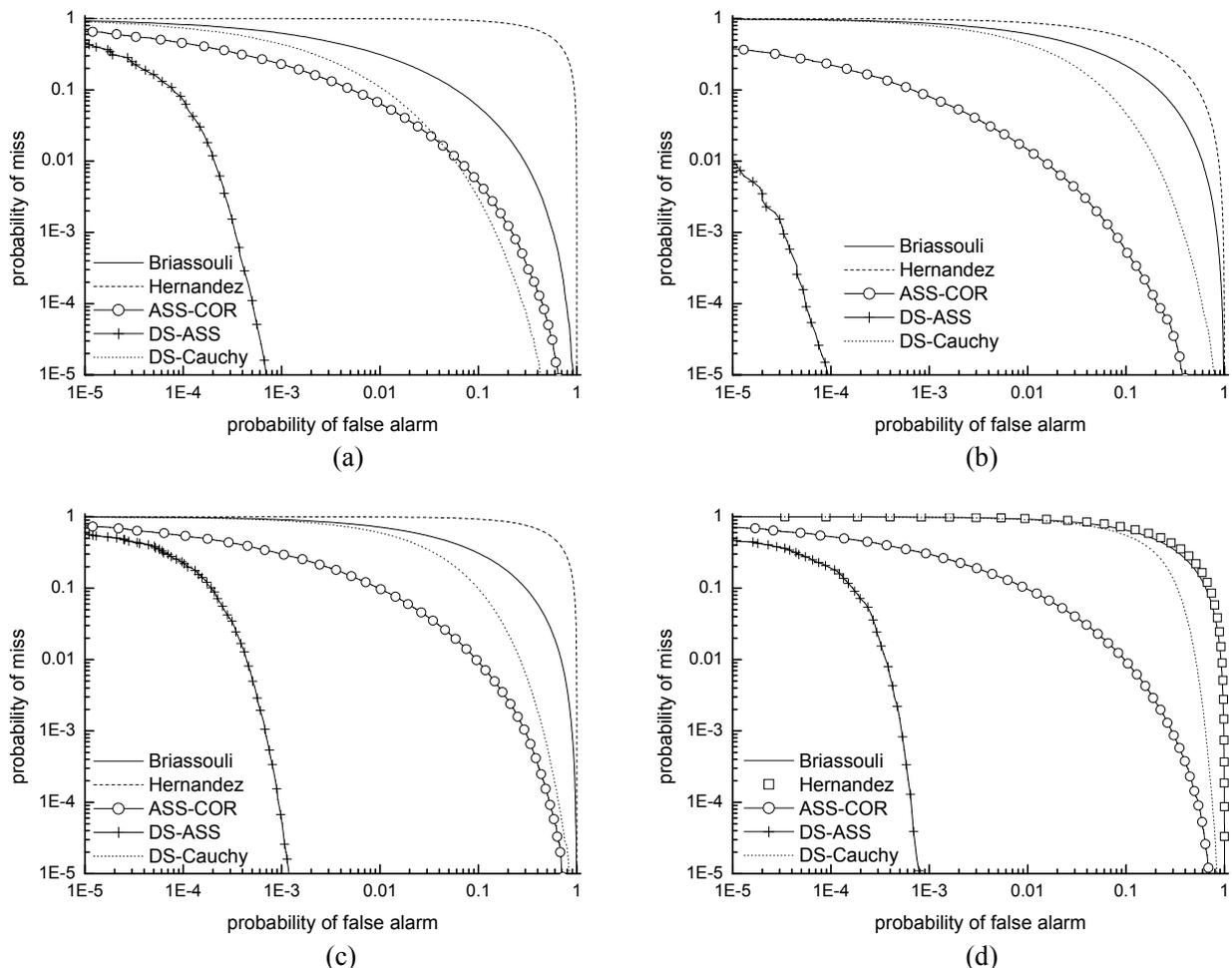

Fig. 8.12. Performance comparisons under JPEG (QF = 50) attacks (with $a = 0.3$, $N = 2000$). (a) Lena. At embedder, $\gamma = 6.69$; at detector, $\gamma = 6.69$ and $c = 0.69$. (b) Peppers. At embedder, $\gamma = 7.35$; at detector, $\gamma = 7.35$ and $c = 1.03$. (c) Boat. At embedder, $\gamma = 8.65$; at detector, $\gamma = 8.65$ and $c = 0.73$. (d) Baboon. At embedder, $\gamma = 15.59$; at detector, $\gamma = 15.59$ and $c = 1.11$.

## 8.7  Conclusions

In this chapter, we further demonstrated the advantage of double-sided schemes over its single-sided counterparts in a scenario where the Watson's perceptual model was implemented to improve the fidelity of the watermarked contents. However, through performance comparisons, we found that DS-ASS is still the most appealing scheme. Even if the contrast masking is not implemented, DS-ASS is not inferior to DS-Cauchy in performance. However, if the contrast masking is implemented, it offers a dramatic performance advantage over DS-Cauchy. Moreover, DS-ASS is widely applicable to almost all kinds of host data whatever their probability of distributions. Thus, it is also applicable to audio and video watermarking.

# Chapter 9   Conclusions and Future Work

## 9.1  Conclusions

The embedder and the detector (or decoder) are the two most important components of the digital watermarking systems. Thus in this work, we discuss how to design a better embedder and detector (or decoder). We first explore the optimum detector or decoder according to a particular probability distribution of the host signals. The optimum detection is not new since it has already been widely investigated in the literature. However, our work offers new insights into its theoretical performance. First, we examined its theoretical performance analytically and experimentally. The theoretical analyses presented in the literature are not correct since their analyses (with the watermark sequence as a random vector) are not in accordance with the prerequisite that the watermark sequence is fixed in their deriving the likelihood ratio tests. Second, we found that for MSS in both DCT and DWT domain, its performance depends on the shape parameter $c$ or $\delta$ of the host signals. Third, without perceptual analysis, ASS also has a performance dependent on the host signals and outperforms MSS at the shape parameter below 1.3.

For spread spectrum schemes, the detector or the decoder's performance is reduced by the host interference. Thus, we came up with a new idea of host-interference rejection idea for MSS scheme. In this work, we call this new host interference rejection scheme EMSS whose embedding rule is tailored to the optimum decision rule. We particularly examined its performance in the DCT domain and found that they produce a nicer performance than the traditional non-rejecting schemes.

Though the host interference schemes enjoy a big performance gain over the traditional spread spectrum schemes, their drawbacks that it is difficult for them to be implemented with the perceptual analysis to improve the fidelity of the watermarked contents discourage its use in real scenarios. Thus, in the final several chapters of this thesis, we introduced a new double-sided idea to combat this difficulty. This idea differs from the host interference scheme in that it does not reject the host interference at the embedder. However, it also utilizes the side host information at the embedder. Though it does not reject the host interference, it can also achieve a great performance enhancement over the tradition spread spectrum schemes. Moreover, it also has a big advantage over the host interference rejection schemes in that we can embed the watermark with a maximum allowable level of perceptual distortion.

**In short, this work contributes in**

1. **Theoretical performance analyses for optimal detectors in the DCT and the DFT domains, and finding that the performance of MSS depends on the shape parameters of the host signals,**

2. **Proposing a host interference rejection scheme whose embedding rule is tailored to the corresponding optimal detection or decoding rules,**

3. **Most important of all, presenting a new watermarking model — double-sided watermark embedding and detection.**

## 9.2  Future work

For double-sided schemes, we have only investigated their performances under noise and JPEG attacks. In real scenarios, the watermarked contents may also suffer from geometrical attacks. The geometrical attacks coupled with, for instance, JPEG attacks, comprise the most severe attacks for the watermarked data. Almost all schemes proposed up till now are not robust enough against these attacks. This constitutes one of the major reasons why the digital watermarking is still in its infancy. **Thus our future work would be directed to investigate the robustness of our double-sided schemes against geometrical attacks.** My very tentative idea is to embed a template to rectify the possible geometrical distortions before watermark detection.

Second, I have mainly examined the spread spectrum schemes in both the DCT and the DFT domains. An interesting problem is that in both domains, we found that the performance of MSS (DS-MSS) depends on the shape parameter $c$ ($\delta$) of the host signals. In the DCT domain, the performance is decided by

$$\mathrm{MVR}(c) = \sqrt{c} \ . \tag{9.1}$$

In the DFT domain,

$$\text{MVR}(\delta) = \delta \,. \tag{9.2}$$

**Thus in which domain should we embed our data?** For the 8×8 DCT coefficients, we found that the average $c$ over a set of 93 standard images is about 1.0. However, if the DCT coefficients are obtained by transforming the whole image, then the average $c$ would be about 1.6, with some even quite close to or larger than 2.0. On this set of images, we found that the average $\delta$ is about 1.7. Thus from both (9.1) and (9.2), we found that a better performance can be achieved in the DFT domain. Therefore, it may be an interesting future task to compare their performances in the real images. These arguments are also applicable to the information hiding in these domains.

Finally, quantization schemes such as QIM or SCS have been a state-of-art technique for information hiding schemes. Many previous works claimed that they greatly improve the performance of information hiding systems. To my viewpoint, this is quite questionable. As to now, most of these works concentrate on a theoretical analysis of their performances. And their advantage over spread spectrum schemes are based on the assumption that the distortion is measured by MSE. Furthermore, it is difficult to implement the perceptual analysis in the quantization schemes to achieve the maximum allowable embedding distortion. An interesting problem is that even without the perceptual analysis, the quantization schemes are not as superior as claimed. For instance, QIM's embedding rule can be formulated as

$$s_i = x_i + \left[q_b(\overline{x}) - \overline{x}\right] w_i \,. \tag{9.3}$$

Suppose that the maximum embedding strength for the host data is $a_{\max}$. Thus, if $|q_b(\overline{x}) - \overline{x}|$ is larger than $a_{\max}$, QIM will result in the perceptible distortion. However, if $|q_b(\overline{x}) - \overline{x}|$ is smaller than $a_{\max}$, QIM will leave a perceptual allowance for the attacker to mount a stronger attacks on the watermarked contents. However, it is quite easy for the spread spectrum schemes to achieve the maximum embedding strength. The perceptual analysis for the spread spectrum schemes can be simply implemented as

$$s_i = x_i + a_{\max} m_i w_i \,. \tag{9.4}$$

From this equation, we may also incorporate the perceptual analysis into the quantization schemes. To see how it works, we first substitute $a$ with $a_{\max}$ in the above equation

$$s_i = x_i + am_i w_i .$$

(9.5)

Projection both sides on **w**, we obtain

$$\overline{s} = \overline{x} + a \frac{1}{N} \sum_{i=1}^{N} m_i .$$

(9.6)

Let $\overline{s} = q_b(\overline{x})$, we immediately have

$$a = \frac{q_b(\overline{x}) - \overline{x}}{\left( \sum_{i=1}^{N} m_i \right) \big/ N}$$

(9.7)

In fact, the above idea is formulated in a recent paper [119]. However, such an implementation of the perceptual analysis has an inherent problem that discourages its use in the real scenarios. In the above equation, $a$ should be smaller than $a_{max}$ to keep the distortion imperceptible. However, if it is smaller than $a_{max}$, it leaves a perceptual allowance for the attacker. Thus, it is still hard to say whether quantization schemes really surpass the spread spectrum schemes in performance. **It remains a very challenging future research direction to compare quantization schemes with spread spectrum schemes.**

# Appendix A

## A.A  Lemma A.1

Lemma A.1. Let $Z$ be a GGD random variable with a pdf given by (2.7). Then $|Z|^c$ has a Gamma distribution with shape parameter $1/c$ and scale parameter $1/\beta^c$.

Proof: Let $U = |Z|^\gamma$ $(\gamma > 0)$. Thus, for $u > 0$,

$$P(U \le u) = p(|Z|^\gamma \le u) = p(-u^{1/\gamma} \le Z \le u^{1/\gamma}) = F_Z(u^{1/\gamma}) - F_Z(-u^{1/\gamma}),$$

where $F_Z$ is the cumulative density function of $Z$. Therefore,

$$
\begin{aligned}
f_U(u) &= \frac{d}{du} P(U \le u) = \frac{1}{\gamma} u^{\frac{1}{\gamma}-1} f_Z(u^{\frac{1}{\gamma}}) + \frac{1}{\gamma} u^{\frac{1}{\gamma}-1} f_Z(-u^{\frac{1}{\gamma}}) \\
&= \frac{1}{\gamma} u^{\frac{1}{\gamma}-1} A \cdot \exp(-|\beta u^{1/\gamma}|^c) + \frac{1}{\gamma} u^{\frac{1}{\gamma}-1} A \cdot \exp(-|-\beta u^{1/\gamma}|^c) \\
&= \frac{2A}{\gamma} u^{\frac{1}{\gamma}-1} \exp(-\beta^c u^{c/\gamma})
\end{aligned}
$$

If $\gamma = c$, $U$ has a Gamma distribution with shape parameter $1/c$ and scale parameter $1/\beta^c$.                                     ∎

## A.B  Corollary A.1

Corollary A.1. Let $Z_1$, $Z_2$, …, $Z_N$ be i.i.d. random variables with pdf given by (2.7). Then $|Z_1|^c + |Z_2|^c + \ldots + |Z_N|^c$ has a Gamma distribution with shape parameter $N/c$ and scale parameter $1/\beta^c$.

Proof: Let $U_i = |Z_i|^c$. Since $U_1$, $U_2$, …, $U_N$ are Gamma random variables (by Lemma A.1), it is thus easy to derive this corollary by using the moment generating function of a sum of independent Gamma random variables (see Example 4.6.8 at page 183 in [107]).                                     ∎

## A.C  Lemma A.2

Lemma A.2. If $Y$ and $Z$ be two independent random variables with $Y \sim$ Gamma $(\theta_1, \delta)$ and $Z \sim$ Gamma $(\theta_2, \delta)$, then $Y/(Y+Z)$ and $Y+Z$ are independent.

Proof: Let $U = Y/(Y+Z)$, and $V = Y+Z$. Therefore, $Y = UV$ and $Z = V - VU$. As a result, the determinant of the

Jacobian matrix is

$$J = \begin{vmatrix} V & U \\ -V & 1-U \end{vmatrix} = V - UV + UV = V .$$

Hence,

$$f_{U,V}(u,v) = |J| f_{Y,Z}(uv, v-uv) = v f_Y(uv) f_Z(v-uv) = \frac{u^{\theta_1 - 1}(1-u)^{\theta_2 - 1}}{\Gamma(\theta_1)\Gamma(\theta_2)\delta^{\theta_1}\delta^{\theta_2}} v^{\theta_1 + \theta_2 - 1} e^{-v/\delta}$$

Thus, $f_{U,V}(u, v)$ can be factored into a function only of $u$ and a function only of $v$. Therefore, $U$ and $V$ are independent (by Lemma 4.2.7 at page 153 in [107]).                                                  ∎

## A.D  Corollary A.2

Corollary A.2. Let $Z_1$, $Z_2$, …, and $Z_N$ be i.i.d. random variables with $Z_i \sim$ Gamma($\theta$, $\delta$). Then $Z_i/(Z_1+Z_2+\ldots+Z_N)$ and $(Z_1+Z_2+\ldots+Z_N)$ are independent.

Proof: Since $Z_i \sim$ Gamma($\theta$, $\delta$) and $Z_1+\ldots+Z_{i-1}+Z_{i+1}+\ldots+Z_N \sim$ Gamma($(N-1)\theta$, $\delta$) (see Example 4.6.8 at page 183 in [107]), $Z_i/(Z_1+Z_2+\ldots+Z_N)$ and $(Z_1+Z_2+\ldots+Z_N)$ are independent by Lemma A.2.                 ∎

## A.E  Theorem A.1

Theorem A.1. Let $Z_1$, $Z_2$, … , and $Z_N$ be i.i.d. random variables, and $Z_i / (Z_1+Z_2+\ldots+Z_N)$ be independent of $(Z_1+Z_2+\ldots+Z_N)$. Then for any $k$ distinct $i_1, i_2, \cdots, i_k \in \{1, 2, \cdots, N\}$,

$$E\left[\frac{Z_{i_1}^{n_1} Z_{i_2}^{n_2} \ldots Z_{i_k}^{n_k}}{\left(\sum_{i=1}^{N} Z_i\right)^m}\right] = \frac{E\left[Z_{i_1}^{n_1} Z_{i_2}^{n_2} \ldots Z_{i_k}^{n_k}\right] E\left[\left(\sum_{i=1}^{N} Z_i\right)^{\sum_{j=1}^{k} n_j - m}\right]}{E\left[\left(\sum_{i=1}^{N} Z_i\right)^{\sum_{j=1}^{k} n_j}\right]}$$

Proof:

$$E\left[\frac{Z_{i_1}^{n_1} Z_{i_2}^{n_2} \dots Z_{i_k}^{n_k}}{\left(\sum_{i=1}^{N} Z_i\right)^m}\right]$$

$$= E\left[\frac{Z_{i_1}^{n_1} Z_{i_2}^{n_2} \dots Z_{i_k}^{n_k} \left(\sum_{i=1}^{N} Z_i\right)^{n_1+n_2+\dots+n_k-m}}{\left(\sum_{i=1}^{N} Z_i\right)^{n_1+n_2+\dots+n_k}}\right]$$

$$= E\left[\left(\frac{Z_{i_1}}{\sum_{i=1}^{N} Z_i}\right)^{n_1} \left(\frac{Z_{i_2}}{\sum_{i=1}^{N} Z_i}\right)^{n_2} \cdots \left(\frac{Z_{i_k}}{\sum_{i=1}^{N} Z_i}\right)^{n_k} \left(\sum_{i=1}^{N} Z_i\right)^{n_1+n_2+\dots+n_k-m}\right]$$

$$\overset{(a)}{=} E\left[\left(\frac{Z_{i_1}}{\sum_{i=1}^{N} Z_i}\right)^{n_1} \left(\frac{Z_{i_2}}{\sum_{i=1}^{N} Z_i}\right)^{n_2} \cdots \left(\frac{Z_{i_k}}{\sum_{i=1}^{N} Z_i}\right)^{n_k}\right] E\left[\left(\sum_{i=1}^{N} Z_i\right)^{n_1+n_2+\dots+n_k-m}\right]$$

$$\tag{A.1}$$

where ($a$) follows from the fact that $Z_i / (Z_1+Z_2+\dots+Z_N)$ is independent of $(Z_1+Z_2+\dots+Z_N)$ and functions of independent random variables are still independent. The first expectation in ($a$) can be evaluated by

$$E\left[Z_{i_1}^{n_1} Z_{i_2}^{n_2} \dots Z_{i_k}^{n_k}\right]$$

$$= E\left[\left(\frac{Z_{i_1}}{\sum_{i=1}^{N} Z_i}\right)^{n_1} \left(\frac{Z_{i_2}}{\sum_{i=1}^{N} Z_i}\right)^{n_2} \cdots \left(\frac{Z_{i_k}}{\sum_{i=1}^{N} Z_i}\right)^{n_k} \left(\sum_{i=1}^{N} Z_i\right)^{n_1+n_2+\dots+n_k}\right]$$

$$= E\left[\left(\frac{Z_{i_1}}{\sum_{i=1}^{N} Z_i}\right)^{n_1} \left(\frac{Z_{i_2}}{\sum_{i=1}^{N} Z_i}\right)^{n_2} \cdots \left(\frac{Z_{i_k}}{\sum_{i=1}^{N} Z_i}\right)^{n_k}\right] E\left[\left(\sum_{i=1}^{N} Z_i\right)^{n_1+n_2+\dots+n_k}\right]$$

$$\tag{A.2}$$

Plugging (A.2) into (A.1), we get Theorem A.1.                                                        ∎

## A.F  Proof of (4.6)

Let $Z_i = |X_i|^\gamma / \sum_{1 \le j \le N} |X_j|^\gamma$. Since $X_i$s are all identically distributed, $Z_i$s are all identically distributed. Therefore, from (2.2), we have

$$E(\eta) = \sum_{i=1}^{N} w_i E(Z_i) = 0 \tag{A.3}$$

$$\sigma_\eta^2 = E(\eta^2) = E\left[\sum_{i=1}^{N} Z_i^2 w_i^2 + \sum_{i=1}^{N} \sum_{j=1, j \ne i}^{N} Z_i Z_j w_i w_j\right] \overset{(a)}{=} N E(Z_1^2) - N E(Z_1 Z_2) \tag{A.4}$$

where ($a$) follows from the fact that all $Z_i$s are identically distributed, and that in the second term the number of negative $w_i w_j$ is ($N^2/2$) and that number for the positive $w_i w_j$ is ($N^2/2-N$). In order to evaluate (A.4), we consider two cases, the first case being $\gamma = c$ and the second $\gamma \neq c$.

## (1) Case $\gamma = c$

By Lemma A.1, Corollary A.2 and Theorem A.1 (with $k = 1$, $n_1 = 2$ and $m = 2$), we have

$$E(Z_1^2) = E\big(|X_1|^{2\gamma}\big) \Big/ E\left[\left(\sum_{i=1}^{N}|X_i|^{\gamma}\right)^2\right] \tag{A.5}$$

Similarly, with $k = 2$, $n_1 = 1$, $n_2 = 1$ and $m = 2$, we obtain

$$E(Z_1 Z_2) = E\big(|X_1|^{\gamma}|X_2|^{\gamma}\big) \Big/ E\left[\left(\sum_{i=1}^{N}|X_i|^{\gamma}\right)^2\right] \tag{A.6}$$

Therefore, substituting (A.5) and (A.6) into (A.4), we get

$$\sigma_{\eta}^2 = \frac{\text{var}(|X|^{\gamma})}{\text{var}(|X|^{\gamma}) + N[E(|X|^{\gamma})]^2} \approx \frac{\text{var}(|X|^{\gamma})}{N[E(|X|^{\gamma})]^2} \tag{A.7}$$

## (2) Case $\gamma \neq c$

The above derivation works for $\gamma = c$. For the case $\gamma \neq c$, Lemma A.2 does not hold since $|X|^{\gamma}$ does not have a Gamma distribution. Thus, we resort to an approximation approach to calculate $E(X/Y)$. Usually it is difficult to calculate $E(X/Y)$. However, if $Y$ is a function of $N$ independent random variables and converges to a constant $C$ in probability as $N$ increases, then

$$E(X \,/\, Y) \rightarrow E(X \,/\, C) = E(X) \,/\, C \tag{A.8}$$

By the weak law of large numbers,

$$\sum_{i=1}^{N}|X_i|^{\gamma} \Big/ N \rightarrow E\big(|X|^{\gamma}\big) \tag{A.9}$$

Thus by Theorem 5.5.4 at page 233 in [107], we obtain

$$\Big(\sum_{i=1}^{N}|X_i|^{\gamma} \Big/ N\Big)^2 \rightarrow [E(|X|^{\gamma})]^2 \tag{A.10}$$

Therefore, with (A.8), we have

$$E(Z_1^2) = \frac{1}{N^2} E\left[\frac{|X_1|^{2\gamma}}{\left(\sum_{i=1}^{N}|X_i|^{\gamma} \Big/ N\right)^2}\right] \approx \frac{E[|X_1|^{2\gamma}]}{N^2[E(|X|^{\gamma})]^2} \tag{A.11}$$

$E(Z_1 Z_2)$ can be evaluated similarly. Finally, we find that (A.7) is still valid, however, in the approximation sense. The accuracy of this approximation method has also been verified through Monte-Carlo simulations (See Fig. 4.5(b) in Chapter 4 where the same approximation technique (A.8) is employed to calculate the embedding distortion (See Appendix B)).

# Appendix B

## B.A  Lemma B.1

Lemma B.1. Let $Z$ be a random variable distributed as Gamma $(\theta, \delta)$. Then $E[Z^r] = \delta^r \Gamma(r + \theta)/\Gamma(\theta)$.

Proof: $\int_0^{+\infty} x^r \dfrac{x^{\theta-1} e^{-x/\delta}}{\Gamma(\theta)\delta^\theta} dx = \dfrac{\Gamma(r+\theta)\delta^r}{\Gamma(\theta)} \int_0^{+\infty} \dfrac{x^{r+\theta-1} e^{-x/\delta}}{\Gamma(r+\theta)\delta^{r+\theta}} dx = \delta^r \Gamma(r+\theta)/\Gamma(\theta)$.  ∎

## B.B  Proof of (4.7)

Proof: $D_w = E[DT(\mathbf{S}, \mathbf{X})] \approx a^2 \sigma_X^2 + (\lambda^2/\gamma^2)\sigma_X^2 \sigma_\eta^2$  (B.1)

Proof: In order to prove (4.7) or (B.1), we disintegrate (B.1) into several parts and obtain

$$
\begin{aligned}
E[DT(\mathbf{S}, \mathbf{X})] &= \frac{1}{N} E\left[ \sum_{i=1}^N X_i^2 (a - \lambda \eta/\gamma)^2 \right] \\
&= \frac{1}{N} E\left[ \sum_{i=1}^N X_i^2 (a^2 - 2\lambda \eta a/\gamma + \lambda^2 \eta^2/\gamma^2) \right] \\
&= a^2 E[X^2] - \frac{2\lambda a}{N\gamma} E\left[ \eta \sum_{i=1}^N X_i^2 \right] + \frac{\lambda^2}{N\gamma^2} E\left[ \eta^2 \sum_{i=1}^N X_i^2 \right]
\end{aligned}
$$
(B.2)

In the above equation, it is easy to see that

$$
\begin{aligned}
E\left( \eta \sum_{i=1}^N X_i^2 \right) &= E\left[ \left( \sum_{i=1}^N |X_i|^\gamma w_i \Big/ \sum_{i=1}^N |X_i|^\gamma \right) \left( \sum_{i=1}^N X_i^2 \right) \right] \\
&= E\left[ \frac{\sum_{i=1}^N |X_i|^{2+\gamma} w_i + \sum_{i=1}^N \sum_{j=1, j\neq i}^N X_i^2 |X_j|^\gamma w_j}{\sum_{i=1}^N |X_i|^\gamma} \right] \\
&\overset{(a)}{=} 0
\end{aligned}
$$
(B.3)

where ($a$) follows from (2.2). In order to evaluate the third expectation in (B.2), we first expand it into several simple terms, and these terms can then be simplified into

$$E(\eta^2 \sum_{i=1}^{N} X_i^2) = E\left[\frac{(\sum_{i=1}^{N} X_i^2)(\sum_{i=1}^{N} |X_i|^\gamma w_i)^2}{(\sum_{i=1}^{N} |X_i|^\gamma)^2}\right]$$

$$= E\left[\frac{(\sum_{i=1}^{N} X_i^2)(\sum_{i=1}^{N} |X_i|^{2\gamma} w_i^2 + \sum_{i=1}^{N}\sum_{j=1, j\neq i}^{N} |X_i|^\gamma |X_j|^\gamma w_i w_j)}{(\sum_{i=1}^{N} |X_i|^\gamma)^2}\right]$$

$$\overset{(a)}{=} NE\left[\frac{|X_1|^{2+2\gamma}}{(\sum_{i=1}^{N} |X_i|^\gamma)^2}\right] + N(N-1)E\left[\frac{|X_1|^2 |X_2|^{2\gamma}}{(\sum_{i=1}^{N} |X_i|^\gamma)^2}\right]$$

$$- 2NE\left[\frac{|X_1|^{2+\gamma} |X_2|^\gamma}{(\sum_{i=1}^{N} |X_i|^\gamma)^2}\right] - N(N-2)E\left[\frac{|X_1|^2 |X_2|^\gamma |X_3|^\gamma}{(\sum_{i=1}^{N} |X_i|^\gamma)^2}\right]$$

(B.4)

where in the above equation $(a)$ we have used the fact that $X_i$s are identically distributed. Care must be taken to derive the above equation since some terms are cancelled due to (2.2). There are two cases to consider in evaluating (B.4), the first case being $\gamma = c$ and the second being $\gamma \neq c$. We first consider the case $\gamma = c$. Let $Z_i = |X_i|^\gamma$. By Lemma A.1, $Z_i$s are Gamma random variables. Thus, by Corollary A.2, $Z_i/(Z_1+Z_2+\ldots+Z_N)$ and $(Z_1+Z_2+\ldots+Z_N)$ are independent. As a result of Theorem A.1 (with $k = 1$, $n_1 = 2+2/\gamma$, $m = 2$), we have

$$E\left[\frac{|X_1|^{2+2\gamma}}{(\sum_{i=1}^{N} |X_i|^\gamma)^2}\right] = \frac{E(|X_i|^{2+2\gamma})E\left[(\sum_{i=1}^{N} |X_i|^\gamma)^{2/\gamma}\right]}{E\left[(\sum_{i=1}^{N} |X_i|^\gamma)^{2+2/\gamma}\right]}$$

(B.5)

We can similarly calculate the other three expectations in (B.4) and finally have

$$E\left[\eta^2 \sum_{i=1}^{N} X_i^2\right] = R \cdot F$$

(B.6)

where

$$R = E\left[(\sum |X_i|^\gamma)^{2/\gamma}\right] \Big/ E\left[(\sum |X_i|^\gamma)^{2+2/\gamma}\right]$$

(B.7)

$$F = NE(|X|^{2+2\gamma}) - 2NE(|X|^{2+\gamma})E(|X|^\gamma) + N(N-1)E(|X|^2)E(|X|^{2\gamma}) - N(N-2)E(|X|^2)[E(|X|^\gamma)]^2$$

(B.8)

Moreover, since $|X|^\gamma \sim Gamma(\theta, \delta)$ at $\gamma = c$, where $\theta = 1/c$, $\delta = 1/\beta^c$, Corollary A.1 and Lemma B.1 lead to

$$R = \frac{\Gamma(N/c + 2/c)}{\Gamma(N/c)\beta^2} \frac{\Gamma(N/c)\beta^{2+2c}}{\Gamma(N/c + 2/c + 2)} = \frac{\beta^{2c}}{(N/c + 2/c)(N/c + 2/c + 1)}$$

(B.9)

The above calculation is exact. In order to prove (B.1), with Lemma A.1 (or directly from (3.34)) and since $2/c << N/c$, we can approximate (B.9) by

$$R \approx 1/[N^2(1/c)^2(1/\beta^c)^2] = 1/\{N^2[E(|X|^\gamma)]^2\} \tag{B.10}$$

Moreover, by discarding some less important terms in (B.8), namely, the first and second terms, and approximating $(N-2)$ and $(N-1)$ by $N$, we get

$$F \approx N^2 E(|X|^2)E(|X|^{2\gamma}) - N^2 E(|X|^2)[E(|X|^\gamma)]^2 = N^2 E(|X|^2)\,\mathrm{var}(|X|^\gamma) \tag{B.11}$$

Substituting both (B.10) and (B.11) into (B.6) will lead to

$$E[\eta^2 \textstyle\sum_{i=1}^{N} X_i^2] \approx E(|X|^2)\,\mathrm{var}(|X|^\gamma)\big/[E(|X|^\gamma)]^2 = N\sigma_X^2\sigma_\eta^2 \tag{B.12}$$

Thus, (B.1) follows easily from the above arguments. The above derivation works for $\gamma = c$. For the case $\gamma \neq c$, Lemma A.2 does not hold since $|X|^\gamma$ does not have a Gamma distribution. Thus, we have to resort to the approximation approach given by (A.8). Hence, as we did for (A.11), we obtain

$$E\left[\frac{|X_1|^{2+2\gamma}}{(\sum_{i=1}^{N}|X_i|^\gamma)^2}\right] \approx \frac{E(|X_1|^{2+2\gamma})}{N^2[E(|X|^\gamma)]^2} \tag{B.13}$$

Other terms in (B.4) can be approximated similarly to reach the same result (B.12).

# Appendix C

In this appendix, we prove (4.14) and (4.15). It is easy to see from (2.2) that $E(\eta \sum_{i=1}^{N} |X_i|^\xi) = 0$ . Thus (4.14) is justified and

$$\sigma_1^2 = E\{[L(\mathbf{S} \mid H_1)]^2\} - a^2 \xi^2 [E(|X|^\xi)]^2 \tag{C.1}$$

We expand the first term in (C-1) (also see (4.9)) as

$$[L(\mathbf{S} \mid H_1)]^2 = T_1 + T_2 + T_3 + T_4 + T_5 + T_6 \tag{C.2}$$

where

$$T_1 = \frac{1}{N^2} (\sum_{i=1}^{N} |X_i|^\xi w_i)^2 \tag{C.3}$$

$$T_2 = \frac{\xi^2 a^2}{N^2} (\sum_{i=1}^{N} |X_i|^\xi)^2 \tag{C.4}$$

$$T_3 = \frac{\lambda^2 \xi^2}{N^2 \gamma^2} (\sum_{i=1}^{N} |X_i|^\xi)^2 (\sum_{i=1}^{N} |X_i|^\gamma w_i / \sum_{i=1}^{N} |X_i|^\gamma)^2 \tag{C.5}$$

$$T_4 = \frac{2\xi a}{N^2} (\sum_{i=1}^{N} |X_i|^\xi w_i)(\sum_{i=1}^{N} |X_i|^\xi) \tag{C.6}$$

$$T_5 = -\frac{2\lambda \xi^2 a}{N^2 \gamma} (\sum_{i=1}^{N} |X_i|^\xi)^2 (\sum_{i=1}^{N} |X_i|^\gamma w_i / \sum_{i=1}^{N} |X_i|^\gamma) \tag{C.7}$$

$$T_6 = -\frac{2\lambda \xi}{N^2 \gamma} (\sum_{i=1}^{N} |X_i|^\xi w_i)(\sum_{i=1}^{N} |X_i|^\xi) \frac{\sum_{i=1}^{N} |X_i|^\gamma w_i}{\sum_{i=1}^{N} |X_i|^\gamma} \tag{C.8}$$

By (2.2), it is easy to see that

$$E(T_4) = E(T_5) = 0 \tag{C.9}$$

Now we begin to evaluate the expectation of other terms in (C.2). Likewise we first consider the case $\gamma = c$. By simple algebraic operations, we obtain,

$$E(T_1) = \frac{1}{N} Var(|X|^\xi) \tag{C.10}$$

$$E(T_2) = \frac{\xi^2 a^2}{N^2} E\left[\left(\sum_{i=1}^{N} |X_i|^\xi\right)^2\right] = \frac{\xi^2 a^2}{N^2}\left\{Var\left(\sum_{i=1}^{N} |X_i|^\xi\right) + \left[E\left(\sum_{i=1}^{N} |X_i|^\xi\right)\right]^2\right\}$$

$$= \frac{\xi^2 a^2}{N} Var\left(|X|^\xi\right) + \xi^2 a^2\left[E\left(|X|^\xi\right)\right]^2 \quad \text{(C.11)}$$

In order to evaluate $E(T_3)$, we first have

$$E\left[\left(\frac{\sum_i |X_i|^\gamma w_i}{\sum_i |X_i|^\gamma}\right)^2\left(\sum_i |X_i|^\xi\right)^2\right] = E\left[\frac{(\sum_i |X_i|^{2\gamma} + \sum_i\sum_{j\neq i}|X_i|^\gamma |X_j|^\gamma w_i w_j)(\sum_i |X_i|^\xi)^2}{(\sum_i |X_i|^\gamma)^2}\right] = T_{3,1} + T_{3,2} + T_{3,3} \quad \text{(C.12)}$$

where

$$T_{3,1} = E\left[\frac{(\sum_i |X_i|^{2\gamma})(\sum_i |X_i|^\xi)^2}{(\sum_i |X_i|^\gamma)^2}\right] \quad \text{(C.13)}$$

$$T_{3,2} = E\left[\frac{(\sum_i\sum_{j\neq i}|X_i|^\gamma |X_j|^\gamma w_i w_j)(\sum_i |X_i|^{2\xi})}{(\sum_i |X_i|^\gamma)^2}\right] \quad \text{(C.14)}$$

$$T_{3,3} = E\left[\frac{(\sum_i\sum_{j\neq i}|X_i|^\gamma |X_j|^\gamma w_i w_j)(\sum_i\sum_{j\neq i}|X_i|^\xi |X_j|^\xi)}{(\sum_i |X_i|^\gamma)^2}\right] \quad \text{(C.15)}$$

Then, we disintegrate (C.13) into several parts, that is,

$$T_{3,1} = E\left[\frac{(\sum_i |X_i|^{2\gamma})(\sum_i |X_i|^{2\xi} + \sum_i\sum_{j\neq i}|X_i|^\xi |X_j|^\xi)}{(\sum_i |X_i|^\gamma)^2}\right]$$

$$= NE\left[\frac{|X_1|^{2\gamma+2\xi}}{(\sum_i |X_i|^\gamma)^2}\right] + N(N-1)E\left[\frac{|X_1|^{2\gamma}|X_2|^{2\xi}}{(\sum_i |X_i|^\gamma)^2}\right]$$

$$+ 2N(N-1)E\left[\frac{|X_1|^{2\gamma+\xi}|X_2|^\xi}{(\sum_i |X_i|^\gamma)^2}\right] + N(N-1)(N-2)E\left[\frac{|X_1|^{2\gamma}|X_2|^\xi |X_3|^\xi}{(\sum_i |X_i|^\gamma)^2}\right]$$

$$= F_{3,1} \cdot R_3 \quad \text{(C.16)}$$

where

$$F_{3,1} = NE\left(|X_1|^{2\gamma+2\xi}\right) + N(N-1)E\left(|X_1|^{2\gamma}|X_2|^{2\xi}\right) + 2N(N-1)E\left(|X_1|^{2\gamma}|X_2|^{\xi}\right)$$
$$+ N(N-1)(N-2)E\left(|X_1|^{2\gamma}|X_2|^{\xi}|X_3|^{\xi}\right) \tag{C.17}$$

$$R_3 = E\left[\left(\sum_{i=1}^N |X_i|^{\gamma}\right)^{2\xi/\gamma}\right] \Bigg/ E\left[\left(\sum_{i=1}^N |X_i|^{\gamma}\right)^{2+2\xi/\gamma}\right] \tag{C.18}$$

Similarly, we obtain

$$T_{3,2} = -2NE\left[\frac{|X_1|^{\gamma+2\xi}|X_2|^{\gamma}}{(\sum_i |X_i|^{\gamma})^2}\right] - N(N-2)E\left[\frac{|X_1|^{\gamma}|X_2|^{\gamma}|X_3|^{2\xi}}{(\sum_i |X_i|^{\gamma})^2}\right] = F_{3,2} \cdot R_3 \tag{C.19}$$

where

$$F_{3,2} = -2NE\left[|X_1|^{\gamma+2\xi}|X_2|^{\gamma}\right] - N(N-2)E\left[|X_1|^{\gamma}|X_2|^{\gamma}|X_3|^{2\xi}\right] \tag{C.20}$$

Finally, we have

$$T_{3,3} = F_{3,3} \cdot R_3 \tag{C.21}$$

where

$$F_{3,3} = -2NE\left(|X_1|^{\gamma+\xi}|X_2|^{\gamma+\xi}\right) - 4N(N-2)E\left(|X_1|^{\gamma+\xi}|X_2|^{\gamma}|X_3|^{\xi}\right)$$
$$- N(N-2)(N-3)E\left(|X_1|^{\gamma}|X_2|^{\gamma}|X_3|^{\xi}|X_4|^{\xi}\right) \tag{C.22}$$

Summarizing all the above arguments, we readily have

$$E\left[\left(\sum_{i=1}^N |X_i|^{\gamma} w_i \Big/ \sum_{i=1}^N |X_i|^{\gamma}\right)^2 \left(\sum_{i=1}^N |X_i|^{\xi}\right)^2\right] = F_3 \cdot R_3 = (F_{3,1} + F_{3,2} + F_{3,3}) \cdot R_3 \tag{C.23}$$

with

$$F_3 = NE\left(|X_1|^{2\gamma+2\xi}\right) + N(N-1)E\left(|X_1|^{2\gamma}|X_2|^{2\xi}\right) + 2N(N-1)E\left(|X_1|^{2\gamma+\xi}|X_2|^{\xi}\right)$$
$$+ N(N-1)(N-2)E\left(|X_1|^{2\gamma}|X_2|^{\xi}|X_3|^{\xi}\right) - 2NE\left(|X_1|^{\gamma+2\xi}|X_2|^{\gamma}\right) - 2NE\left(|X_1|^{\gamma+\xi}|X_2|^{\gamma+\xi}\right)$$
$$- N(N-2)E\left(|X_1|^{\gamma}|X_2|^{\gamma}|X_3|^{2\xi}\right) - 4N(N-2)E\left(|X_1|^{\gamma+\xi}|X_2|^{\gamma}|X_3|^{\xi}\right)$$
$$- N(N-2)(N-3)E\left(|X_1|^{\gamma}|X_2|^{\gamma}|X_3|^{\xi}|X_4|^{\xi}\right)$$
$$\approx N^3[E(|X|^{\xi})]^2\{E(|X|^{2\gamma}) - [E(|X|^{\gamma})]^2\} \tag{C.24}$$

Since $|X|^{\gamma} \sim Gamma(\theta,\delta)$ (By Lemma A.1), where $\theta = 1/c$, $\delta = 1/\beta^c$, Corollary A.1 and Lemma B.1

lead to

$$R_3 = \frac{\Gamma(N\theta + 2\xi/\gamma)\delta^{2\xi/\gamma}}{\Gamma(N\theta)} \frac{\Gamma(N\theta)}{\Gamma(N\theta + 2\xi/\gamma + 2)\delta^{2+2\xi/\gamma}}$$

$$= \frac{1}{\delta^2(N\theta + 2\xi/\gamma + 1)(N\theta + 2\xi/\gamma)} \approx \frac{1}{N^2\theta^2\delta^2} = \frac{1}{N^2[E(|X|^\gamma)]^2}$$

(C.25)

Therefore,

$$E(T_3) = \frac{\lambda^2\xi^2}{N^2\gamma^2} F_3 \cdot R_3$$

$$\approx \frac{\lambda^2\xi^2}{N^2\gamma^2} \frac{N^3 E(|X|^{2\gamma})[E(|X|^\xi)]^2 - N^3[E(|X|^\gamma)]^2[E(|X|^\xi)]^2}{N^2[E(|X|^\gamma)]^2}$$

$$\approx \frac{\lambda^2\xi^2}{N\gamma^2} \frac{[E(|X|^\xi)]^2 \operatorname{var}(|X|^\gamma)}{[E(|X|^\gamma)]^2}$$

(C.26)

Please note that in computing $T_{3,3}$, we have used the fact that

$$(\sum_i \sum_{j\neq i} |X_i|^\gamma |X_j|^\gamma w_i w_j)(\sum_i \sum_{j\neq i} |X_i|^\xi |X_j|^\xi)$$

$$= \sum_i \sum_{j\neq i} |X_i|^\gamma |X_j|^\gamma w_i w_j \sum_k \sum_{w\neq k} |X_k|^\xi |X_w|^\xi$$

$$= \sum_i \sum_{j\neq i} |X_i|^\gamma |X_j|^\gamma w_i w_j |X_i|^\xi |X_j|^\xi \quad (k=i, w=j)$$

$$+ \sum_i \sum_{j\neq i} |X_i|^\gamma |X_j|^\gamma w_i w_j |X_i|^\xi \sum_{w\neq i,j} |X_w|^\xi \quad (k=i, w\neq i,j)$$

$$+ \sum_i \sum_{j\neq i} |X_i|^\gamma |X_j|^\gamma w_i w_j |X_j|^\xi |X_i|^\xi \quad (k=j, w=i)$$

$$+ \sum_i \sum_{j\neq i} |X_i|^\gamma |X_j|^\gamma w_i w_j |X_j|^\xi \sum_{w\neq i,j} |X_w|^\xi \quad (k=j, w\neq i,j)$$

$$+ \sum_i \sum_{j\neq i} |X_i|^\gamma |X_j|^\gamma w_i w_j \sum_{k\neq i,j} |X_k|^\xi |X_i|^\xi \quad (k\neq i,j, w=i)$$

$$+ \sum_i \sum_{j\neq i} |X_i|^\gamma |X_j|^\gamma w_i w_j \sum_{k\neq i,j} |X_k|^\xi |X_j|^\xi \quad (k\neq i,j, w=j)$$

$$+ \sum_i \sum_{j\neq i} |X_i|^\gamma |X_j|^\gamma w_i w_j \sum_{k\neq i,j} |X_k|^\xi \sum_{w\neq i,j,k} |X_w|^\xi \quad (k\neq i,j, w\neq i,j,k)$$

(C.27)

The above combinations of sums can be expressed by the following graph to prevent errors occurring in the calculation process.

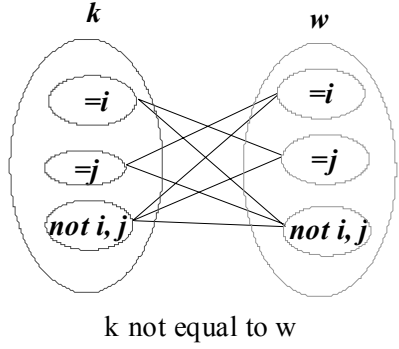

Fig. C.1. Matching graph in summation.

The above equation (C.27) can be simplified into

$$
(\sum_i \sum_{j \neq i} |X_i|^\gamma |X_j|^\gamma w_i w_j)(\sum_i \sum_{j \neq i} |X_i|^\xi |X_j|^\xi)
$$

$$
= \sum_i \sum_{j \neq i} |X_i|^{\gamma+\xi} |X_j|^{\gamma+\xi} w_i w_j + \sum_i \sum_{j \neq i} |X_i|^{\gamma+\xi} |X_j|^\gamma w_i w_j \sum_{w \neq i,j} |X_w|^\xi
$$

$$
+ \sum_i \sum_{j \neq i} |X_i|^{\gamma+\xi} |X_j|^{\gamma+\xi} w_i w_j + \sum_i \sum_{j \neq i} |X_i|^\gamma |X_j|^{\gamma+\xi} w_i w_j \sum_{w \neq i,j} |X_w|^\xi
$$

$$
+ \sum_i \sum_{j \neq i} |X_i|^{\gamma+\xi} |X_j|^\gamma w_i w_j \sum_{k \neq i,j} |X_k|^\xi + \sum_i \sum_{j \neq i} |X_i|^\gamma |X_j|^{\gamma+\xi} w_i w_j \sum_{k \neq i,j} |X_k|^\xi
$$

$$
+ \sum_i \sum_{j \neq i} |X_i|^\gamma |X_j|^\gamma w_i w_j \sum_{k \neq i,j} |X_k|^\xi \sum_{w \neq i,j,k} |X_w|^\xi
$$

(C.28)

Now we begin to compute $E(T_6)$. As we did to calculate $E(T_3)$, we have

$$
E\left[ (\sum |X_i|^\xi w_i)(\sum |X_i|^\xi)(\sum |X_i|^\gamma w_i / \sum |X_i|^\gamma) \right]
$$

$$
= E\left[ \frac{(\sum_i |X_i|^{\xi+\gamma} w_i^2 + \sum_i \sum_{j \neq i} |X_i|^\xi |X_j|^\xi w_i w_j)(\sum_i |X_i|^\xi)}{\sum_i |X_i|^\gamma} \right]
$$

$$
= E\left[ \frac{\sum_i |X_i|^{2\xi+\gamma} + \sum_i \sum_{j \neq i} |X_i|^{\xi+\gamma} |X_j|^\xi}{\sum_i |X_i|^\gamma} \right]
$$

$$
+ E\left[ \frac{\sum_i \sum_{j \neq i} |X_i|^\xi |X_j|^\gamma w_i w_j (|X_i|^\xi + |X_j|^\xi) + \sum_i \sum_{j \neq i} \sum_{k \neq i,j} |X_i|^\xi |X_j|^\gamma |X_k|^\xi w_i w_j}{\sum_i |X_i|^\gamma} \right]
$$

(C.29)

Thus,

$$
E(T_6) = -\frac{2\lambda_5 \xi}{N^2 \gamma} F_6 \cdot R_6
$$

(C.30)

where

$$F_6 = NE(|X|^{2\xi+\gamma}) + N(N-1)E(|X|^{\xi+\gamma})E(|X|^\xi) - NE(|X|^{2\xi})E(|X|^\gamma)$$
$$\quad - NE(|X|^\xi)E(|X|^{\gamma+\xi}) - N(N-2)E(|X|^\xi)E(|X|^\xi)E(|X|^\gamma)$$
$$\approx N^2 E(|X|^\xi)[E(|X|^{\gamma+\xi}) - E(|X|^\xi)E(|X|^\gamma)] \tag{C.31}$$

$$R_6 = E\left[ \left(\sum_{i=1}^{N}|X_i|^\gamma\right)^{2\xi/\gamma} \right] \Big/ E\left[ \left(\sum_{i=1}^{N}|X_i|^\gamma\right)^{1+2\xi/\gamma} \right] \tag{C.32}$$

Since $|X|^\gamma \sim Gamma(\theta, \delta)$ (By Lemma A.1), where $\theta = 1/c, \quad \delta = 1/\beta^c$, Corollary A.1 and Lemma B.1 lead to

$$R_6 = \frac{\Gamma(N\theta + 2\xi/\gamma)\delta^{2\xi/\gamma}}{\Gamma(N\theta)} \frac{\Gamma(N\theta)}{\Gamma(N\theta + 2\xi/\gamma + 1)\delta^{1+2\xi/\gamma}}$$
$$= \frac{1}{\delta(N\theta + 2\xi/\gamma + 1)} \approx \frac{1}{N\theta\delta} \approx \frac{1}{NE(|X|^\gamma)} \tag{C.33}$$

We readily have

$$E(T_6) \approx -\frac{2\lambda\xi}{N\gamma} \frac{E(|X|^\xi)[E(|X|^{\gamma+\xi}) - E(|X|^\xi)E(|X|^\gamma)]}{E(|X|^\gamma)} \tag{C.34}$$

Combining all above arguments, we obtain the desired result (4.15). For the case $\gamma \neq c$, we resort to the approximation approach given by (A.8) and obtain the same result.

# Appendix D

## D.A  Proof of Lemma 6.1

Lemma 6.1. If $U$ is a random variable, and $Z$ is defined by $Z = U + a$ if $U > 0$ and $Z = U - a$ if $U \leq 0$, where $a$ is a const. Then $Z$ has a pdf given by

$$f_Z(z) = \begin{cases} f_U(z-a), & \text{if } z > a \\ 0, & \text{if } -a < z \leq a \\ f_U(z+a), & \text{if } z \leq -a \end{cases} \tag{D.1}$$

Proof:  Let $I$ be an indicator defined as $I = 1$ if $U > 0$ and $I = 0$ if $U \leq 0$. Thus we obtain

$$f_{U|I=1}(u \mid I=1) = \frac{d}{du} P(U \leq u \mid I=1) = \frac{d}{du}[P(U \leq u, I=1)/P(I=1)] = \frac{1}{P(I=1)} \cdot \frac{d}{du} P(0 < U \leq u)$$

$$= f_U(u)/P(I=1), \quad \text{where } u > 0 \tag{D.2}$$

Similarly, we have $f_{U|I=0}(u \mid I=0) = f_U(u)/P(I=0)$, where $u \leq 0$. Since $(Z \mid I=1) = [(U+a) \mid I=1] = (U \mid I=1) + a$ and $(Z \mid I=0) = (U \mid I=0) - a$, we have

$$f_{Z|I=1}(z \mid I=1) = \begin{cases} f_U(z-a)/P(I=1), & \text{if } z > a \\ 0 & \text{if } z \leq a \end{cases} \tag{D.3}$$

$$f_{Z|I=0}(z \mid I=0) = \begin{cases} f_U(z+a)/P(I=0), & \text{if } z \leq -a \\ 0 & \text{if } z > -a \end{cases} \tag{D.4}$$

Since $f_Z(z) = f_{Z|I=0}(z \mid I=0)P(I=0) + f_{Z|I=1}(z \mid I=1)P(I=1)$, (D.3) and (D.4) thus lead to (D.1).    ∎

## D.B  Proof of lemma 6.2

Lemma 6.2. For any real $k > 0$, $Q(z+k)/Q(z)$ is a decreasing function of $z$.

Proof: Let $u(z) = Q(z+k)/Q(z)$. Thus the first derivative is

$$u'(z) = \frac{Q'(z+k)Q(z) - Q(z+k)Q'(z)}{Q^2(z)} = \frac{-\exp[-(z+k)^2/2]Q(z) + Q(z+k)\exp(-z^2/2)}{\sqrt{2\pi}Q^2(z)}$$

Now we prove that $-\exp[-(z+k)^2/2]Q(z) + Q(z+k)\exp(-z^2/2) < 0$, that is,

$$\exp[-(z+k)^2/2]/Q(z+k) > \exp(-z^2/2)/Q(z) \tag{D.5}$$

Let $g(z) = \exp(-z^2/2)/Q(z)$. Thus the first derivative of $g(z)$ is

$$g'(z) = \frac{-z\exp(-z^2/2)Q(z) - \exp(-z^2/2)Q'(z)}{Q^2(z)} = \frac{-z\exp(-z^2/2)Q(z) + \exp(-z^2/2)\exp(-z^2/2)/\sqrt{2\pi}}{Q^2(z)}$$

$$= \frac{\exp(-z^2/2)[-zQ(z) + \exp(-z^2/2)/\sqrt{2\pi}]}{Q^2(z)}$$

$$> 0$$

since $Q(z) < \exp(-z^2/2)/(z\sqrt{2\pi})$ for $z > 0$ (see Equation (71) at page 39 in [105]). Therefore $g(z)$ is an increasing function of $z$ and (D.5) holds. Thus, $u'(z) < 0$ and $u(z)$ is a decreasing function of $z$ for any given positive $k$.                                                                                                                  ∎

### D.C  Proof of Lemma 6.3

Lemma 6.3: Prove that $\sum_{1 \le i \le N} |X_i|^\xi$ is independent of the indicator $I$, where $I$ is defined in (6.5).

Proof: Let $T = \sum_{1 \le i \le N} |X_i|^\xi$. Since $N$ is an even integer, we may suppose that $N = 2m$, where $m$ is also an integer. For the convenience of the proof, let $w_1 = w_2 = \ldots = w_m = 1$, and $w_{m+1} = w_{m+2} = \ldots = w_{2m} = -1$. For any given $t \ge 0$, let $\Omega_1 = \{(x_1, x_2, \ldots, x_N): \bar{x} > 0 \text{ and } \sum_{1 \le i \le N} |x_i|^\xi < t\}$ and $\Omega_0 = \{(x_1, x_2, \ldots, x_N): \bar{x} < 0 \text{ and } \sum_{1 \le i \le N} |x_i|^\xi < t\}$, where $\bar{x}$ is defined in (6.3). Since all $X_i$s are independent, we have

$$P(T = \sum_{i=1}^{N} |X_i|^\xi < t, I = 1) = \int_{\Omega_1} f_{\mathbf{X}}(\mathbf{x})d\mathbf{x} = \int_{\Omega_1} f_X(x_1)f_X(x_2)\cdots f_X(x_N)dx_1 dx_2 \cdots dx_N \qquad \text{(D.6)}$$

$$P(T = \sum_{i=1}^{N} |X_i|^\xi < t, I = 0) = \int_{\Omega_0} f_{\mathbf{X}}(\mathbf{x})d\mathbf{x} = \int_{\Omega_0} f_X(x_1)f_X(x_2)\cdots f_X(x_N)dx_1 dx_2 \cdots dx_N \qquad \text{(D.7)}$$

Define a permutation $\Phi$ from $\Omega_1$ to $\Omega_0$ such that $y_1 = x_{m+1}$, $y_2 = x_{m+2}$, …, $y_m = x_{2m}$, $y_{m+1} = x_1$, $y_{m+2} = x_2$, …, $y_{2m} = x_{m+1}$, where $(x_1, x_2, \ldots, x_{2m}) \in \Omega_1$ and $(y_1, y_2, \ldots, y_{2m}) \in \Omega_0$. It is well defined since $\bar{y} = (\sum_{1 \le i \le N} y_i w_i)/N = (\sum_{1 \le i \le m} y_i - \sum_{m+1 \le i \le 2m} y_i)/N = (\sum_{m+1 \le i \le 2m} x_i - \sum_{1 \le i \le m} x_i)/N < 0$ and $\sum_{1 \le i \le N} |y_i|^\xi = \sum_{1 \le i \le N} |x_i|^\xi < t$. Moreover, $\Phi$ is one-to-one and onto for it is a simple permutation. Since the absolute value of the Jacobian $|\partial(x_1, x_2, \ldots, x_N)/\partial(y_1, y_2, \ldots, y_N)|$ is 1, the change of variables theorem (See page 460 in [106]) results in

$$\int_{\Omega_0} f_X(x_1)f_X(x_2)\cdots f_X(x_{2m})dx_1 dx_2 \cdots dx_{2m}$$

$$= \int_{\Omega_1} f_X(y_{m+1})\cdots f_X(y_{2m})f_X(y_1)\cdots f_X(y_m)dy_1 dy_2 \cdots dy_{2m}$$

$$= \int_{\Omega_1} f_X(x_1)f_X(x_2)\cdots f_X(x_{2m})dx_1 dx_2 \cdots dx_{2m}$$

Therefore, we have $P(T < t, I = 1) = P(T < t, I = 0)$ from (D.6) and (D.7). Since $P(T < t, I = 1) + P(T < t, I = $

$0) = P(T < t)$, $P(T < t, I = 1) = P(T < t, I = 0) = 0.5 \cdot P(T < t)$. Since $P(I = 1) = P(I = 0) = 0.5$, we have $P(T < t, I = 1) = P(I = 1)P(T < t)$ and $P(T < t, I = 0) = P(I = 0)P(T < t)$. Thus, $T$ is independent of $I$.                    ∎

# Appendix E   Important notes for experiments in this work

In the literature, there are two kinds of setups for experiments to evaluate the detector's performance. In the first setup (adopted in this work), $w_i$ is fixed and also satisfies the restriction (2.2). In the second setup, the watermark sequence is taken as a sequence of independent Bernoulli random variables with $P(w_i = 1) = P(w_i = -1) = 0.5$.

## E.A  Case 1

First we consider the first setup where (2.2) is satisfied. For the convenience of discussions, $p_{m,1}^{\mathbf{X}}$ stands for the probability of miss for the case where $\mathbf{X}$ is a random vector and $\mathbf{w}$ is fixed; $p_{m,1}^{\mathbf{X},\mathbf{W}}$ for the case where both $\mathbf{X}$ and $\mathbf{W}$ are random; $p_{m,1}^{\mathbf{W}}$ for the case where only $\mathbf{W}$ is random. Since for most cases in this work, $p_{m,1}^{\mathbf{X}}$ does not depend on $\mathbf{w}$ (due to (2.2)), we obtain

$$
\begin{aligned}
p_{m,1}^{\mathbf{X},\mathbf{W}} &= P\{L(\mathbf{S}\,|\,H_1) < \psi\} \\
&= \sum_{\mathbf{w}} P\{L(\mathbf{S}\,|\,H_1, \mathbf{W} = \mathbf{w}) < \psi\}P\{\mathbf{W} = \mathbf{w}\} \\
&= \sum_{\mathbf{w}} p_{m,1}^{\mathbf{X}} \cdot P\{\mathbf{W} = \mathbf{w}\} \\
&= p_{m,1}^{\mathbf{X}}
\end{aligned}
\tag{E.1}
$$

$$
\begin{aligned}
p_{m,1}^{\mathbf{X},\mathbf{W}} &= P\{L(\mathbf{S}\,|\,H_1) < \psi\} \\
&= \int_{\mathbf{x}} \left\{ \sum_{\mathbf{w}} P[L(\mathbf{S}\,|\,H_1, \mathbf{W} = \mathbf{w}, \mathbf{X} = \mathbf{x}) < \psi]P(\mathbf{W} = \mathbf{w}) \right\} \cdot f_{\mathbf{X}}(\mathbf{x})d\mathbf{x} \\
&= \int_{\mathbf{x}} p_{m,1}^{\mathbf{W}} \cdot f_{\mathbf{X}}(\mathbf{x})d\mathbf{x}
\end{aligned}
\tag{E.2}
$$

Thus, we see that

$$
p_{m,1}^{\mathbf{X}} = \int_{\mathbf{x}} p_{m,1}^{\mathbf{W}} \cdot f_{\mathbf{X}}(\mathbf{x})d\mathbf{x}
\tag{E.3}
$$

Hence we know that $p_{m,1}^{\mathbf{W}}$ cannot be larger than $p_{m,1}^{\mathbf{X}}$ since if $p_{m,1}^{\mathbf{W}} > p_{m,1}^{\mathbf{X}}$, we would have

$$
\int_{\mathbf{x}} p_{m,1}^{\mathbf{W}} \cdot f_{\mathbf{X}}(\mathbf{x})d\mathbf{x} > \int_{\mathbf{x}} p_{m,1}^{\mathbf{X}} \cdot f_{\mathbf{X}}(\mathbf{x})d\mathbf{x} = p_{m,1}^{\mathbf{X}}
\tag{E.4}
$$

Similarly, $p_{m,1}^{\mathbf{W}}$ cannot be smaller than $p_{m,1}^{\mathbf{X}}$. Since $p_{m,1}^{\mathbf{W}}$ depends on a given $\mathbf{x}$, we thus conclude that for some

**x**s, $p_{m,1}^{\mathbf{W}}$ is larger than $p_{m,1}^{\mathbf{X}}$; for other **x**s, $p_{m,1}^{\mathbf{W}}$ is not larger than $p_{m,1}^{\mathbf{X}}$.

## E.B  Case 2

In this subsection, $p_{m,2}^{\mathbf{X}}$ stands for the probability of miss for the case where **X** is a random vector and **w** is fixed; $p_{m,2}^{\mathbf{X},\mathbf{W}}$ for the case where both **X** and **W** are random; $p_{m,2}^{\mathbf{W}}$ for the case where only **W** is random. However, in these definitions, **w** is not necessary to satisfy the restriction (2.2). Since (2.2) is not met, $p_{m,2}^{\mathbf{X}}$ does depend on **w** and thus $p_{m,2}^{\mathbf{X},\mathbf{W}} \neq p_{m,2}^{\mathbf{X}}$. However, we still have

$$p_{m,2}^{\mathbf{X},\mathbf{W}} = \int_{\mathbf{x}} p_{m,2}^{\mathbf{W}} \cdot f_{\mathbf{X}}(\mathbf{x}) d\mathbf{x} \tag{E.5}$$

Let $\Omega_2 = \{+1, -1\}^N$, namely all binary sequence of length $N$, and $\Omega_1 = \{ \mathbf{w}: \mathbf{w} \in \Omega_2$ and $\sum_{1 \leq i \leq N} w_i = 0 \}$. It is easy to see that $\Omega_1 \subset \Omega_2$ and therefore

$$\begin{aligned} p_{m,2}^{\mathbf{W}} &= \sum_{\mathbf{w} \in \Omega_2} P[L(\mathbf{S} \mid H_1, \mathbf{W} = \mathbf{w}, \mathbf{X} = \mathbf{x}) < \psi] P(\mathbf{W} = \mathbf{w}) \\ &> \sum_{\mathbf{w} \in \Omega_1} P[L(\mathbf{S} \mid H_1, \mathbf{W} = \mathbf{w}, \mathbf{X} = \mathbf{x}) < \psi] P(\mathbf{W} = \mathbf{w}) \\ &= p_{m,1}^{\mathbf{W}} \end{aligned} \tag{E.6}$$

Similarly, we have

$$p_{m,2}^{\mathbf{X},\mathbf{W}} = \int_{\mathbf{x}} p_{m,2}^{\mathbf{W}} \cdot f_{\mathbf{X}}(\mathbf{x}) d\mathbf{x} > \int_{\mathbf{x}} p_{m,1}^{\mathbf{W}} \cdot f_{\mathbf{X}}(\mathbf{x}) d\mathbf{x} = p_{m,1}^{\mathbf{X},\mathbf{W}} \tag{E.7}$$

## E.C  Comparisons between the first and second setup

We now compare the performance of the first setup $p_{m,1}^{\mathbf{X}}$ with that of the second setup $p_{m,2}^{\mathbf{W}}$. If $p_{m,1}^{\mathbf{W}} > p_{m,1}^{\mathbf{X}}$, since (E.6), we have

$$p_{m,2}^{\mathbf{W}} > p_{m,1}^{\mathbf{W}} > p_{m,1}^{\mathbf{X}}.$$

The second setup would produce a poorer performance. If $p_{m,1}^{\mathbf{W}} < p_{m,1}^{\mathbf{X}}$, then it is hard to compare $p_{m,2}^{\mathbf{W}}$ with $p_{m,1}^{\mathbf{X}}$. We now consider an average case. If **x** is fixed and **W** is random, then the decision statistic $L(\mathbf{S}|H_0)$ and $L(\mathbf{S}|H_1)$ is often expressed as a sum of $g_0(x_i, W_i)$, $g_1(x_i, W_i)$, that is

$$L(\mathbf{S}\,|\,H_0) = \frac{1}{N}\sum_i g_0(x_i, W_i), \quad L(\mathbf{S}\,|\,H_1) = \frac{1}{N}\sum_i g_1(x_i, W_i)\,.$$

For instance, the decision statistic in [69] is given by

$$g_0(x_i, W_i) = |x_i|^c - |x_i - a_i W_i|^c, \quad g_1(x_i, W_i) = |x_i + a_i W_i|^c - |x_i|^c\,.$$

Therefore, we have

$$m_0 = \frac{1}{2N}\sum_i [g_0(x_i, 1) + g_0(x_i, -1)]\,,$$

$$\sigma_0^2 = \frac{1}{N^2}\sum_i Var[g_0(x_i, W_i)] = \frac{1}{N^2}\sum_i E\{[g_0(x_i, W_i)]^2\} - \{E[g_0(x_i, W_i)]\}^2$$

$$= \sum_i \frac{[g_0(x_i, 1)]^2 + [g_0(x_i, -1)]^2}{2N^2} - \frac{[g_0(x_i, 1) + g_0(x_i, -1)]^2}{4N^2}$$

Thus by the weak law of large numbers, we have

$$m_0 \to \frac{E[g_0(X, 1)] + E[g_0(X, -1)]}{2}$$

$$\sigma_0^2 \to \frac{E\{[g_0(X, 1)]^2\} + E\{[g_0(X, -1)]^2\}}{2N} - \frac{E\{[g_0(X, 1) + g_0(X, -1)]^2\}}{4N}$$

that is, the mean and variance converge to fixed values in probability. Similarly, $m_1$ and $\sigma_1^2$ also converge to fixed values in probability. Therefore, if $N$ is large enough, $p_{m,2}^{\mathbf{W}}$ can be roughly taken as a fixed value (invariant to $\mathbf{x}$) with a large probability. It thus follows from (E.1) and (E.7) that

$$p_{m,1}^{\mathbf{X}} = p_{m,1}^{\mathbf{X},W} < p_{m,2}^{\mathbf{X},\mathbf{W}} = \int_{\mathbf{x}} p_{m,2}^{\mathbf{W}} \cdot f_{\mathbf{X}}(\mathbf{x})d\mathbf{x} \approx p_{m,2}^{\mathbf{W}} \qquad (E.8)$$

Therefore, this asserts that in most cases, the first setup would produce a better performance than the second setup does. That is, assuming that $\mathbf{x}$ is fixed and $\mathbf{W}$ random would underestimate the true performance of the detector. Since (2.2), we can similarly argue by the weak law of large numbers that $p_{m,1}^{\mathbf{W}}$ converges to a fixed value in probability and thus

$$p_{m,1}^{\mathbf{W}} \approx p_{m,1}^{\mathbf{X},\mathbf{W}} = p_{m,1}^{\mathbf{X}} \qquad (E.9)$$

### E.D  Hints for experiments on real images

In real scenarios, it is not possible to have a large database of host data with the same shape parameter and standard deviation. In this work, we instead permute the host signals to generate new test data $\mathbf{x}$s. For instance, if the host data are of size $M$, we randomly permute this data and take the previous $N$ data for experiments. It is important that $N$ should be smaller than $M$. The smaller the $N$, the better the effects we can achieve. If $M = N$, the permutation of host data is equivalent to the second setup that we are using different watermark sequences. In such a case, we know from (E.9) that using different watermark sequences can reflect the true performance $p_{m,1}^{\mathbf{x}}$ with a large probability. If however the watermark sequence $\mathbf{w}$ is a random sequence without the restraint (2.2) being satisfied, such an experimental setup is equivalent to the second setup discussed in the previous sections. Thus, due to (E.8), the experimental results cannot well predict the detector's theoretical performance.